\documentclass[reqno,11pt]{article}
\pdfoutput=1
\usepackage{jheppub}
\graphicspath{ {./graphs1/} }
\usepackage{epsfig}
\usepackage{amssymb}
\usepackage{verbatim}
\usepackage{amsmath}
\usepackage{hyperref}
\usepackage{dsfont}
\usepackage{slashed}
\usepackage[dvipsnames]{xcolor}
\usepackage{epstopdf}
\usepackage{graphicx}
\def\o{\omega}

\def\p{\partial}
\def\l{\lambda}
\def\O{\Omega}
\def\L{\Lambda}

\def\s{\sigma}
\def\half{{1 \over 2}}
\def\a{\alpha}
\def\b{\beta}

\def\ss{\mathsf{s}}
\def\q{\mathsf{q}}
\newcommand{\be}{\begin{eqnarray}}
\newcommand{\en}{\end{eqnarray}}

\newcommand{\bc}{\begin{center}}
\newcommand{\ec}{\end{center}}

\title{CHY Loop Integrands from Holomorphic Forms}
\author[a,b]{Humberto Gomez,}
\author[b,c]{Sebastian Mizera,}
\author[b,c]{and Guojun Zhang}
\affiliation[a]{Facultad de Ciencias Basicas,  Universidad Santiago de Cali,\\
Calle 5 $N^\circ$  62-00 Barrio Pampalinda, Cali, Valle, Colombia}
\affiliation[b]{Perimeter Institute for Theoretical Physics,\\31 Caroline Street N, Waterloo, ON N2L 2Y5, Canada}
\affiliation[c]{Department of Physics \& Astronomy, University of Waterloo,\\ Waterloo, ON N2L 3G1, Canada}

\emailAdd{humgomzu@gmail.com, smizera, gzhang2@pitp.ca}

\abstract{Recently, the Cachazo--He--Yuan (CHY) approach for calculating scattering amplitudes has been extended beyond tree level. In this paper, we introduce a way of constructing CHY integrands for $\Phi^3$ theory up to two loops from holomorphic forms on Riemann surfaces. We give simple rules for translating Feynman diagrams into the corresponding CHY integrands. As a complementary result, we extend the $\L$-algorithm, originally introduced in \href{https://arxiv.org/abs/1604.05373}{arXiv:1604.05373}, to two loops. Using this approach, we are able to analytically verify our prescription for the CHY integrands up to seven external particles at two loops. In addition, it gives a natural way of extending to higher-loop orders.}

\begin{document}

\maketitle


\section{Introduction}\label{sect:intro}

On-shell methods for the computation of scattering amplitudes have been intensively studied during the last decade, since the seminal work of Witten \cite{Witten:2003nn} on the ${\cal N}=4$ super Yang--Mills theory. Among these methods, the Cachazo--He--Yuan (CHY)  prescription \cite{Cachazo:2013gna, Cachazo:2013hca,Cachazo:2013iaa,Cachazo:2013iea} stands out for being applicable in arbitrary dimension and, more importantly, for a large family of interesting theories, including scalars, gauge bosons, gravitons and mixing interactions among them \cite{Cachazo:2014xea, Cachazo:2014nsa, Cachazo:2016njl}. The proposal is to write the tree-level S-matrix in terms of integrals  localized over solutions of the so-called {\it scattering equations} \cite{Cachazo:2013gna} on the moduli space of $n$-punctured Riemann spheres. Other approaches that use the same moduli space include the Witten--RSV \cite{Witten:2003nn,Roiban:2004yf}, Cachazo--Geyer \cite{Cachazo:2012da}, and Cachazo--Skinner \cite{Cachazo:2012kg} constructions, but are special to four dimensions.

The CHY formalism has already been verified to reproduce well-known results, such as the soft limits of various theories \cite{Cachazo:2013hca}, the Kawai--Lewellen--Tye relations \cite{Kawai:1985xq} between gauge and gravity amplitudes \cite{Cachazo:2013gna}, as well as the correct Britto--Cachazo--Feng--Witten \cite{Britto:2005fq} recursion relations in Yang--Mills and bi-adjoint $\Phi^3$ theories \cite{Dolan:2013isa}.

Although the application of the prescription is quite straightforward, direct evaluation of the amplitudes for higher multiplicities has proven to be difficult. Several methods have been developed during the last year to deal with the integration over the Riemann sphere at the solutions of the scattering equations. These attempts include the study of solutions at particular kinematics and/or dimensions \cite{Cachazo:2013iea, Kalousios:2013eca, Lam:2014tga, Cachazo:2013iaa, Cachazo:2016sdc, He:2016vfi, Cachazo:2015nwa, Cachazo:2016ror}, encoding the solutions to the scattering equations in terms of linear transformations \cite{Kalousios:2015fya,Dolan:2014ega, Huang:2015yka, Cardona:2015ouc, Cardona:2015eba, Dolan:2015iln, Sogaard:2015dba, Bosma:2016ttj, Zlotnikov:2016wtk}, or the formulation of integration rules in terms of the polar structures \cite{Baadsgaard:2015ifa,Baadsgaard:2015voa,Huang:2016zzb,Cardona:2016gon}.

The CHY formalism has been generalized to loop level in different but equivalent ways. Using the ambitwistor string \cite{Mason:2013sva}, a proposal was made in \cite{Geyer:2015bja, Geyer:2015jch} which have been extended by the same authors to two loops very recently \cite{Geyer:2016wjx}. In \cite{Cachazo:2015aol,He:2015yua}, a parallel approach has been proposed, by performing a forward limit on the scattering equations for massive particles formulated previously in \cite{Naculich:2014naa, Dolan:2013isa} and a generalization of this approach to higher loops has been considered in \cite{Feng:2016nrf}. In addition, recent works at one-loop level have been published, where differential operators on the moduli space were developed  \cite{Chen:2016fgi,Chen:2017edo}.

One of the current authors made an independent proposal by generalizing the double-cover formulation, the so-called {\it $\Lambda$-algorithm}, made at tree level in \cite{Gomez:2016bmv} to the one-loop case by embedding the torus in a $\mathbb{CP}^2$ through an elliptic curve \cite{Cardona:2016bpi} and used it to reproduce the $\Phi^3$ theory at one loop \cite{Cardona:2016wcr}.

In this work, we study the CHY formulation for $\Phi^3$ theory up to two loops from a new perspective. We propose a construction for {\it CHY integrands} based on the holomorphic forms on Riemann surfaces. We show how it reproduces cubic Feynman diagrams up to two loops. 

Following the approach of \cite{Geyer:2015bja,Geyer:2015jch,Geyer:2016wjx}, at one loop we first consider the torus embedded in $\mathbb{CP}^2$, which can be described by an elliptic curve $y^2=z(z-1)(z-\lambda)$. The prescription of obtaining the correct field-theory limit, corresponding to the CHY formulation at one loop, is to consider the pinching of the torus. This yields a nodal Riemann sphere with two punctures, $\sigma_{\ell^+}$ and $\sigma_{\ell^-}$, identified. The two punctures correspond to the loop momentum $\ell$. The advantage of this approach is that one can work with similar objects as at tree level.

Using this prescription, one can consider reducing the holomorphic form $dz/y$ living on a torus to the following one-form on the nodal Riemann sphere,
\be
\o_\sigma\, d\s :=\left(\frac{1}{\s-\s_{\ell^+}}-\frac{1}{\s-\s_{\ell^-}}\right)d\s.
\en
We review how to obtain this geometrical object from pinching the A-cycle on a torus in section \ref{oneloop}. The one-form $\o_\sigma$ is an essential building block for CHY integrands of the symmetrized $n$-gon Feynman diagram \cite{Geyer:2015bja,Geyer:2015jch,He:2015yua,Cardona:2016bpi}. In order to satisfy the ${\rm PSL}(2,\mathbb{C})$ invariance, it enters the integrand as a quadratic
differential $\q_a := \o_{\sigma_a}^2$. More specifically, we have:
\be
\parbox[c]{14em}{\includegraphics[scale=0.25]{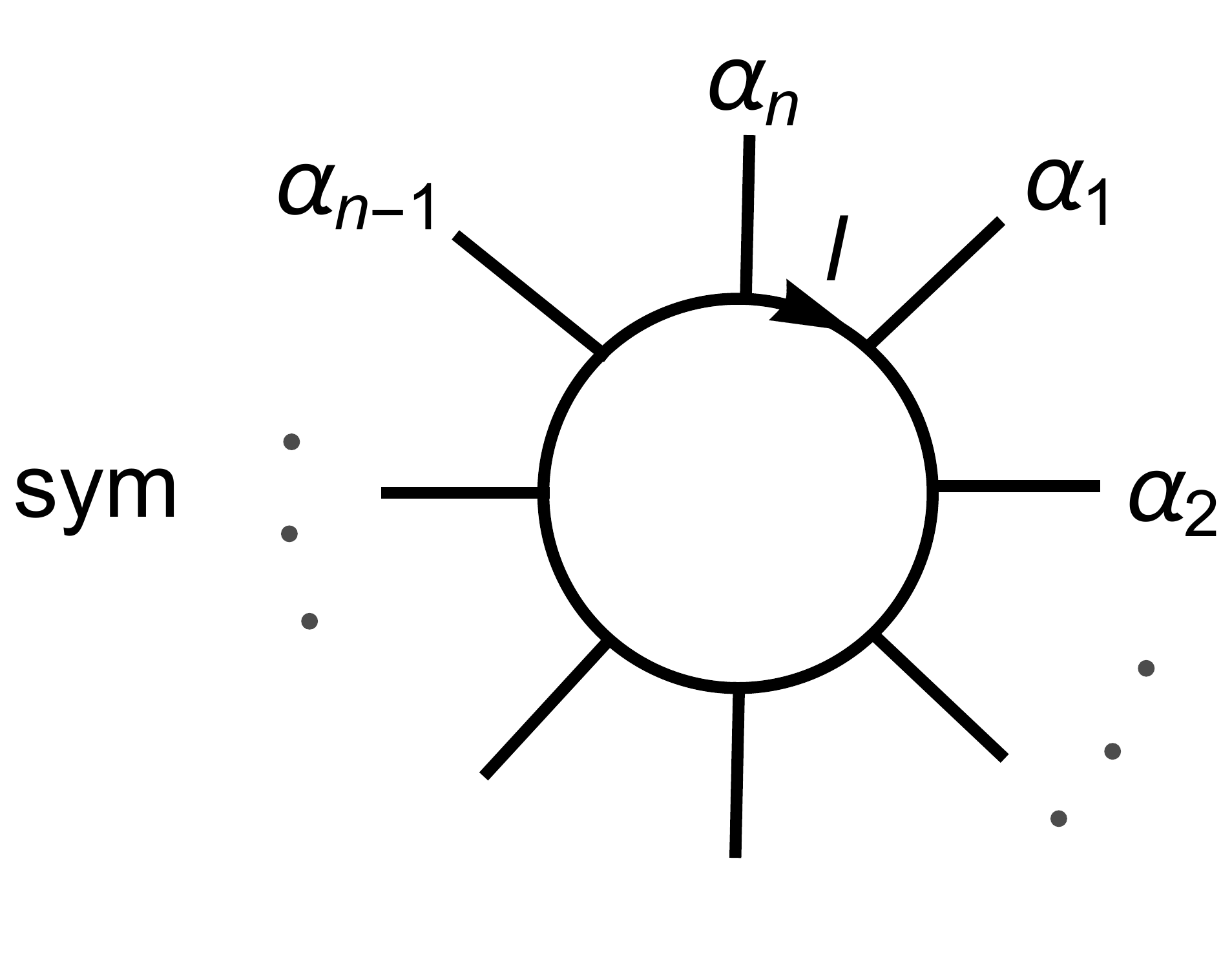}} \sim\quad \frac{1}{(\sigma_{\ell^+ \ell^-})^4}\, \prod_{i=1}^{n}\, \q_{\alpha_i}.\label{intro_1loop}
\en

Here, the right hand side represents a CHY integrand and the left hand side shows the corresponding Feynman diagram that such integrand computes. It is important to emphasize that the CHY integrals always compute the answer in the so-called \emph{Q-cut representation} \cite{Baadsgaard:2015twa}, which is equivalent to the standard Feynman diagram evaluation after using partial fraction identities and shifts of loop momenta. We illustrate this procedure with many examples throughout this work. In the above equations, symmetrization denoted by symbol $\mathsf{sym}$ means a sum over all permutations of external legs.

We propose a similar construction at two loops. The elliptic curve generalizes to the hyperelliptic curve $y^2 = (z-a_1) (z-a_2)(z-\l_1)(z-\l_2)(z-\l_3)$ embedded in $\mathbb{CP}^2$. On this hyperelliptic curve there are two global holomorphic forms,  which we have chosen to be $(z-a_1)\, dz/y$ and $(z-a_2)\, dz/y$, where $a_1\neq a_2$.  These objects induce two meromorphic forms over the sphere,
\be
\o^r_\sigma\, d\s :=\left(\frac{1}{\s-\s_{\ell_r^+}}-\frac{1}{\s-\s_{\ell_r^-}}\right)d\s, \qquad r=1,2,
\en
where we associate the punctures $\sigma_{\ell_1^+}$ and $\sigma_{\ell_1^-}$ with the loop momentum $\ell_1$, and similarly for the other momentum, $\ell_2$.

On a double torus, related with the hyperelliptic curve there are three A-cycles which are dependent on each other. We use the corresponding one-forms, $\o^1_\sigma,\, \o^2_\sigma$ and $\o^1_\sigma-\o^2_\sigma$ to define the following quadratic
differentials:
\be
\q^1_a =\o^1_a (\o^1_a -\o^2_a),\qquad
\q^2_a=\o^2_a (\o^2_a - \o^1_a),\qquad
\q^3_a=\o^1_a\,\o^2_a,
\en
with $\o_a^r := \o_{\sigma_a}^r,\,r=1,2$. The main result of this paper is that these three quadratic
differentials are enough to construct CHY integrands.

In analogy with \eqref{intro_1loop}, we propose that the symmetrized two-loop planar Feynman diagrams are given by the CHY integrand:
\be
\parbox[c]{14em}{\includegraphics[scale=0.25]{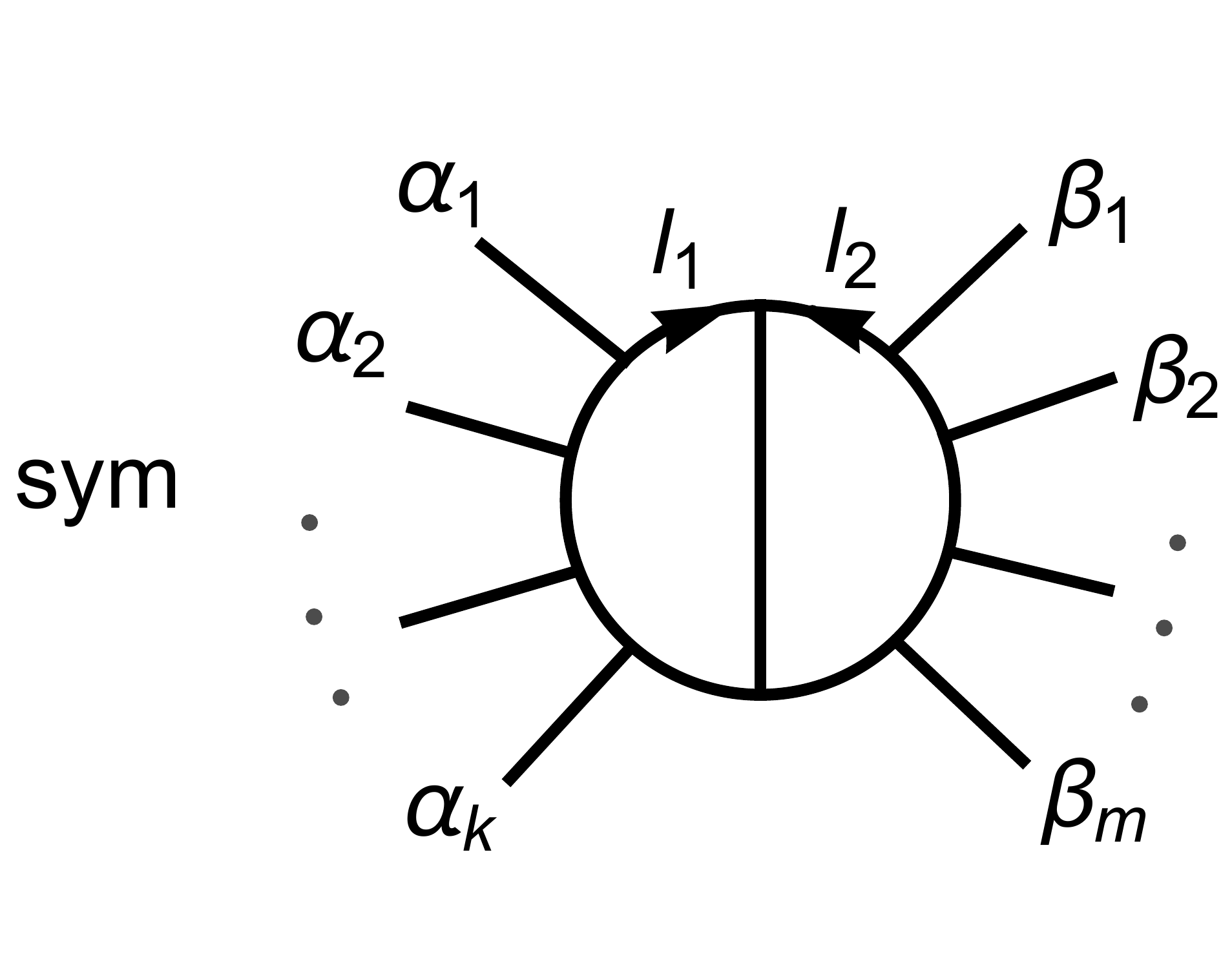}} \sim\quad \frac{1}{(\ell_1^+ \ell_2^+ \ell_2^- \ell_1^-)(\ell_2^+ \ell_1^+ \ell_2^- \ell_1^-)}\, \prod_{i=1}^k \q^1_{{\alpha_i}} \,\prod_{j=1}^m \q^2_{\beta_j}.\label{intro_2loop_planar}
\en
Here, in the denominator we have used a shorthand for a Parke-Taylor factor $(abcd) = (\sigma_a -\sigma_b)(\sigma_b - \sigma_c)(\sigma_c - \sigma_d)(\sigma_d - \sigma_a)$. The symmetrization on the left hand side is done for the sets $\{\alpha_i\}$ and $\{\beta_j\}$ separately. This object is structurally very similar to the one-loop case \eqref{intro_1loop}.

Similarly, we can utilize the remaining quadratic
differential, $\q^3_a$ in order to define a non-planar version of the two-loop diagram:
\be
\parbox[c]{14em}{\includegraphics[scale=0.25]{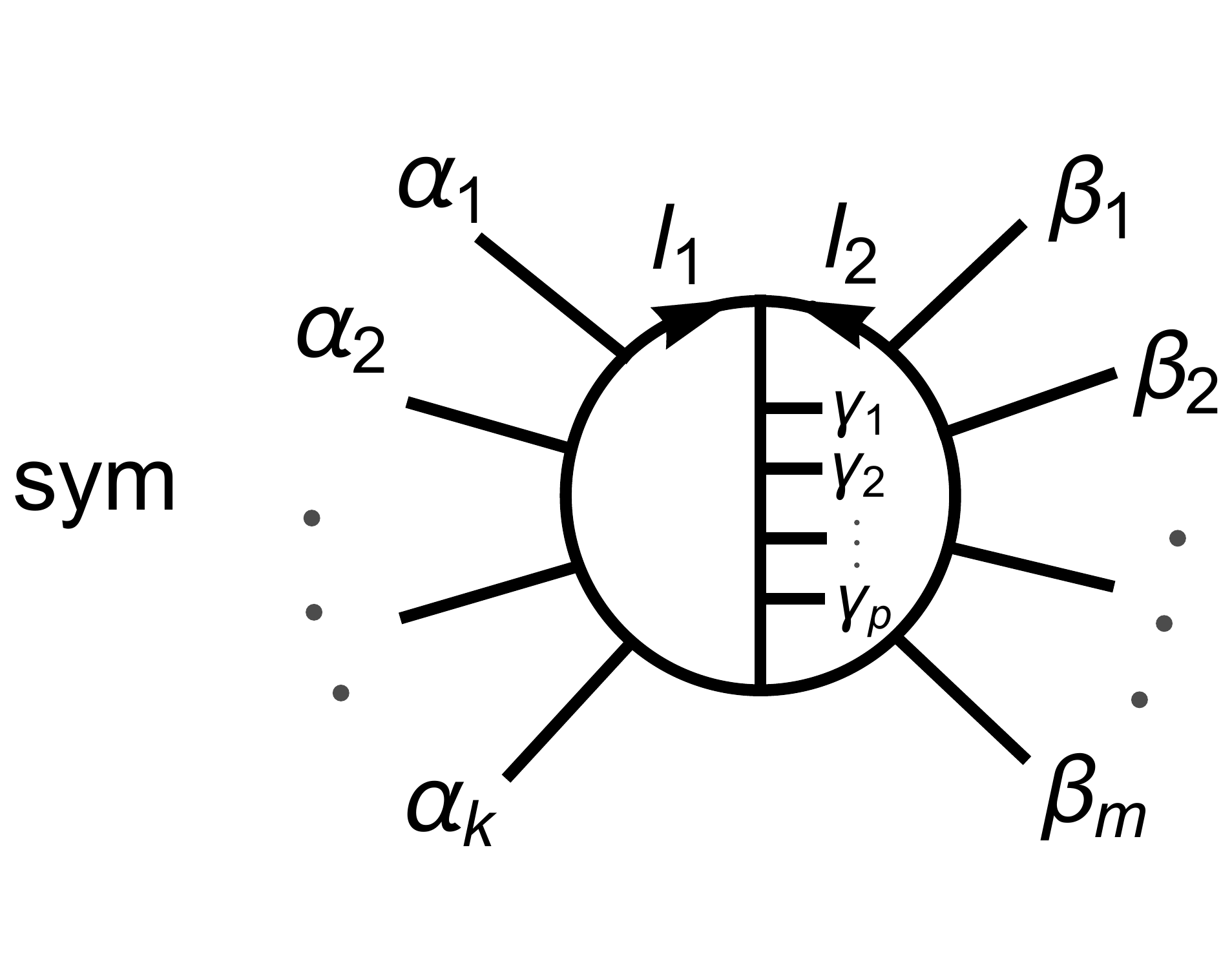}} \sim\quad \frac{1}{(\ell_1^+ \ell_2^+ \ell_2^- \ell_1^-)^2}\, \prod_{i=1}^k \q^1_{{\alpha_i}}\, \prod_{j=1}^m \q^2_{{\beta_j}}\, \prod_{l=1}^p \q^3_{{\gamma_l}}.\label{intro_2loop_nonplanar}
\en
Now the symmetrization proceeds over the three sets of external legs separately.

We give more details about these building blocks in section \ref{building-blocks}. In section \ref{gluesection} we propose a scheme for reconstructing more general CHY graphs from the building blocks mentioned earlier using simple gluing rules, and give several examples of its application in appendix \ref{GluingL}.

As a complementary result, we have generalized the $\Lambda$-algorithm \cite{Gomez:2016bmv} introduced by one of the authors to the two-loop case. Just as at tree level \cite{Gomez:2016bmv} and one loop \cite{Cardona:2016bpi,Cardona:2016wcr}, the $\Lambda$-algorithm allows us to analytically evaluate arbitrary CHY integrals using simple graphical rules. We summarize these rules in section \ref{lambdarules} and give more details in appendices \ref{Sec2} and \ref{Ltwoloop}.

We have combined our proposal for the CHY integrands together with the $\Lambda$-rules for their evaluation, in order to check many explicit examples in section \ref{examplesphi3}. They are verified both analytically and numerically up to seven external particles at two loops.

In section \ref{conclusions} we discuss some of the future research directions, including the extensions to higher-loop orders and to other theories. We comment on the prospects of summing the diagrams into compact expressions for the full integrand, along the lines of \cite{Geyer:2015bja, Geyer:2015jch, Cachazo:2015aol}.

\subsection*{Outline}

This paper is structured as follows. In section \ref{holo} we begin by discussing the holomorphic forms on a torus and a double torus. Using these forms we construct the basic building blocks for CHY integrands at one and two loops in section \ref{building-blocks}. In section \ref{gluesection} we demonstrate how to reconstruct arbitrary Feynman diagrams up to two loops using a gluing procedure. In section \ref{lambdarules} we explain the $\Lambda$-rules and provide many examples  for a direct computation of the diagrams constructed in section \ref{examplesphi3}. We conclude in section \ref{conclusions} with a discussion of future directions. This paper comes with three appendices. We give examples of the gluing operation at loop levels in appendix \ref{GluingL}. In appendix \ref{Sec2} we review the $\Lambda$-algorithm at tree level, and in appendix \ref{Ltwoloop} we generalize it to two loops.

\section{Holomorphic Forms at One and Two Loops}\label{holo}

The main purpose of this section, besides giving a brief review of the CHY formalism at one loop, is to rewrite the one-loop CHY integrand in terms of a fundamental mathematical object, the global holomorphic form over the torus. 

Afterwards we will generalize these ideas to the Riemann surface of genus $g=2$. There we realize a similar analysis, where we first find the holomorphic forms which satisfy the required physical properties and subsequently construct the CHY integrands by gluing together several building blocks.

\subsection{One--loop Holomorphic Form}\label{oneloop}

On the elliptic curve (torus) there is only one $(1,0)-$form given by
\begin{equation}
\Omega(z)dz := \frac{dz}{y} ,\qquad y^2=z(z-1)(z-\l).
\end{equation}
At the nodal singularity, i.e., pinching the A-cycle ($\l=0$), this holomorphic form becomes
\begin{equation}
i\,\Omega(z)\Big|_{\l=0} dz=\left\{\frac{1}{2(z-z_{\ell^+})}\left(\frac{y^{\rm t}_{\ell^+}}{y^{\rm t}}+1  \right) -\frac{1}{2(z-z_{\ell^-})}\left(\frac{y^{\rm t}_{\ell^-}}{y^{\rm t}} +1  \right) \right\}\,dz =: \o_z\,dz,
\end{equation}
where $i=\sqrt{-1}$,  $(z_{\ell^+},y^{\rm t}_{\ell^+} )=-(z_{\ell^-},y^{\rm t}_{\ell^-} )=(0,i)$ and $(y^{\rm t})^2=z-1$. In other words, one can say that the puncture $z_{\ell^+}=0$ is on the upper sheet, $y^{\rm t}_{\ell^+}=i$,  and  the puncture $z_{\ell^-}=0$ is on the lower sheet, $y^{\rm t}_{\ell^-}=-i$, over a double cover of the sphere given by the quadratic curve $(y^{\rm t})^2=z-1$. In order to obtain an expression over a single cover, we use the transformation $z=\s^2+1$, so
\begin{equation}
\o_z \,dz= \frac{i \,\, dz}{z \,\,y^{\rm t}} = \left(\frac{1}{\s-i}-\frac{1}{\s+i}\right)d\s = \left(\frac{1}{\s-\s_{\ell^+}}-\frac{1}{\s-\s_{\ell^-}}\right)d\s=:\o_\s\,d\s,
\end{equation}
where the puncture $(z_{\ell^+},y^{\rm t}_{\ell^+})=(0,i) $ has been mapped to  $\s_{\ell^+}=i$ and the puncture $(z_{\ell^-},y^{\rm t}_{\ell^-})=(0,-i) $ to  $\s_{\ell^-}=-i$. As it was shown in \cite{Geyer:2015bja,Geyer:2015jch,Cardona:2016bpi}, the momentum associated to the punctures $\{\s_{\ell^+},\s_{\ell^-} \}$ are $\{ (\ell^+)^\mu , (\ell^-)^\mu  \}:=\{ +\ell^\mu , -\ell^\mu  \}$, where $\ell^\mu$ is the loop momentum, i.e. it is off-shell ($\ell^2\neq 0$).  

\subsubsection{Geometric Interpretation}
After figuring out the reduction from the holomorphic form on a torus to the meromorphic form on a nodal Riemann sphere, we will give it a geometric interpretation. Before that let us clarify the notation of {\it CHY graphs}. On a Riemann sphere, it is convenient to represent the factor ${1 \over \s_{ab}}$ as a {\it line} and the factor $\s_{ab}$ as a dotted line that we call the {\it anti-line}:
\begin{align}
	&{1\over \s_{ab}}\,\leftrightarrow\,a ~\overline{~~~~~~~~~~~~~}~ b ~~~{\rm (line)},  \\
	& \s_{ab}~\leftrightarrow\, a\,- \, - \,  - \, -\, b ~~~     {\rm  (anti {\rm -} line)},
\end{align}
 In this way, CHY integrands have graphical description as CHY graphs. We will use this notation to represent the meromorphic form $\omega_\sigma$ and also any CHY integrands in the remainder of the paper.
\begin{figure}
\centering
\hspace{-5em}   \includegraphics[scale=0.50]{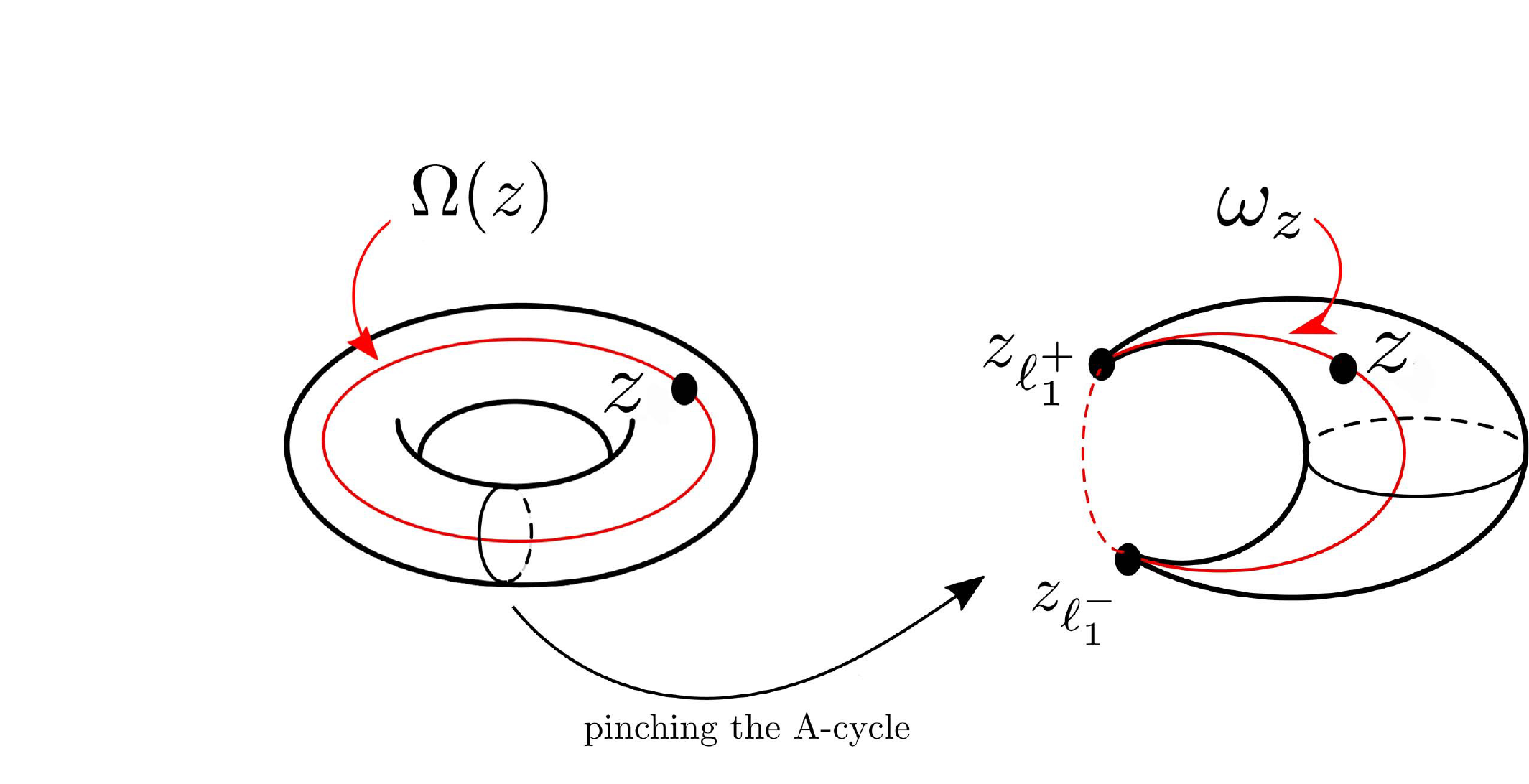}
       \caption{Geometrical interpretation of reducing from the holomorphic form $\O(z)$ on a torus to the meromorphic form $\o_z$ on a Riemann sphere.} \label{H-form}
\end{figure}

On the torus, as shown on the left of figure \ref{H-form}, the holomorphic form $\Omega(z)$  connects a puncture with itself by a line around the B-cycle \cite{Cardona:2016wcr}. Obviously, this object does not have an analogy at tree level. However, after pinching the A-cycle and separating the node, the torus becomes a nodal Riemann sphere. In this way, the holomorphic form $\O(z)$ becomes the meromorphic form $\o_\sigma$ on the Riemann sphere, whose CHY graph representation of this form is given on the right side in figure \ref{H-form}.

Note that the meromorphic form inherited from the torus
\begin{equation}
\o_\s \,d\s=\left(\frac{1}{\s-\s_{\ell^+}}-\frac{1}{\s-\s_{\ell^-}}\right)d\s=\frac{\s_{\ell ^+\ell^-}}{(\s-\s_{\ell^+})(\s-\s_{\ell^-})} d\s,
\end{equation}
has only simple poles at $\s = \s_{\ell^+}$ and $\s = \s_{\ell^-}$, with residues $+1$ and $-1$, respectively. In addition, this form vanishes when $\s_{\ell^+} = \s_{\ell^-}$, namely the factorization channel corresponding to a divergent contribution $\sim {1\over (\ell - \ell)^2}$, is not allowed. This important fact drives us to think that this is a fundamental and natural object to build CHY integrands.

\subsection{Holomorphic Forms over the Double Torus}\label{holo2loop}

In the previous section, we have shown that one can  build physical  CHY integrands at one loop using a natural mathematical object, the global holomorphic form on the torus. This idea may be generalized to Riemann surfaces of higher genus. And here we realize a similar analysis at two loops.  

Let us consider a Riemann surface of genus $2$ as a hyperelliptic curve embedded in $\mathbb{CP}^2$, namely
\begin{equation}\label{hyper}
y^2 = (z-a_1) (z-a_2)(z-\l_1)(z-\l_2)(z-\l_3),
\end{equation}
where $(\lambda_1, \lambda_2, \lambda_3)$ parametrize the curve, and $(a_1,a_2)$ are two fixed branch points such that $a_1\neq a_2$. 
Since we will be ultimately interested in the degeneration of the curve near $\lambda_1 = a_1$ and $\lambda_2 = a_2$, we denote A$_1$-cycle the one that goes around $\lambda_1 = a_1$ in the degeneration limit. The A$_2$-cycle is defined analogously.

Note that this curve has several singular points, but we are only interested in those where the two A-cycles are pinching at different points, i.e., singularities where the Riemann surface degenerates to a sphere with four extra punctures.  Furthermore, as it will be shown below, many of the others singularities cancel out after computing the CHY integrals.

\begin{figure}
\centering
   \includegraphics[scale=0.45]{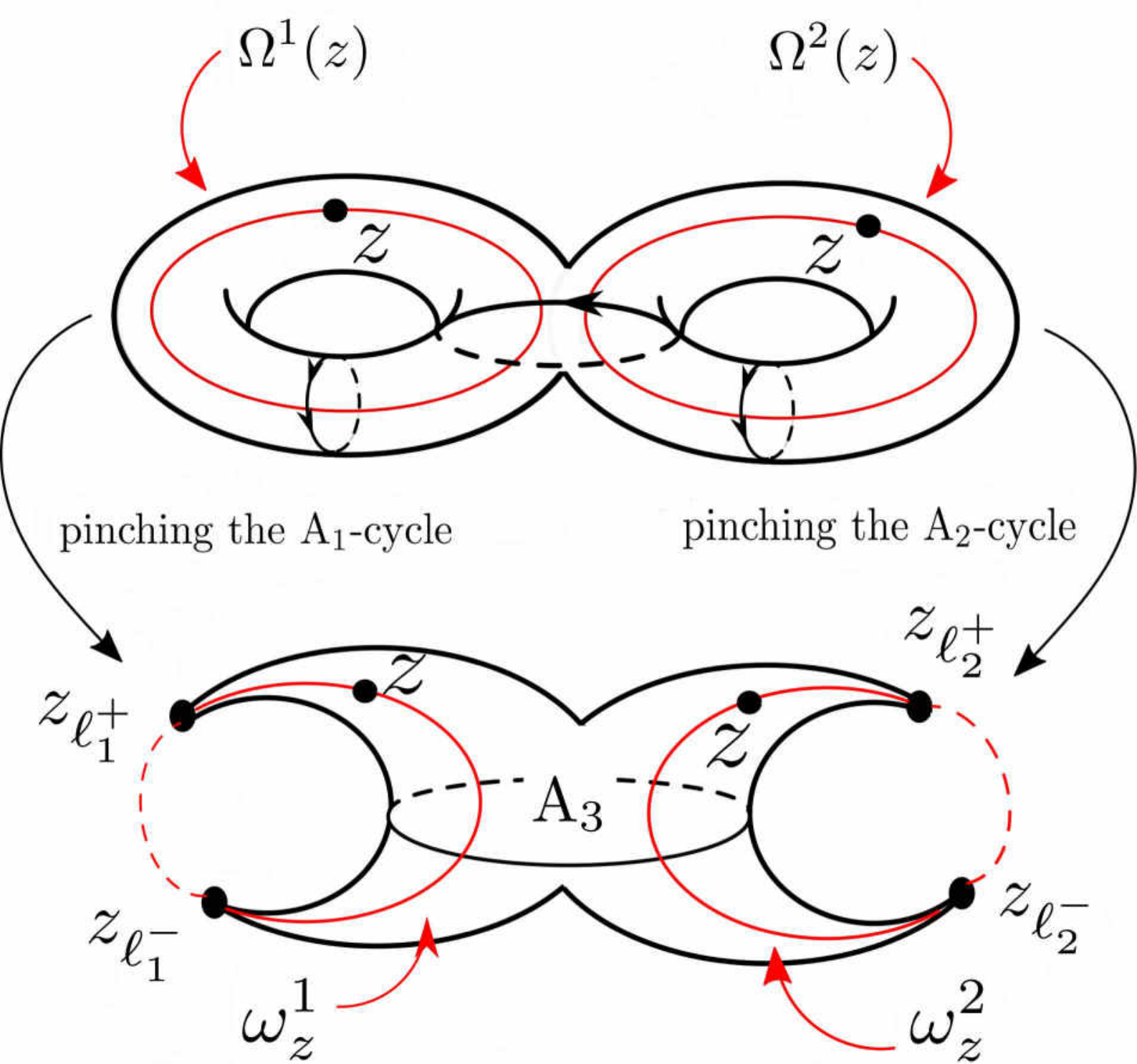}
       \caption{Geometrical meaning of the two holomorphic forms on the double torus.} \label{H-form_2-loop}
\end{figure}

It is well-known that over the algebraic curve given in \eqref{hyper} there are just two global holomorphic forms, which can be written as 
\begin{align}
\Omega^1(z)dz &:= \frac{z-a_2}{y}dz,\\
\Omega^2(z)dz &:= \frac{z-a_1}{y}dz.
\end{align}
In order to pinch the A-cycles, we take, without loss of generality, the parameters  $\l_1=a_1$ and $\l_2=a_2$. Thus the curve in \eqref{hyper}  becomes $y=(z-a_1)(z-a_2)y^{\rm t}$ where $y^{\rm t}$ is a double cover sphere given by $(y^{\rm t})^2 = z-\l_3$. Under this degeneration of the Riemann surface the holomorphic forms turn into
\begin{align}
(a_1-\l_3)^{1/2}\Omega^1(z)\Big|_{\l_1=a_1 \atop \l_2= a_2}dz&= \frac{(a_1-\l_3)^{1/2}}{(z-a_1)\,y^{\rm t}}dz=:\o^1_z\, dz,\\
(a_2-\l_3)^{1/2}\Omega^2(z)\Big|_{\l_1=a_1 \atop \l_2= a_2}dz&= \frac{(a_2-\l_3)^{1/2}}{(z-a_2)\,y^{\rm t}}dz=:\o^2_z\, dz,
\end{align}
where we have included normalization factors.
Note that $\o^1_z\,dz$ and $\o^2_z\,dz$ are now meromorphic forms over a sphere defined by the quadratic curve $(y^{\rm t})^2 = z-\l_3$. It is straightforward to see that $\o^1_z\,dz$ has only two simple poles (one on upper sheet and the other one on the lower sheet), which are associated with the A$_1$-cycle, i.e.,
\begin{align}
\oint_{A_1} \o^1_z\,\,dz = \pm 1,   \qquad
\oint_{A_2} \o^1_z\,\,dz = 0, 
\end{align}
 where $``+"$ is the residue on the upper sheet and $``-"$ is the residue on the lower sheet. In an analogous way, for $\o^2_z\,dz$ one has 
\begin{align}
\oint_{A_1} \o^2_z\,\,dz = 0 ,   \qquad
\oint_{A_2} \o^1_z\,\,dz = \pm 1. 
\end{align}
Therefore, as it is shown in figure \ref{H-form_2-loop}, 
the holomorphic form $\Omega^1(z)$ is related with the A$_1$-cycle and so its CHY interpretation means that it connects a puncture with itself by a line around the B$_1$-cycle. In a similar way, the holomorphic form $\Omega^2(z)$ is related with the A$_2$-cycle and so it connects a puncture with itself by a line around of the B$_2$-cycle.

So far, we have found the two meromorphic forms on the double cover sphere ($\o^1_z\,$ and $\o^2_z\,$), which are related with the A$_1$ and A$_2$ cycles on the double torus.  In order to obtain an expression over a single cover, we use the transformation $z=\s^2+\l_3$, so
\begin{align}
\o^1_\s\,d\s &= \left( \frac{1}{\s - \s_{\ell_1^+}} -\frac{1}{\s - \s_{\ell_1^-}} \right) d\s=  \frac{\s_{\ell_1^+\ell_1^-}}{(\s - \s_{\ell_1^+})(\s - \s_{\ell_1^-})}  d\s,\\
\o^2_\s\,d\s &= \left( \frac{1}{\s - \s_{\ell_2^+}} -\frac{1}{\s - \s_{\ell_2^-}} \right) d\s=  \frac{\s_{\ell_2^+\ell_2^-}}{(\s - \s_{\ell_2^+})(\s - \s_{\ell_2^-})}  d\s,
\end{align}
where $\s_{\ell_{1}^{\pm}}= \pm (a_1-\l_3)^{1/2}$ and  $\s_{\ell_{2}^{\pm}}= \pm (a_2-\l_3)^{1/2}$ and the momenta related with these four punctures are $\ell_{1}$ and $\ell_{2}$ respectively.

The meromorphic forms $\o^1_\s$ and $\o^2_\s$ were previously used in \cite{Geyer:2016wjx} to find the scattering equations at two loops. In addition, in the appendix \ref{Ltwoloop} we used these scattering equations to obtain the $\L-$rules.

\subsection{Physical Requirements}

In the same way as the one-loop case, the meromorphic forms $\o^r_\sigma$ over the sphere vanish when $\s_{\ell_r^+}=\s_{\ell_r^-}$, but this fact occurs independently, namely $\o^1_\s\,$ does not feel anything about what is happening with $\o^2_\s\,$ and vice versa.  This suggests that $\o^1_\s\,$ and $\o^2_\s\,$ are the fundamental objects to construct CHY integrands for amplitudes where the two-loop Feynman diagram can be cut into two one-loop diagrams.

In order to describe one-particle irreducible diagrams (1PI), we must consider a third cycle A$_3$ which connects the two holes of the Riemann surface of genus $2$, as it is shown in figure \ref{H-form_2-loop}.  Nevertheless, it is straightforward to see that this cycle can be written as a linear combination of $\{{\rm A}_1,{\rm A}_2\}$, i.e.,  ${\rm A}_3={\rm A}_1-{\rm A}_2$. Therefore, the dual holomorphic form to A$_3$ is $\Omega^1(z) - \Omega^2(z)$, which after pinching A$_1$ and A$_2$ becomes $\o_\sigma^1-\o_\sigma^2$.  Finally, our proposal to obtain 1PI Feynman diagrams at two loops is to add a third meromorphic form $\o_\sigma^1-\o_\sigma^2$.

Careful analysis of this proposal would require the knowledge of embedding the Riemann surface into the ambitwistor space \cite{Mason:2013sva}. We leave this approach for future work. In addition, it is useful to  remark  that the third meromorphic form,  which must be used to build  CHY integrands describing 1PI Feynman diagrams, depends on the $\a-$parameter introduced by \cite{Geyer:2016wjx} to fomulate the scattering equations at two loops. To be more precise, the meromorphic form that must be used is
\begin{equation}
\o_\sigma^1 - \a\, \o_\sigma^2.
\end{equation}
Since we are choosing $\a=1$ to obtain the integration rules (appendix \ref{Ltwoloop}), this form is written as  $\o_\sigma^1-\o_\sigma^2$.

In the next section we construct building blocks in order to compute any $\Phi^3$ Feynman diagram at one and two loops. We will assemble the meromorphic forms found in this section into quadratic
differentials so that the building blocks can be written in a compact way.

\section{Building Blocks of CHY Integrands at One and Two Loops for $\Phi^3$}\label{building-blocks}

\begin{figure}[!h]
\centering
   \includegraphics[scale=0.25]{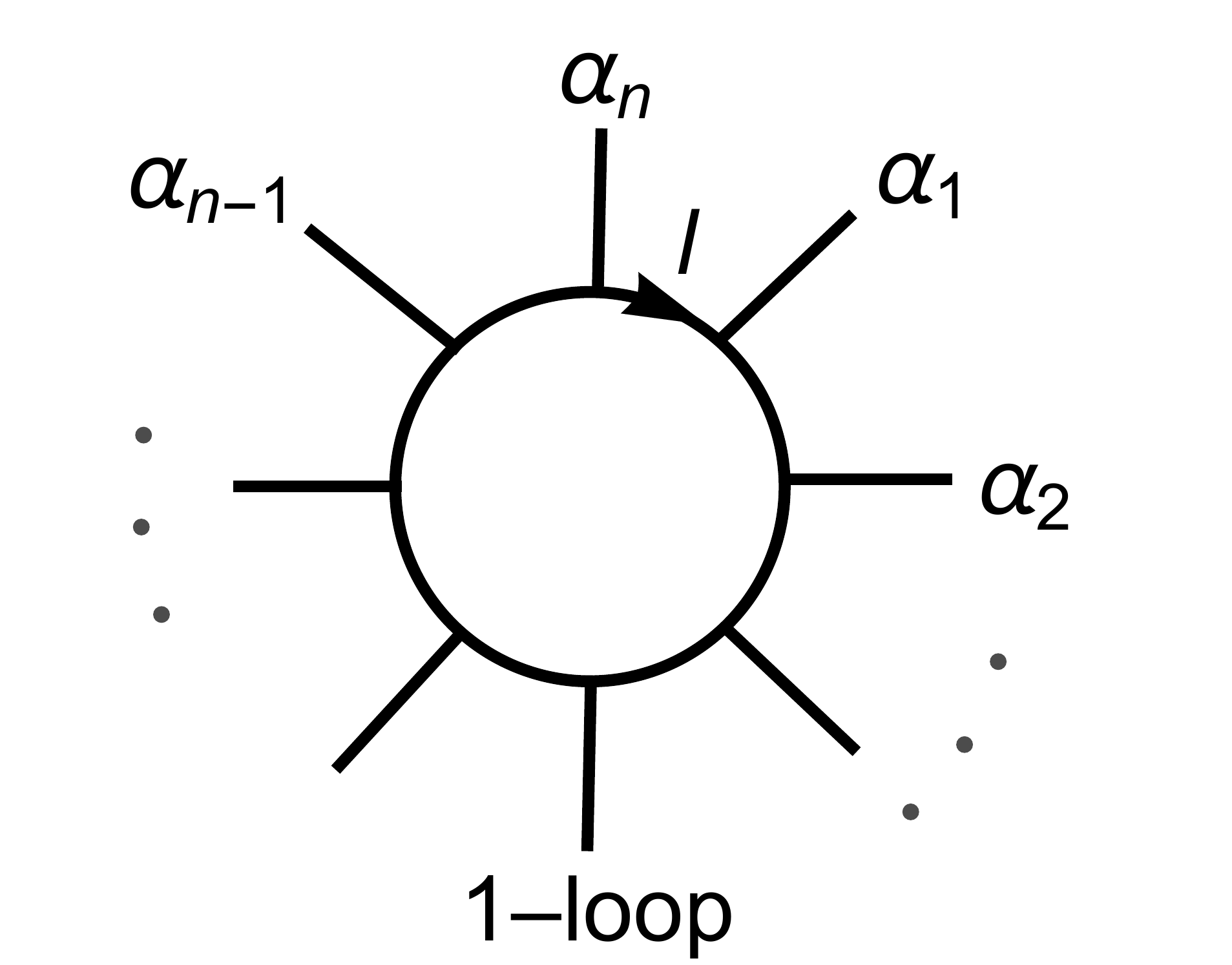}
        \includegraphics[scale=0.25]{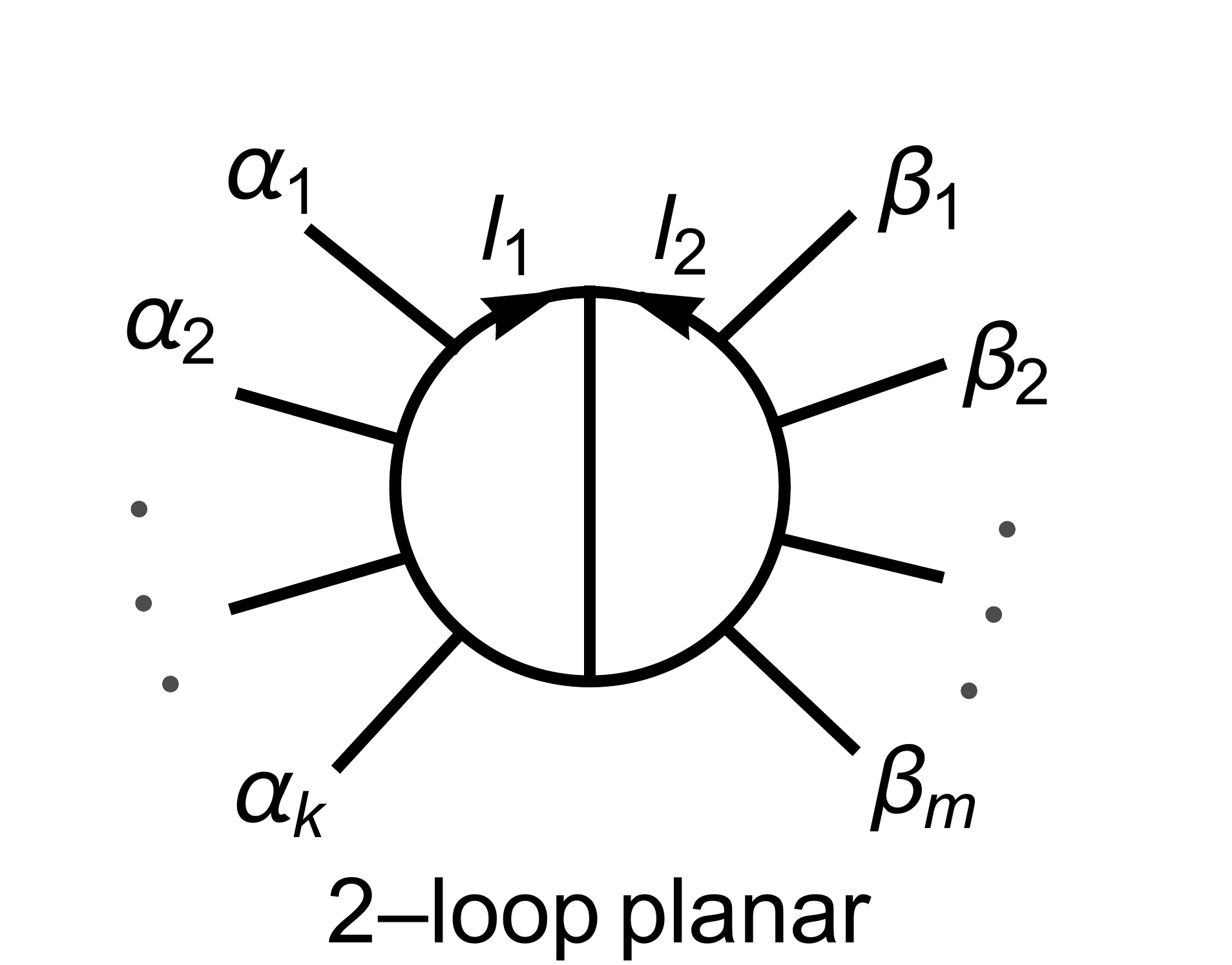}
       \includegraphics[scale=0.25]{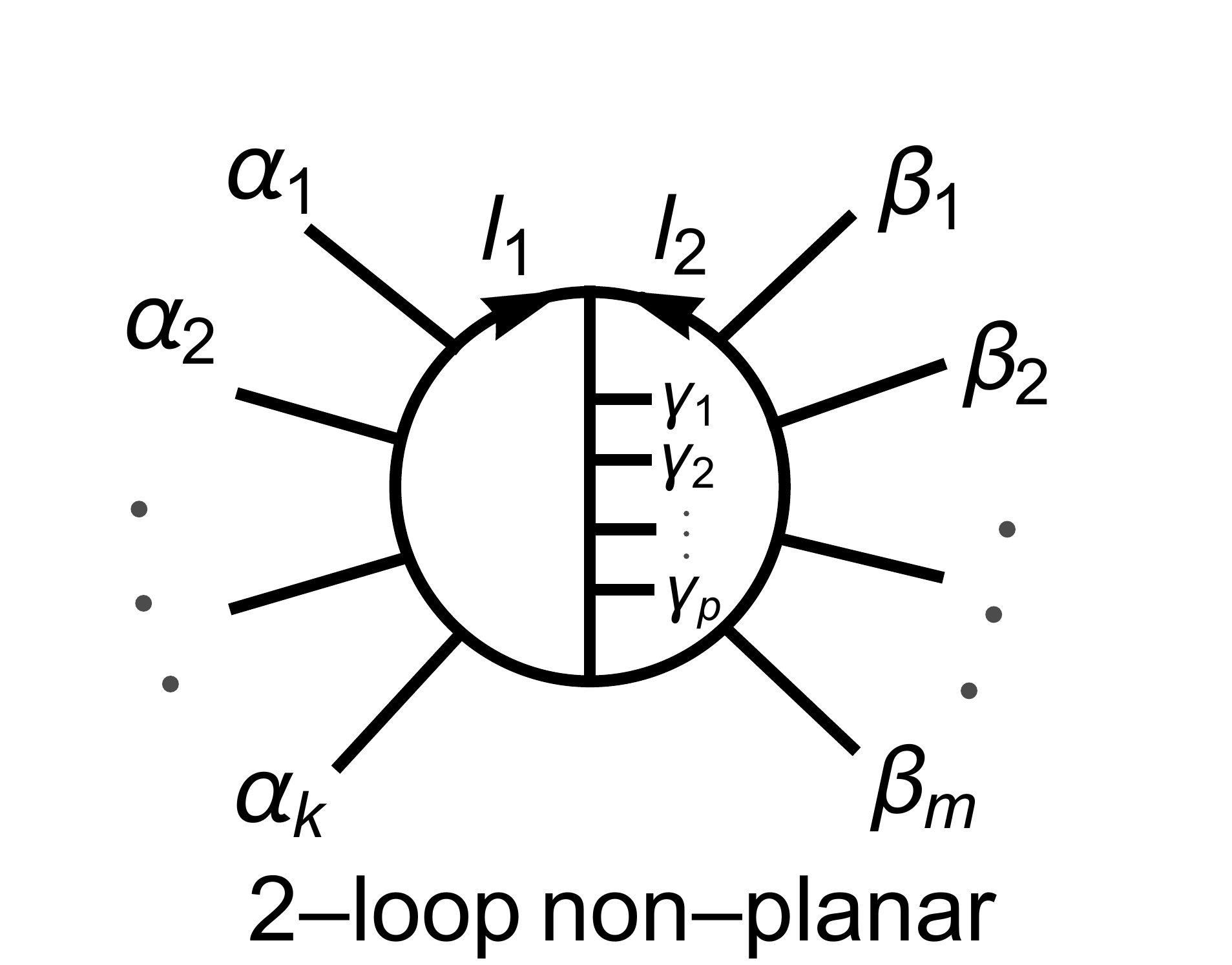}
       \caption{Three building blocks for two-loop Feynman diagrams for $\Phi^3$. We will use these to construct more complicated 1PI diagrams.} \label{bb}
\end{figure}

In this section, we will give a general definition of the building blocks for CHY graphs from the meromorphic forms  $\o^r_\s$ obtained in section \ref{holo}. There are three building blocks for Feynman diagrams at one and two loops as it is shown in figure \ref{bb}. We want to consider how the corresponding CHY integrands look like. 

The general construction is as follows. For a given topology of the graph, see figure \ref{bb}, we first assign a skeleton factor. Similarly, we assign each external leg a factor which depends on the place the leg is attached. For example, in the planar two-loop topology, legs connected to the left and right loops come with distinct coefficients. The CHY integrand for a given graph is then simply given by a product of the skeleton and leg factors.

For the purpose of this section we assume that $\{\alpha_i, \beta_i, \gamma_i\}$ are off-shell particles. In section \ref{gluesection} we will introduce a set of gluing rules that allow extending this construction to arbitrary Feynman graphs.

\subsection{One-Loop Building Block}

At one loop there is only one meromorphic form, which we denote as $\o_a:=\omega_{\sigma_a}$. Now, it is natural to build an integrand as
\begin{align}
&{\cal I}^{\rm n-gon-CHY}_{\rm sym}=\frac{1}{\ell^2}  \int d\mu^{\rm 1-loop} \,\,\mathbf{I}^{\rm n-gon-CHY}_{\rm sym},\\
&\mathbf{I}^{\rm n-gon-CHY}_{\rm sym} := {\rm \ss^{1-loop}}\left(\o_{1}\,\o_{2}\, \cdots \o_{n}\right)^2
= {\rm \ss}^{\rm 1-loop}\prod_{a=1}^{n} \q_{a},\nonumber
\end{align}
where we have defined
\begin{align}
&{\rm \ss}^{\rm 1-loop}:=\frac{1}{(\ell^+, \ell^- )^2},\qquad \q_a=(\o_a)^2, \\
&d\mu^{\rm1-loop}:=\frac{1}{{\rm Vol}\,\,({\rm PSL}(2,\mathbb{C}))}\times \frac{d\s_{\ell^+}}{E^{\rm 1-loop}_{\ell^+}}\times\frac{d\s_{\ell^-}}{E^{\rm 1-loop}_{\ell^-}}\times \prod_{a=1}^n \frac{d\s_{a}}{E^{\rm 1-loop}_{a}},
\end{align}
with the Parke-Taylor factor
\begin{equation}\label{chainS}
	(a_1,a_2,\ldots, a_p):= \s_{a_1a_2}\s_{a_2a_3}\cdots \s_{a_{p-1}a_p}\s_{a_pa_1}.
\end{equation}
and the $\{ E^{\rm 1-loop}_a,E^{\rm 1-loop}_{\ell^\pm} \}$ one-loop scattering equations
\begin{align}
&E_a^{\rm 1-loop}:= \sum^n_{b=1\atop b\neq a}\frac{k_a\cdot k_b}{\s_{ab}} +  \frac{k_a\cdot \ell^+}{\s_{a\ell^+}} + \frac{k_a\cdot \ell^-}{\s_{a\ell^-}}=0,\\
& E_{\ell ^{\pm}}^{\rm 1-loop}:= \sum^n_{b=1}\frac{\ell^{\pm}\cdot k_b}{\s_{\ell^{\pm}b}}=0, 
\qquad  (\ell^+)^\mu=-(\ell^-)^\mu := \ell^\mu, \quad \ell^2\neq 0. 
\end{align}
Note that we have introduced the factor ${\rm \ss^{1-loop}}$ in order to obtain the proper ${\rm PSL}(2,\mathbb{C})$ transformation. We call the ${\rm \ss^{1-loop}}$ factor a {\it skeleton}.
 
\begin{figure}[t]
  \centering
   \raisebox{.65\height}{\includegraphics[scale=.2]{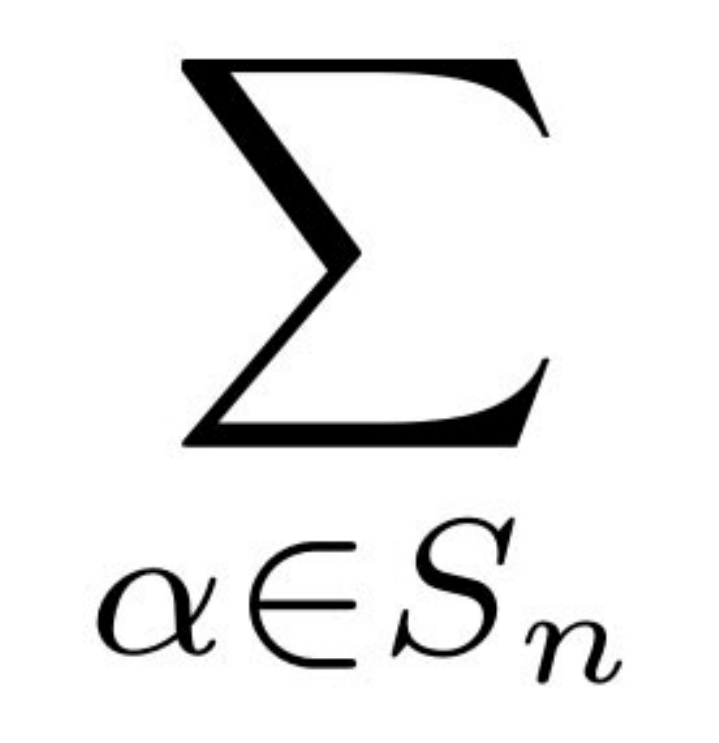}}
         \includegraphics[width=2in]{0_0_1lb}                                               
         \includegraphics[width=1.5in]{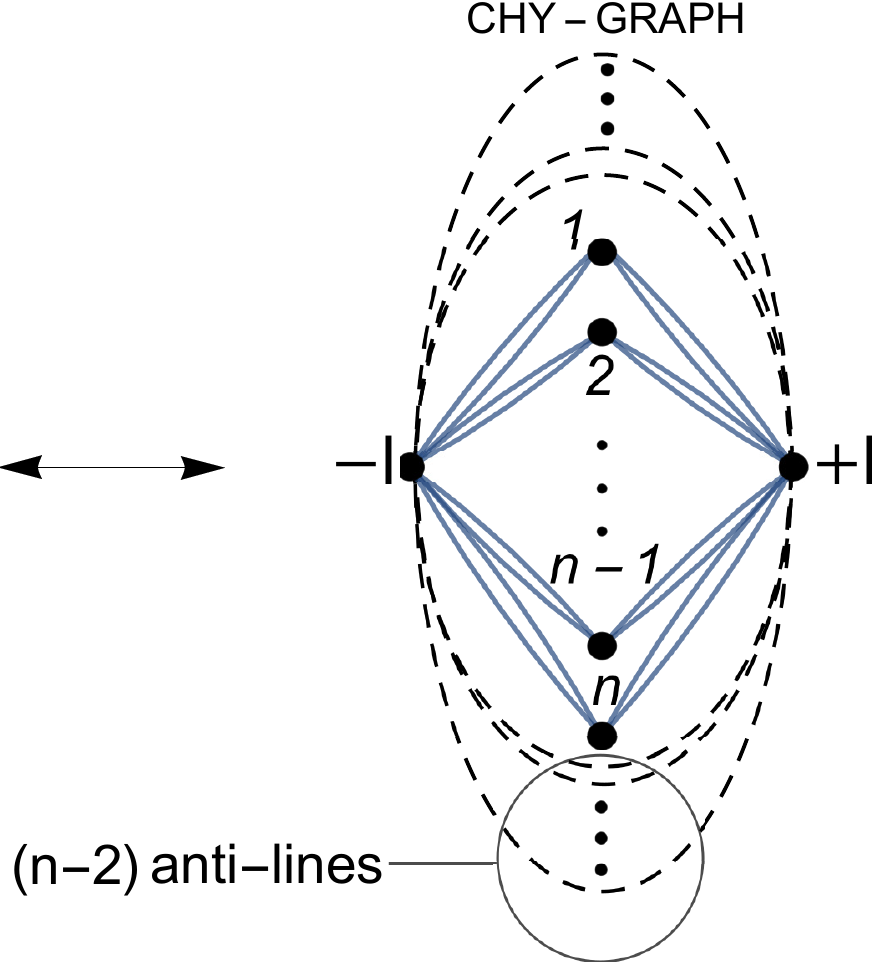}\\
  \caption{Correspondence between the $\Phi^3$ Feynman diagrams ($n$-gon symmetrized)  and the $\mathbf{I}^{\rm n-gon-CHY}_{\rm sym}$ CHY-graphs. $S_n$ is the permutation group.}\label{CHY-pgon-tree}
\end{figure}

It is well-known \cite{Geyer:2015bja,Geyer:2015jch,He:2015yua} that the ${\cal I}^{\rm n-gon-CHY}_{\rm sym}$ loop integrand corresponds to the $\Phi^3$ Feynman integrand of the  symmetrized $n$-gon, as it is represented in figure \ref{CHY-pgon-tree}.
Nevertheless, it is important to recall that the correspondence between Feynman integrand at one-loop and the CHY integrand can be realized after using the {\bf partial fraction identity (${\rm p.f}$)}  \cite{Geyer:2015bja}
\begin{equation}\label{partialfrac}
{1\over \prod_{i=1}^n D_i}=\sum_{i=1}^{n}{1\over D_i\prod_{j\neq i}(D_j-D_i)}\, ,
\end{equation}
and {\bf shifting (${\rm S}$)} the loop momentum $\ell^\mu$. In addition,  one must suppose that the integral  $\int d^D\ell$ is invariant under that transformation, i.e.
\begin{equation}
{\cal I}^{\rm n-gon-FEY}_{\rm sym}\Big|_{{\rm p.f}\atop {\rm S}} =\frac{1}{2^{n-1}}\, \,{\cal I}^{\rm n-gon-CHY}_{\rm sym},
\end{equation}
where ${\cal I}^{\rm n-gon-FEY}_{\rm sym}$ is the Feynman integrand for the $\Phi^3-$diagram in figure \ref{CHY-pgon-tree} and
$n$ is the number of particles. Here the factor $2^{-n+1}$ comes from the convention of using $k_a\cdot k_b$  instead of $2k_a\cdot k_b$ in the numerators of the scattering equations. In a general $l$-loop case, this factor is $2^{-(n+2l-3)}$ due to the ${\rm PSL}(2,\mathbb{C})$ symmetry of scattering equations and the number of puncture locations.

\subsection{Two-loop Building Blocks}
\label{form}

Next let us focus on the two-loop building blocks, including the planar and non-planar cases. At two loops, in section \ref{holo2loop}, we have found that the  meromorphic forms, $\o^1_{\s}$ and  $\o^2_{\s}$, are interpreted as circles going around the B-cycles, in a disjoint way. Namely, $\o^1_\s$ does not feel $\o^2_\s$  and vice versa. In addition, we have also argued that the linear combination, $\o_\sigma^1-\o_\sigma^2$, is related with the 1PI Feynman diagram at two loops.

In the previous section, we wrote the one-loop integrand for a symmetrized $n-$gon of $\Phi^3$, as a product of quadratic differentials living on the torus. Following this idea,  one can easily observe that the quadratic diferrentials $(\o^1_\s)^2$ and $(\o^2_\s)^2$  generate CHY integrands for Feynman diagrams with two separated loops, which we are going to show in sections \ref{gluesection}  and \ref{examplesphi3}. However, in order to construct general CHY integrands 
associated to 1PI Feynman diagrams at two loops, we should define the following quadratic differentials
\begin{align}
\q^1_a &=\o^1_a (\o^1_a -\o^2_a), \label{q1}\\
\q^2_a &=\o^2_a (\o^2_a - \o^1_a),\label{q2}\\
\q^3_a &=\o^1_a\,\o^2_a,  \label{q3}
\end{align}
where we are using the notation
\begin{equation}
\o^r_a:=\frac{\s_{\ell^+_r\ell^-_r}}{(\s_a-\s_{\ell_r^+})(\s_a-\s_{\ell_r^-})}, \qquad r=1,2.
\end{equation}
For the two-loop planar building block, we have the following proposal
\begin{align}
&{\cal I}^{\rm planar}_{\rm CHY} :=\frac{1}{\ell_1^2 \,\, \ell_2^2}\int d\mu^{\rm 2-loop} \,\,\,\mathbf{I}^{\rm planar}_{\rm CHY}\label{Chy_planar},
\\
&\mathbf{I}^{\rm planar}_{\rm CHY} :={\rm \ss^{planar}}\prod_{i=1}^k \q^1_{{\a_1}} \prod_{j=1}^m \q^2_{{\b_j}},\\
&{\rm \ss^{ planar}}:=\frac{1}{(\ell_1^+,\ell_2^+,\ell_2^-,\ell_1^-)(\ell_2^+,\ell_1^+,\ell_2^-,\ell_1^-)},
\\
&d\mu^{\rm 2-loop}:=\frac{1}{{\rm Vol}\,\,({\rm PSL}(2,\mathbb{C}))}\times \prod_{r=1}^2
\frac{d\s_{\ell^+_r}}{E^{\rm 2-loop}_{\ell^+_r}}\times\frac{d\s_{\ell^-_r}}{E^{\rm 2-loop}_{\ell^-_r}}\times \prod_{a=1}^n \frac{d\s_{a}}{E^{\rm 2-loop}_{a}},
\end{align}
where the two-loop scattering equations \cite{Geyer:2016wjx}, $\{ E^{\rm 2-loop}_a, E^{\rm2-loop}_{\ell^\pm_1} , E^{\rm2-loop}_{\ell^\pm_2} \}$, are given in appendix \ref{Ltwoloop}, and $(\ell_r^{+})^\mu=-(\ell_r^{+})^\mu:=(\ell_r)^\mu, \,\, \ell_r^2\neq 0, \,\, r=1,2$.

\begin{figure}[h]
  \centering
   \raisebox{.7\height}{\includegraphics[scale=0.18]{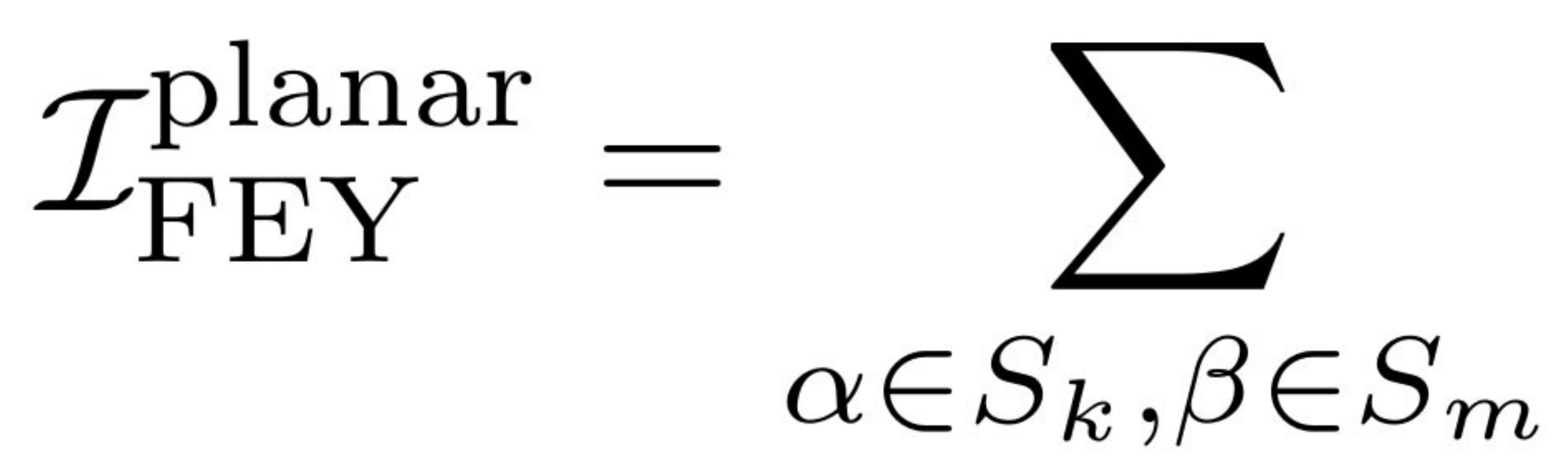}}\quad
         \includegraphics[width=2.0in]{0_0_pb}                                                      
  \caption{The $\Phi^3$ planar Feynman diagrams we want to compare with the CHY-graphs. $S_n$ is the permutation group.}\label{Fey_Planar}
\end{figure}

It is straightforward to check that  the CHY integrand, ${\bf I}^{\rm planar}_{\rm CHY}$, is invariant under permutations over $\{\a_1,\ldots, \a_k\}$ and $\{\b_1,\ldots, \b_n\}$. Therefore, in order to compare with the Feynman diagram results, we consider the symmetrization of the planar two-loop diagrams, as it is shown in figure \ref{Fey_Planar}.

We conjecture that the  Feynman integrand, ${\cal I}^{\rm planar}_{\rm FEY}$, of the $\Phi^3$ diagram given in figure \ref{Fey_Planar} actually corresponds to the CHY integrand given in \eqref{Chy_planar}, i.e.,
\\
\begin{center}
\begin{tabular}{| l |}
 \cline{1-1}  
\\
 \hspace{0.5cm}${\cal I}^{\rm planar}_{\rm FEY}\Big|_{\rm p.f \atop S} = \frac{1}{2^{n+1}}\,\,\,{\cal I}^{\rm planar}_{\rm CHY}\;\, ,$\hspace{3cm}\\
\\
\cline{1-1}
\end{tabular}
\end{center}
where $n$ is the number of particles, for this case it means $n=k+m$. The equality is given after using the partial fraction identity \eqref{partialfrac} over the loop integrand, and keep in mind we have defined  
\begin{align*}
{\rm p.f}& := {\rm patial \, \, fraction \, \, identity},\\
{\rm S} & := {\rm Shifting\,\, the\,\, loop\,\, momentum}.
\end{align*}
This conjecture has been checked analytically up to seven points, using a computer implementation of the $\L-$algorithm described in appendix \ref{Ltwoloop}.

\begin{figure}[h]
  \centering
   \raisebox{.6\height}{\includegraphics[scale=.18]{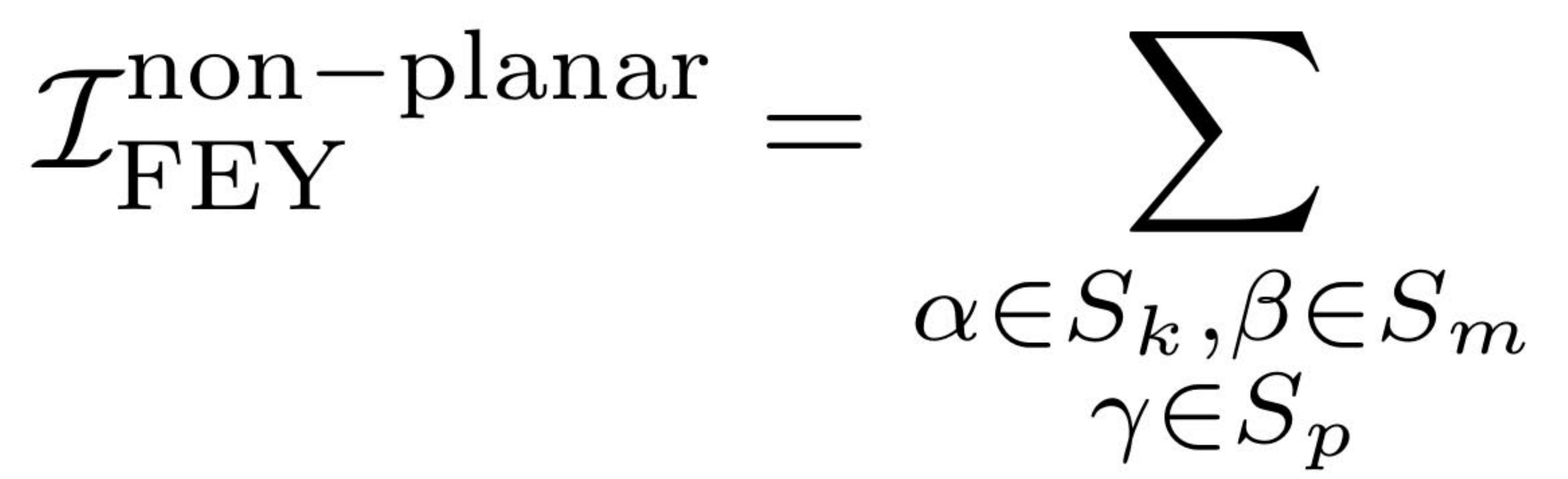}}\quad
         \includegraphics[width=2.2in]{0_0_npb}                                                      
  \caption{The $\Phi^3$ non-planar Feynman diagrams we want to compare with the CHY graphs (up to ${1\over \ell^2}$ overall factor). }\label{Fey_nonplanar}
\end{figure}

Note that unlike the one-loop case, here the number of CHY graphs is $2^{k+m}$ by expanding \eqref{Chy_planar}. In section \ref{examplesphi3}, we give some illustrative examples where the computations are done in detail. Finally, for the two-loop non-planar case, as in the third graph of figure \ref{bb}, our proposal for the CHY integrand reads:
\begin{align}
&{\cal I}^{\rm non-planar}_{\rm CHY} :=\frac{1}{\ell_1^2 \,\, \ell_2^2}\int d\mu^{\rm 2-loop} \,\,\,\mathbf{I}^{\rm non-planar}_{\rm CHY}, \label{Chy_nonplanar}\\
&{\bf I}^{\rm non-planar}_{\rm CHY}= {\rm \ss^{non-planar}}\prod_{i=1}^k \q^1_{{\alpha_i}} \prod_{j=1}^m \q^2_{{\beta_j}}\prod_{l=1}^p \q^3_{{\gamma_l}} , \\
&{\rm \ss^{non-planar}}=\frac{1}{(\ell_1^+,\ell_2^+,\ell_2^-,\ell_1^-)^2}.
\end{align}
It is simple to see that this CHY integrand  is invariant under the permutation over $\{\alpha_1,\ldots, a_k\}$, $\{\beta_1,\ldots, \b_m\}$ and $\{\gamma_1,\ldots, \gamma_p\}$. In order to compare it with the loop integrand, we consider the symmetrization of the non-planar two-loop diagrams, as it is shown in figure \ref{Fey_nonplanar}.

We conjecture that the Feynman integrand, ${\cal I}^{\rm non-planar}_{\rm FEY}$,  of the non-planar $\Phi^3$ diagram given in figure \ref{Fey_nonplanar} corresponds to the CHY integrand given in \eqref{Chy_nonplanar}, namely
\\
\begin{center}
\begin{tabular}{| l |}
 \cline{1-1}  
\\
 \hspace{0.5cm}${\cal I}^{\rm non-planar}_{\rm FEY}\Big|_{\rm p.f \atop S}= \frac{1}{2^{n+1}}\,\, {\cal I}^{\rm non-planar}_{\rm CHY}.\;\,$ \hspace{3cm}\\
\\
\cline{1-1}
\end{tabular}
\end{center}
As in the previous case, this conjecture has been checked analytically up to seven points. In section \ref{examplesphi3}, we give an illustrative example with the computation done in detail.

It is interesting to remark that, as it is well-known, at two loops there are only three independent holomorphic quadratic differentials, which are chosen to be $\q^1_\s, \q^2_\s$ and $\q^3_\s$. As it is simple to notice, for a 1PI two-loop diagram, the $\q^1_\s$ quadratic differential is related with the external legs on the loop momentum $\ell_1$, in a similar way $\q^2_\s$ with $\ell_2$ and $\q^3_\s$ with $\ell_1+\ell_2$. It would be interesting to generalize this idea to explore what the quadratic
differentials are beyond two loops.

\section{Constructing CHY Graphs from Gluing  Building Blocks}\label{gluesection}

\begin{figure}[!h]
 \centering
         \includegraphics[scale=0.25]{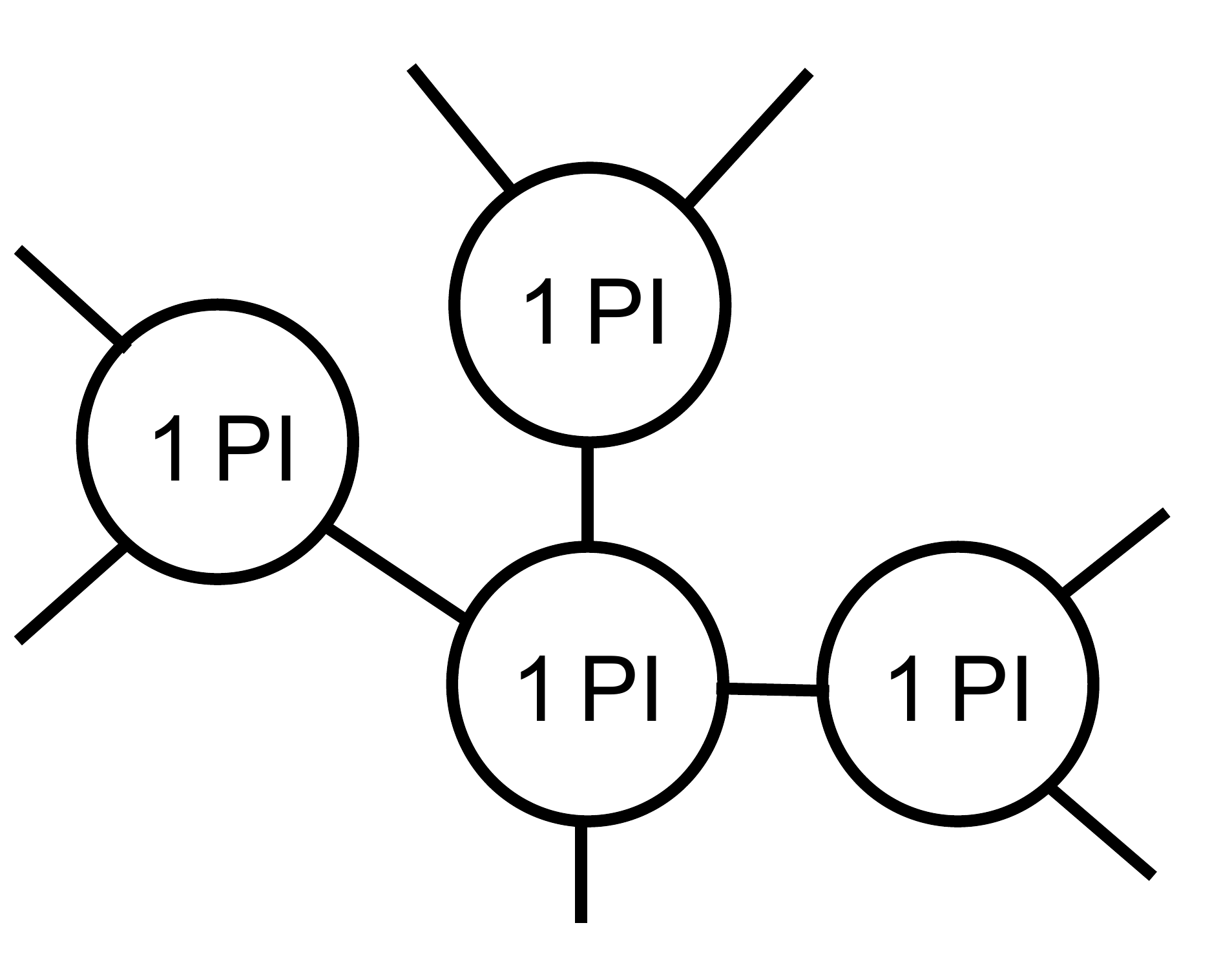}
  \caption{The Feynman diagram we are able to construct using building blocks. The 1PI subdiagrams can be up to two loops.}\label{1pi}
\end{figure}

After finding a construction for all the building blocks up to two loops, we are ready to glue them together to find more general examples. We conjecture the CHY graphs for all the Feynman diagrams shown in figure \ref{1pi} can be constructed using the gluing rules that we are going to illustrate in this section\footnote{At one and two loops (the total number) the conjecture can be verified but no so unless the CHY measure for beyond two loops can be found.}. 

\subsection{Tree-level Gluing}

\begin{figure}[!h]
 \centering
         \includegraphics[scale=0.3]{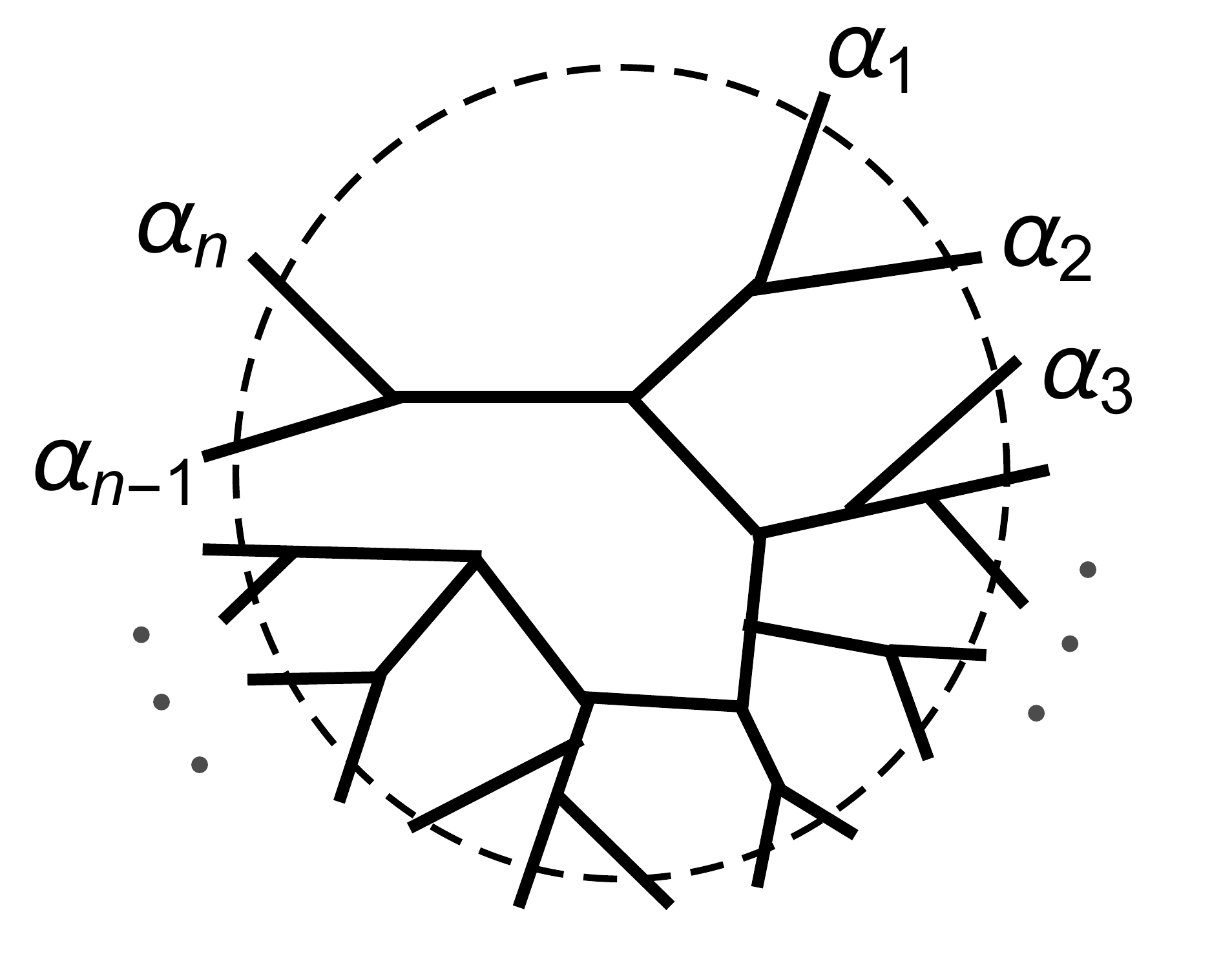}
  \caption{One possible ordering $\alpha$ compatible with a Feynman diagram.}\label{od}
\end{figure}

For the $\Phi^3$ theory at tree level, the gluing procedure was previously considered in \cite{Baadsgaard:2015ifa}. We start by reviewing this procedure in a language that will be useful in generalizing it to higher loops. First, notice that each Feynman diagram $F$ has one or more {\it compatible planar orderings} $\alpha(F)$, i.e., the possible orderings of particles of fitting the Feynman diagram into a circle, as shown in figure \ref{od}. Since each trivalent vertex can be flipped, in general there are more than one compatible orderings. 

\begin{figure}[!h]
  \centering
        \includegraphics[scale=0.5]{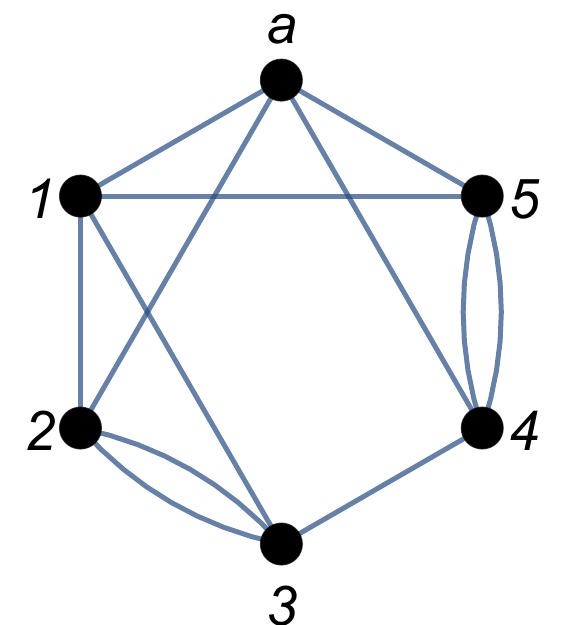}    \qquad\qquad    
         \includegraphics[scale=0.2]{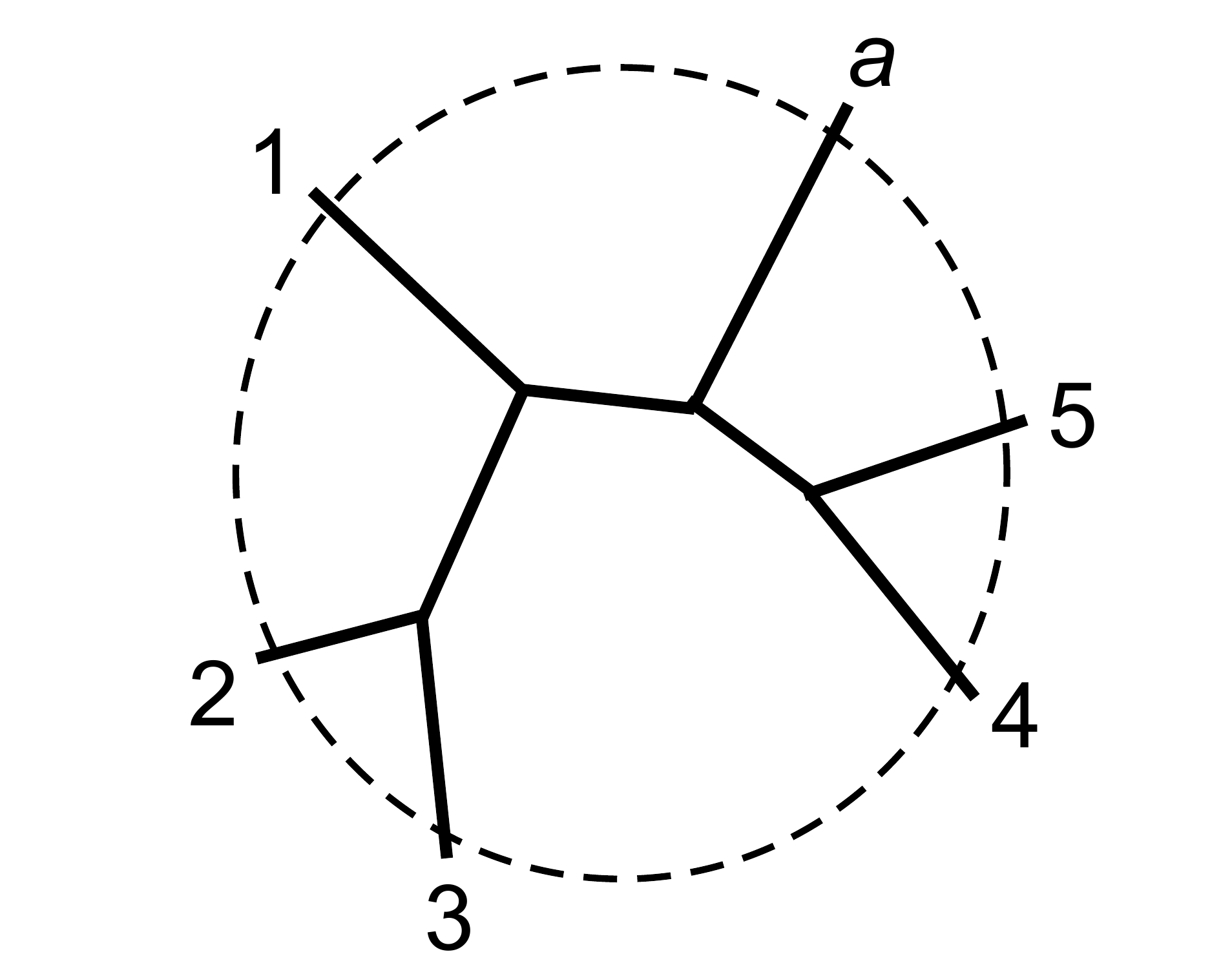}
  \caption{Left: a tree-level CHY graph $G$. Right: a corresponding compatible ordering $\alpha(F(G))$.}\label{egoe}
\end{figure}

On the other hand, the corresponding CHY graph $G$\footnote{There are generally more than one such CHY graphs but choosing any one does not influence the result.} is four-valent: for each graph, every node has four edges. And we define the {\it edge set} ${\rm Edge}(a, G)$ of a node $a$ in the graph $G$ as:
\be
{\rm Edge}(a, G)=\textrm{The set of nodes connected to $a$ by edges in the CHY graph $G$.}
\en
Notice here ${\rm Edge}(a, G)$ may have repeated elements which happens when $a$ is connected with another node by two edges. In order to show the dependence on the compatible ordering, it is necessary to sort $Edge(a,G)$ to define a new object that we call {\it ordered edge set}:
\be
{\rm OE}(a, G)=\textrm{${\rm Edge}(a,G)$ sorted which preserves the ordering $\alpha(F(G))$,}
\en
where the notation $F(G)$ means the Feynman diagram related to the CHY graph $G$.  To understand the definition of ${\rm OE}(a,G)$ better, we give an example in figure \ref{egoe}. From the left and right graphs, it is easy to read ${\rm OE}(a, G)=\{1,2,4,5\}$. Moreover, ${\rm OE}(a, G)=\{1,2,5,4\}$ is also a possible ordered edge set since $\{1,2,3,5,4,a\}$ is another compatible planar ordering $\alpha(F(G))$.  Since there is no $\alpha(F(G))$ as $\{1,5,3,2,4,a\}$, we conclude that ${\rm OE}(a, G)=\{1,5,2,4\}$ is not allowed. Although there could be many choices of ${\rm OE}(a, G)$, we propose the gluing operations that will be defined later are equivalent up to a global sign.

\begin{figure}[!h]
  \centering
        \includegraphics[scale=0.22]{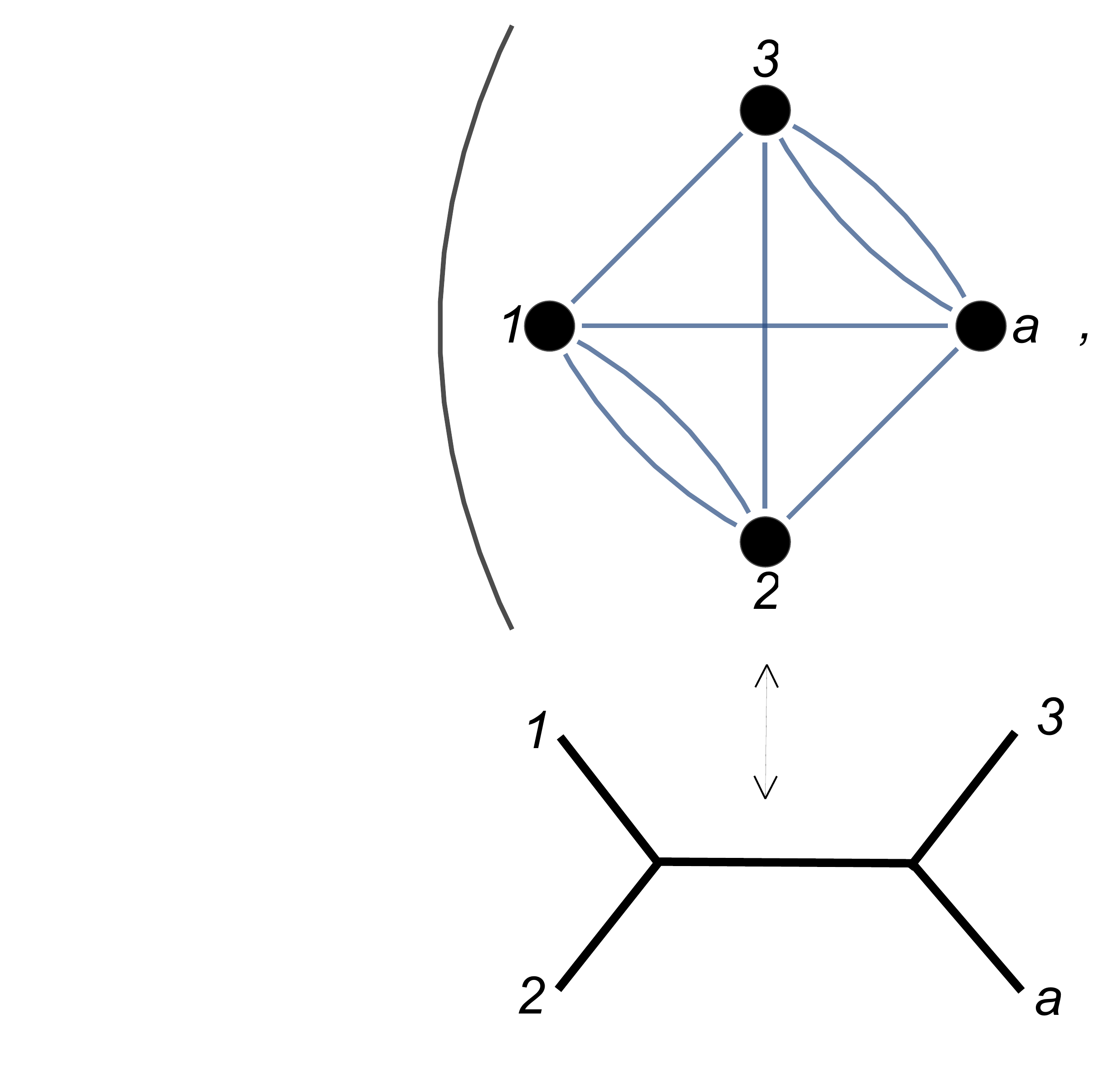}
        \includegraphics[scale=0.22]{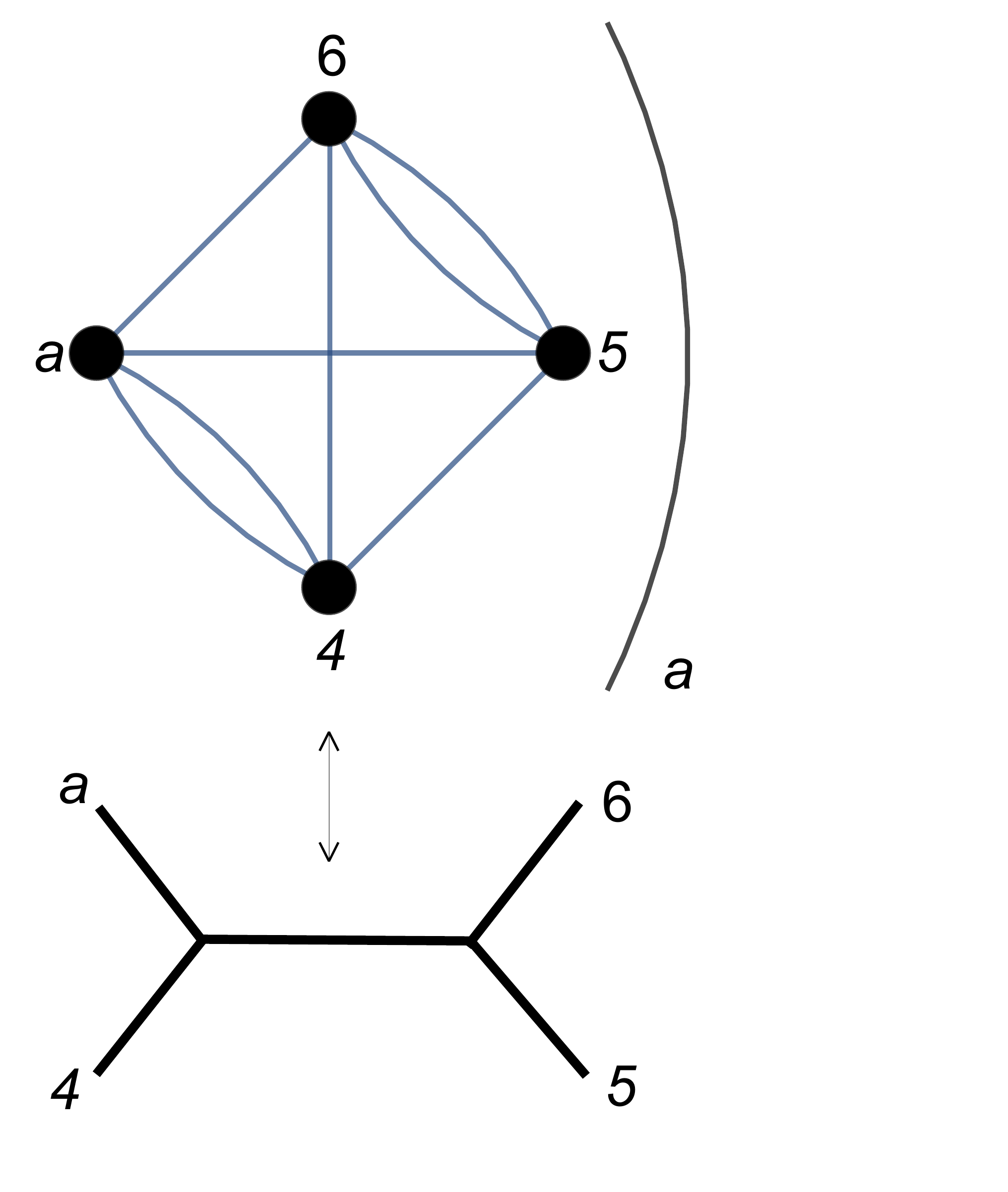}
  \caption{An example of the gluing operation. We draw the CHY graphs in the first line and the corresponding Feynman diagrams in the second.}\label{chyandfey}
\end{figure}

\begin{figure}[!h]
  \centering
        \includegraphics[scale=0.25]{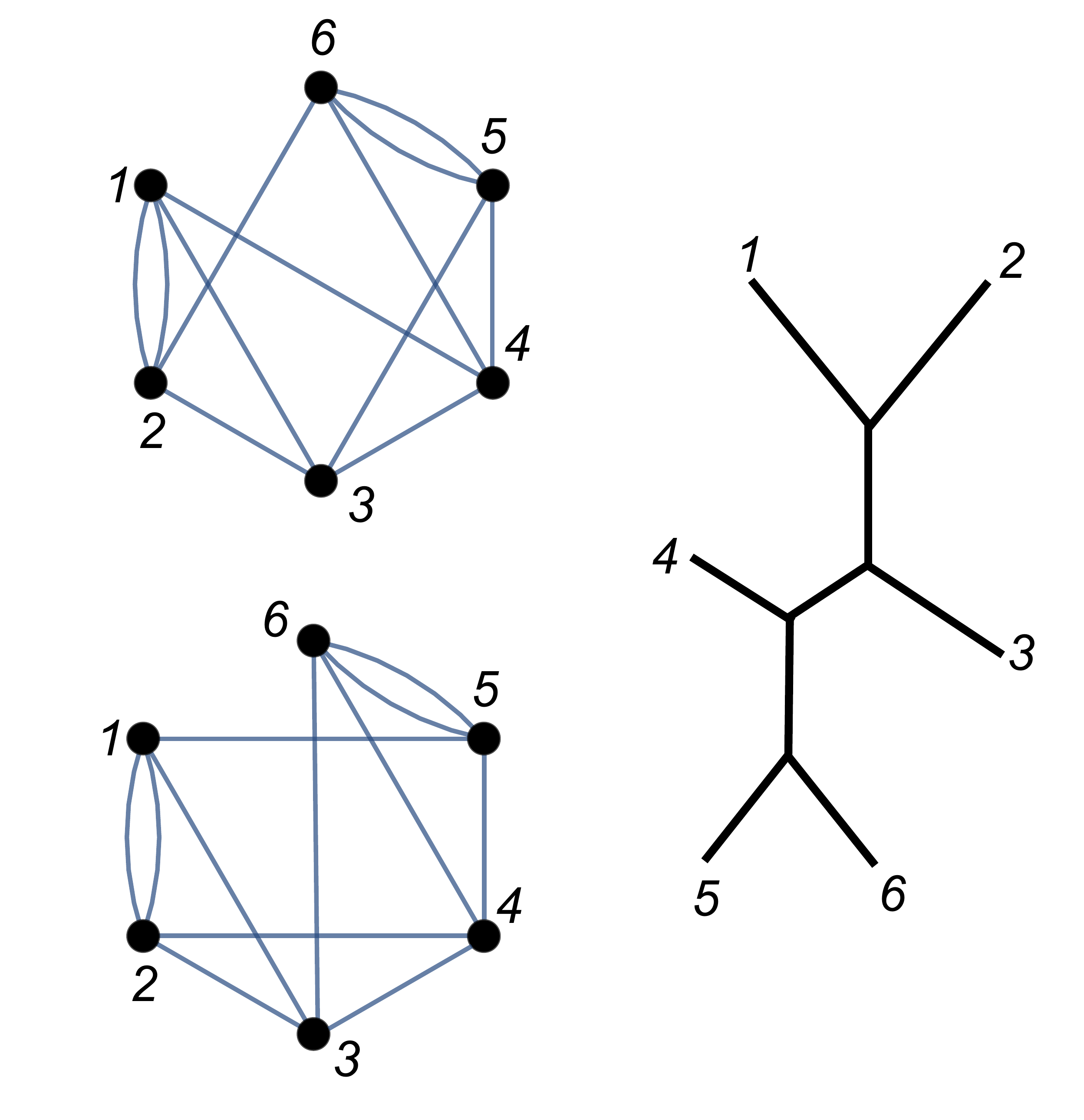}
  \caption{Two equivalent CHY graphs after gluing. They correspond to the same Feynman diagram shown on the right.}\label{egtree}
\end{figure}

Equipped with the ordered edge set, we are ready to define the gluing operation $({\bf\cdot},{\bf\cdot})_a$:
\be
(G_1,G_2)_a=\frac{(\prod_{i \in \alpha} \sigma_{a i})(\prod_{j \in \beta } \sigma_{a j})}{\sigma_{\alpha_1 \beta_1}\sigma_{\alpha_2 \beta_3}\sigma_{\alpha_3 \beta_2}\sigma_{\alpha_4 \beta_4}}G_1 G_2,\;\alpha={\rm OE}(a,G_1),\;\beta={\rm OE}(a,G_2).
\en
\begin{figure}[!h]
  \centering
       \includegraphics[scale=0.2]{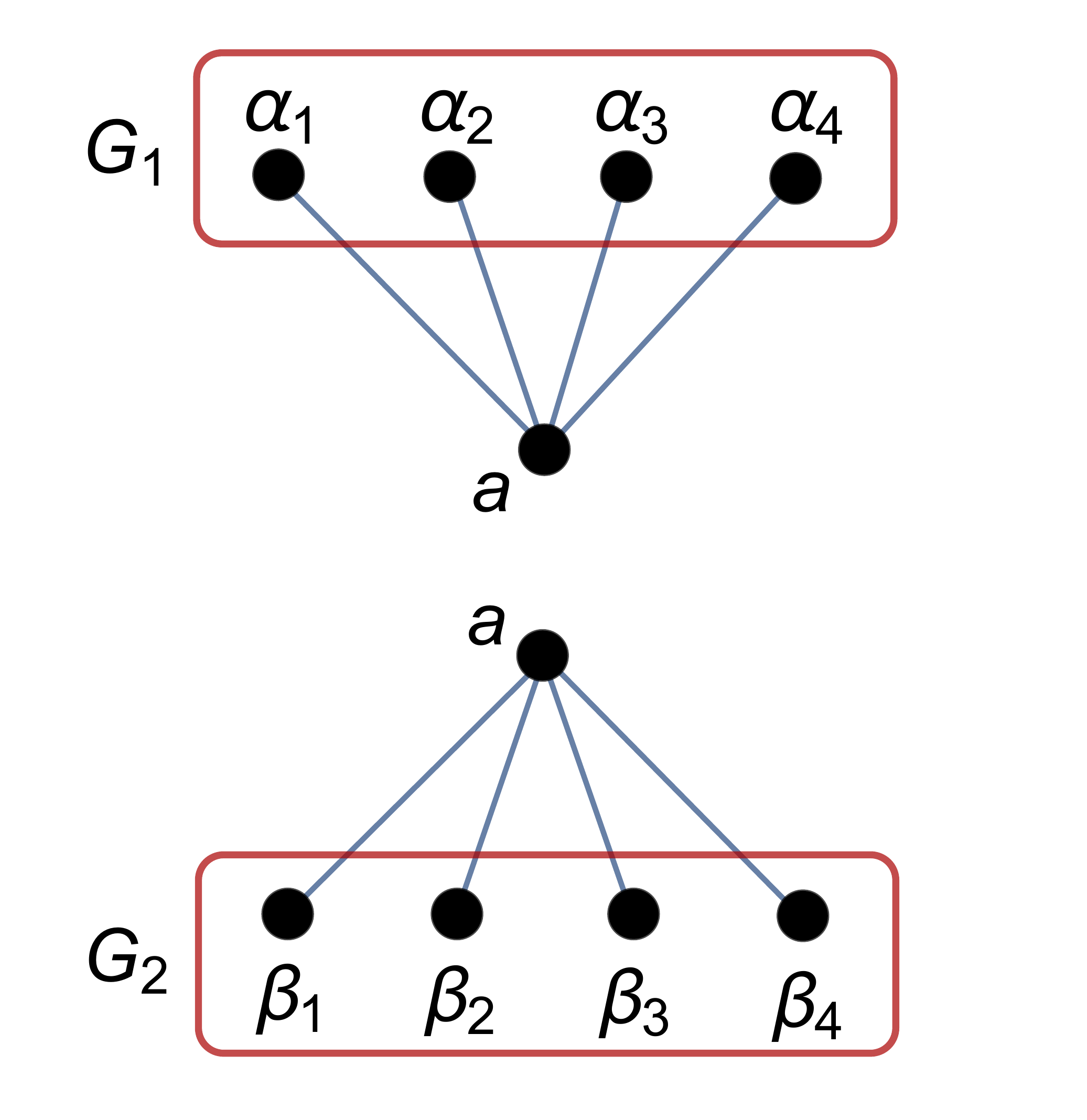}
        \includegraphics[scale=0.2]{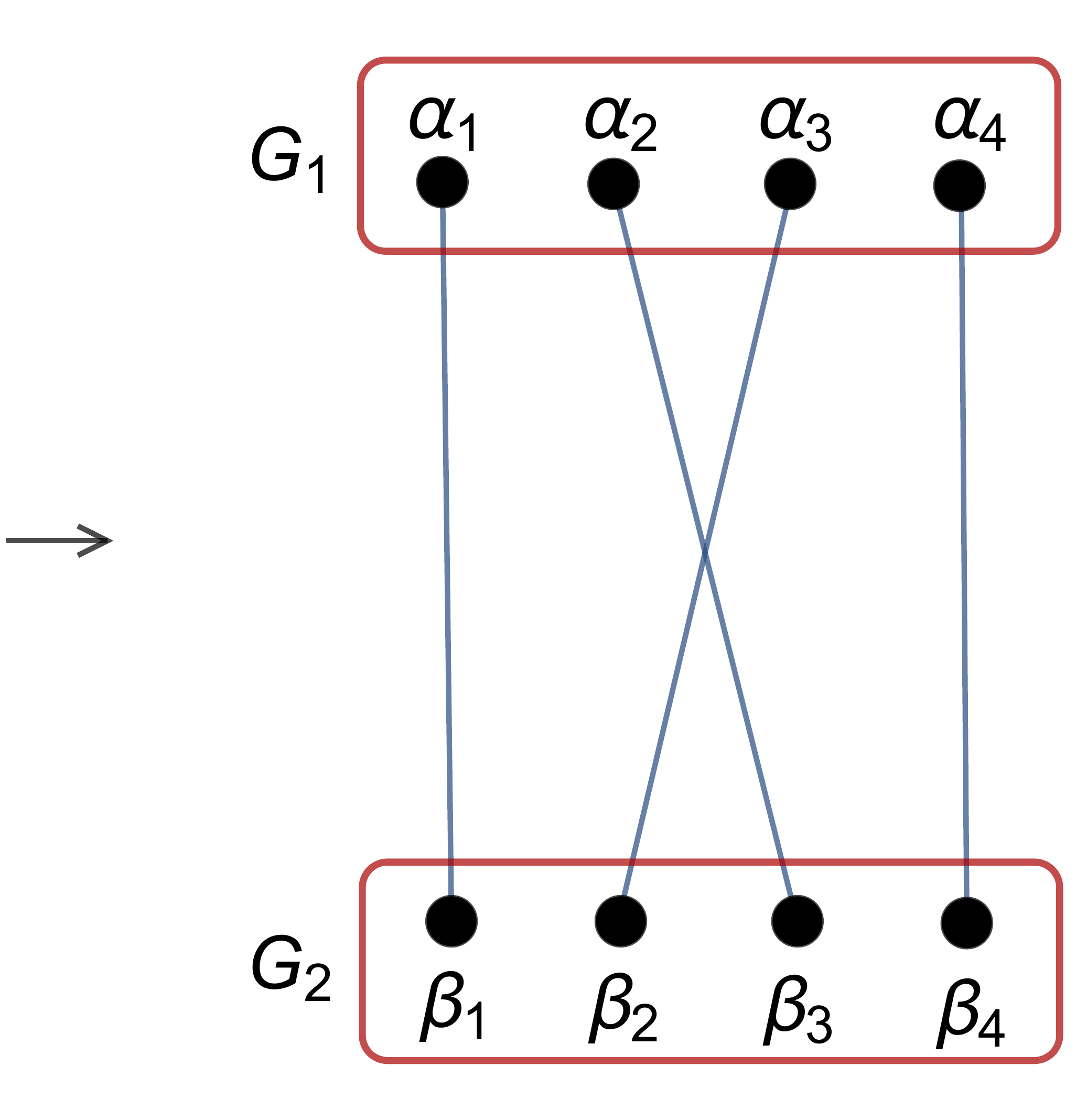}  
  \caption{The gluing operation. Here ${\rm OE(a,G_1)=\{\alpha_1,\alpha_2,\alpha_3,\alpha_4\}}$ and ${\rm OE(a,G_2)=\{\beta_1,\beta_2,\beta_3,\beta_4\}}$. The pairs glued are $(\alpha_1,\beta_1),\,(\alpha_2,\beta_3),\,(\alpha_3,\beta_2)$ and$\,(\alpha_4,\beta_4)$.}\label{glue}
\end{figure}
Figure \ref{glue} gives a graphic interpretation of it. In general, there could be many ways to glue two graphs since ${\rm OE}(a,G)$ may have more than one choice. For example, for the graphs shown in figure \ref{chyandfey}, there are two possible choices: ${\rm OE}(a,G_1)=\{2,1,3,3\},\,{\rm OE}(a,G_2)=\{4,4,5,6\}$; and ${\rm OE}(a,G_1)=\{2,1,3,3\},\,{\rm OE}(a,G_2)=\{6,5,4,4\}$. They give different CHY graphs but the corresponding Feynman diagram is the same, as shown in figure \ref{egtree}.

\begin{figure}[!h]
  \centering
        \includegraphics[scale=0.28]{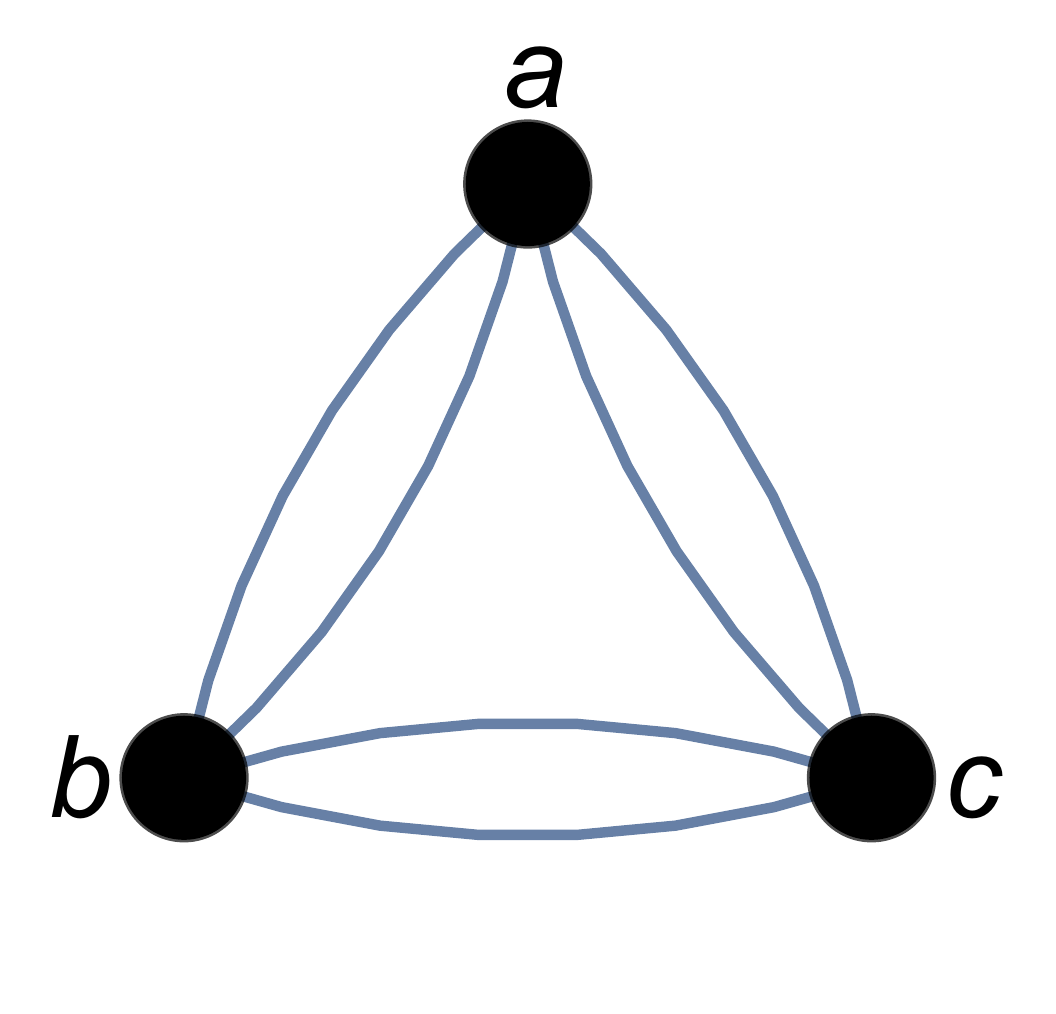}\qquad\quad\qquad
        \includegraphics[scale=0.28]{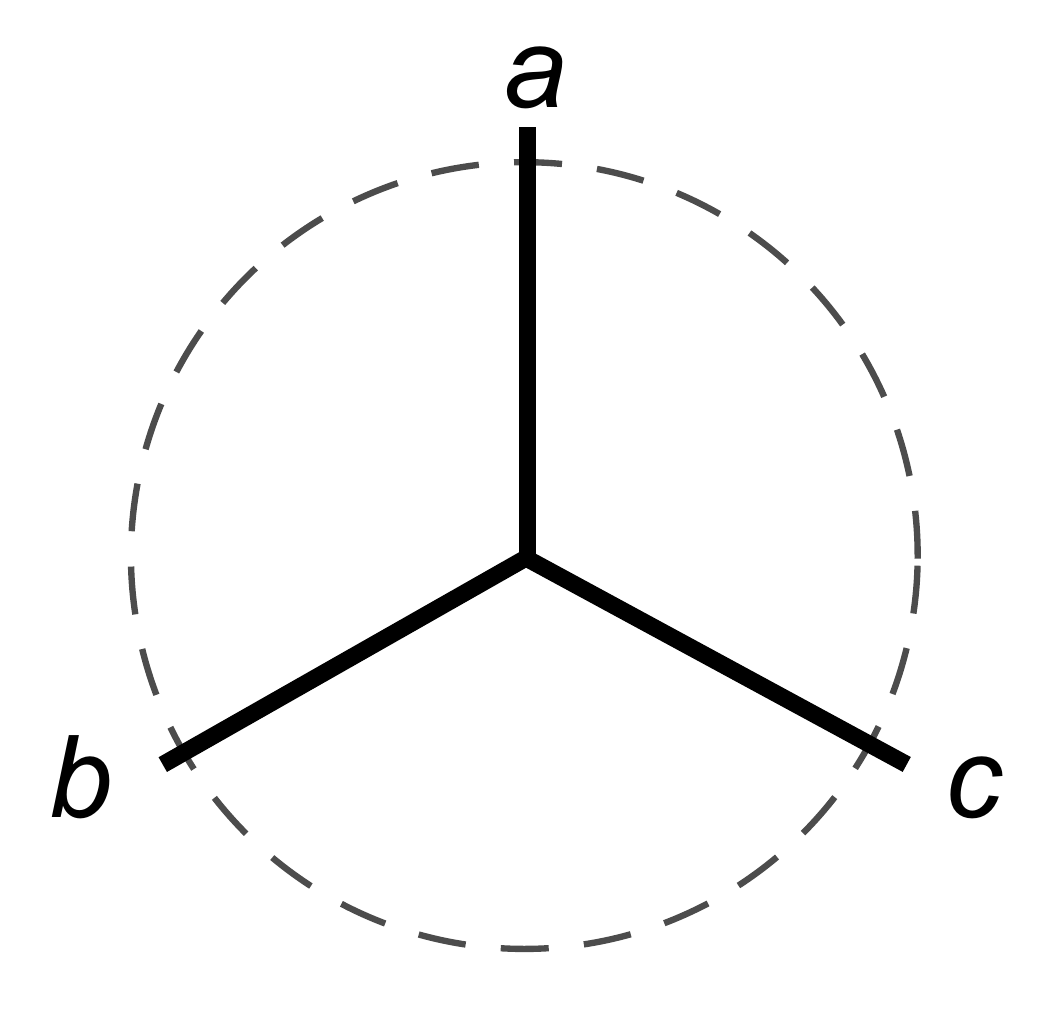}  
  \caption{The tree-level building block. The left is the Feynman diagram and the right the corresponding CHY graph.}\label{tbb}
\end{figure}

Finally, we are able to use the three-point tree-level building block, shown in figure \ref{tbb}, and the gluing operation to generate any tree-level CHY graph, namely:
\be
\textrm{tree-level CHY graph}=((\dots (B_1,B_2)_a,B_3)_b,\dots),
\en
where $B_i$'s are the three-point building blocks.

\subsection{One-loop Level Gluing}

Next let us consider how to build the CHY graph for the Feynman diagram figure \ref{1pi} where the 1PI subdiagrams are up to one-loop level.  Different from the tree-level case, the definition of the ordered edge set ${\rm OE}(a,G)$ should be modified since ${\rm Edge}(a, G)$ may contain loop momenta.

\begin{figure}
  \centering
        \includegraphics[scale=0.25]{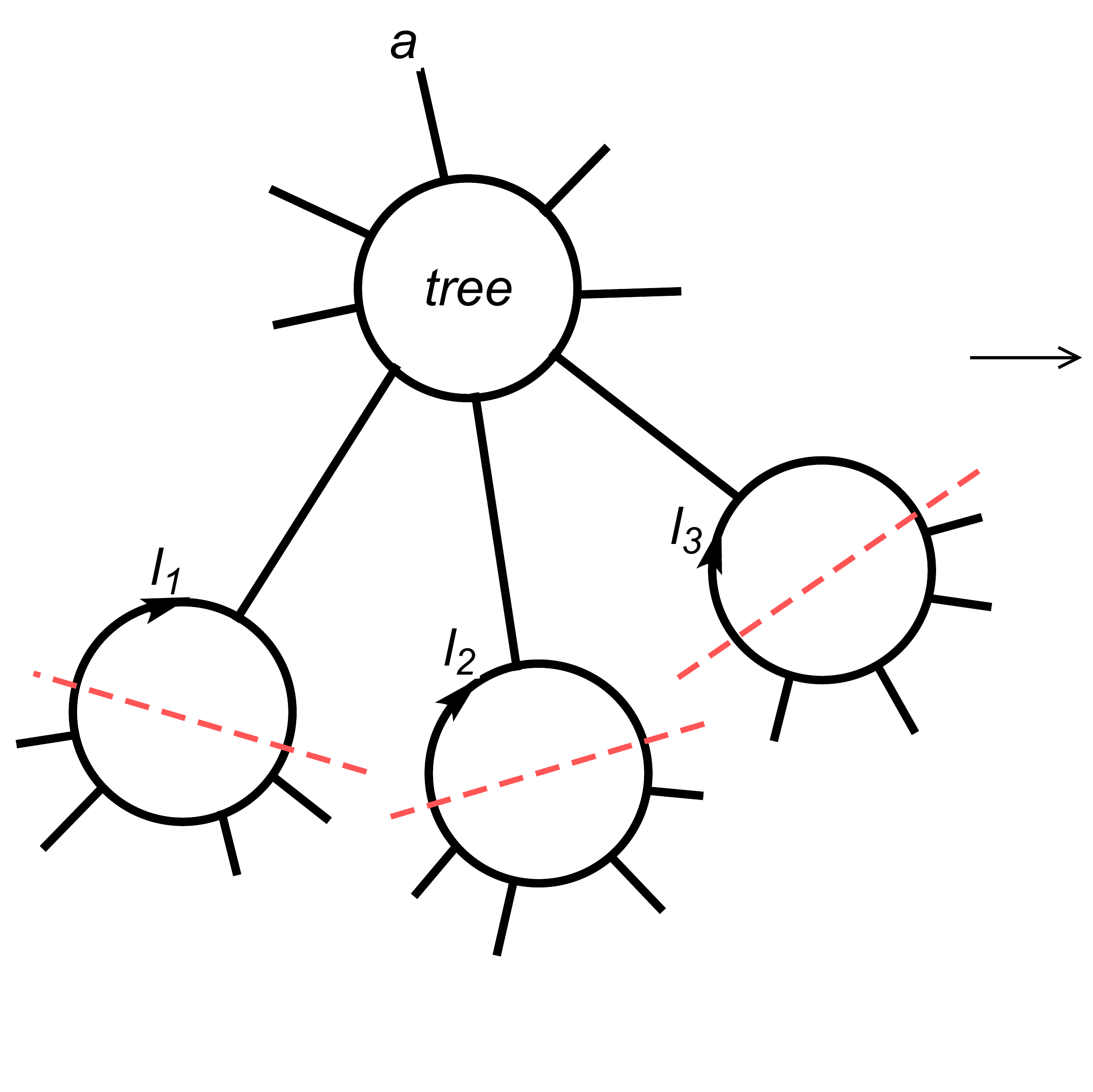}
          \includegraphics[scale=0.25]{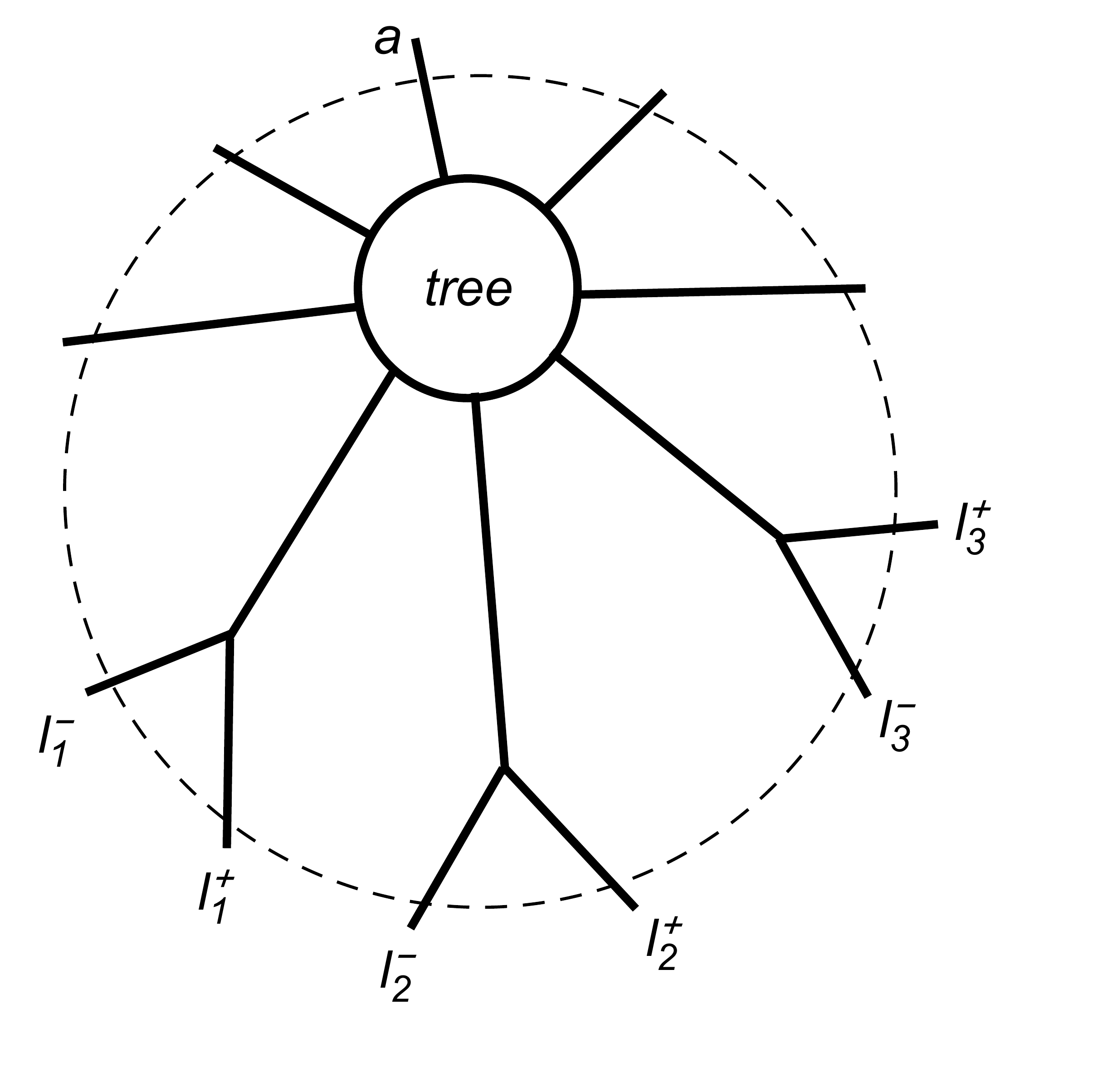}
  \caption{The definition of $F_1(a, F)$, each loop attached is replaced by a pair of loop momenta.}\label{1lc}
\end{figure}

The way out is to study the {\it partially cut Feynman diagram} for $a$, denoted as $F_1(a, F)$, which is defined in figure \ref{1lc}. For each loop $l_r$, one replaces the loop part with a pair $l_r^+,l_r^-$. Thus we define the one-loop ordered edge set:
\be
{\rm OE_1}(a, G)=\textrm{${\rm Edge}(a,G)$ sorted which preserves the ordering $\alpha( F_1(a,F(G)))$,}
\en
The gluing operation is similarly defined as the tree-level case:
\be
(G_1,G_2)_a=\frac{(\prod_{i \in \alpha} \sigma_{a i})(\prod_{j \in \beta } \sigma_{a j})}{\sigma_{\alpha_1 \beta_1}\sigma_{\alpha_2 \beta_3}\sigma_{\alpha_3 \beta_2}\sigma_{\alpha_4 \beta_4}}G_1 G_2,\;\alpha={\rm OE_1}(a,G_1),\;\beta={\rm OE_1}(a,G_2).
\en

\subsection{Two-loop Level Gluing}
\label{2lg}
Finally it is possible to generalize the gluing operation to two-loop level. Other than the problem we meet at one loop, at two loops, an additional obstacle is that for one CHY integrand, there could be more than one CHY graphs that contribute. For example, in equations \ref{q1} and \ref{q2}, each $\q^1_a$ and $\q^2_a$ contains two terms and the whole expansion will yield $2^{k+m}$ CHY graphs.  And in order to figure out the right ordering in ${\rm Edge}(a,G)$, one needs to define a few more objects.

\begin{figure}[!h]
	\centering
	\includegraphics[scale=0.28]{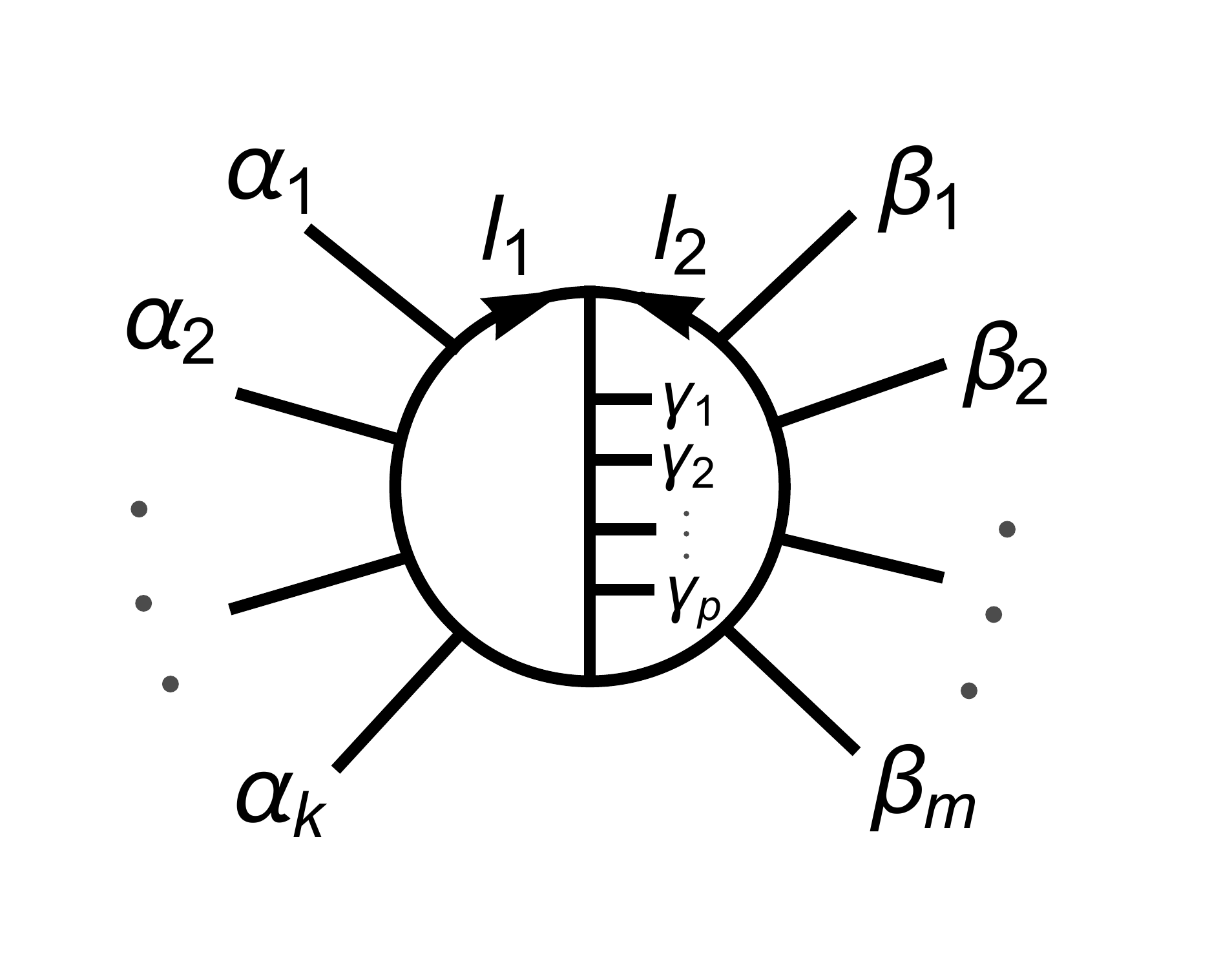}
	\includegraphics[scale=0.28]{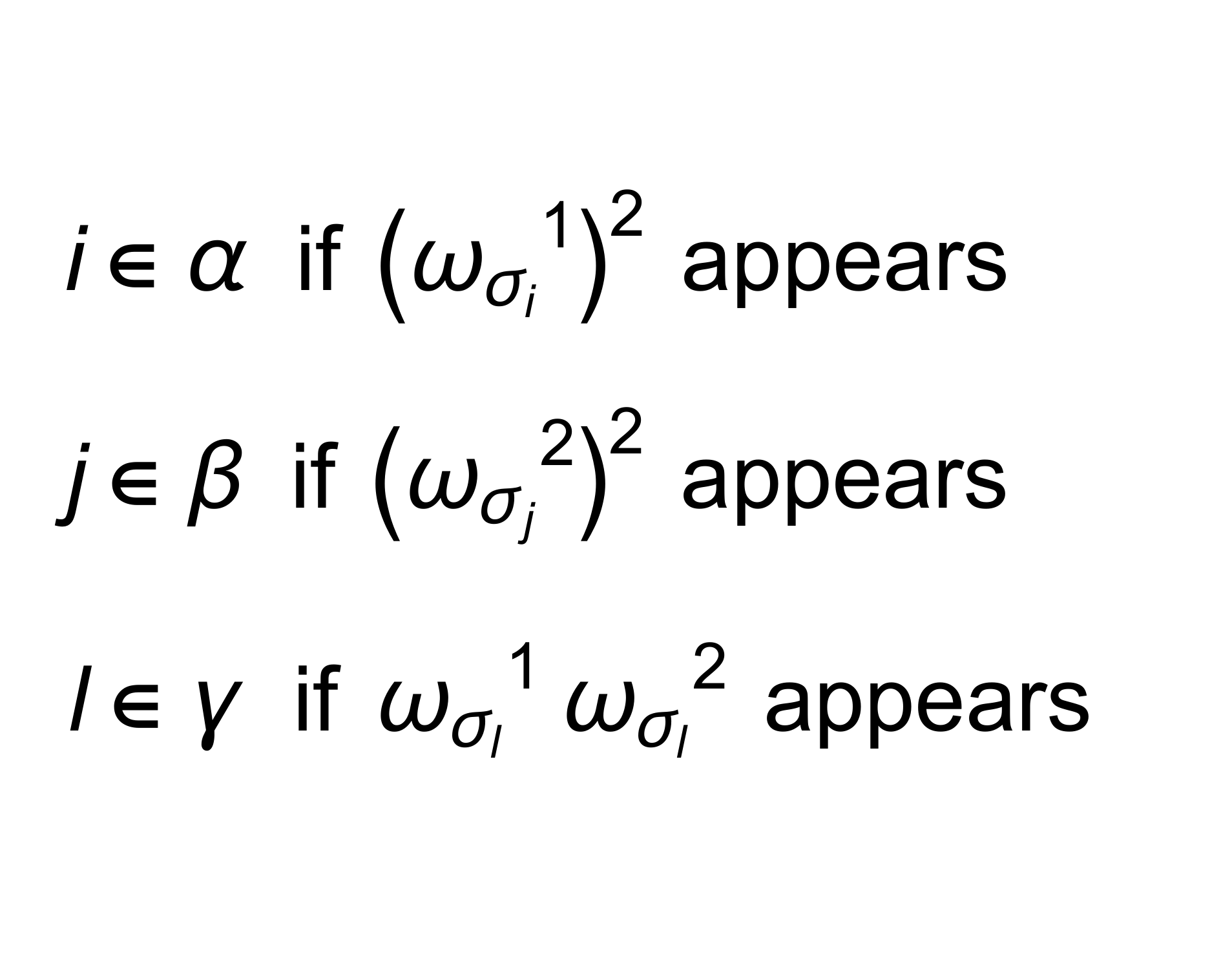}
	\caption{The definition of $\tilde{F}(G)$.}\label{fg}
\end{figure}

First we define $\tilde{F}(G)$ for a CHY two-loop building block $G$, as in figure \ref{fg}. Keep in mind that it is in general different from Feynman diagrams at two loops. For instance, for a planar CHY graph $G$, a non-planar $\tilde{F}(G)$ could arise which is seen from \eqref{q1} and \eqref{q2}. $\tilde{F}(G)$ is obtained by gluing smaller building blocks together. In this way, each CHY graph at two loops has one-to-one correspondence with the $\tilde{F}(G)$, once the gluing operation is determined.

\begin{figure}[t]
  \centering
        \includegraphics[scale=0.25]{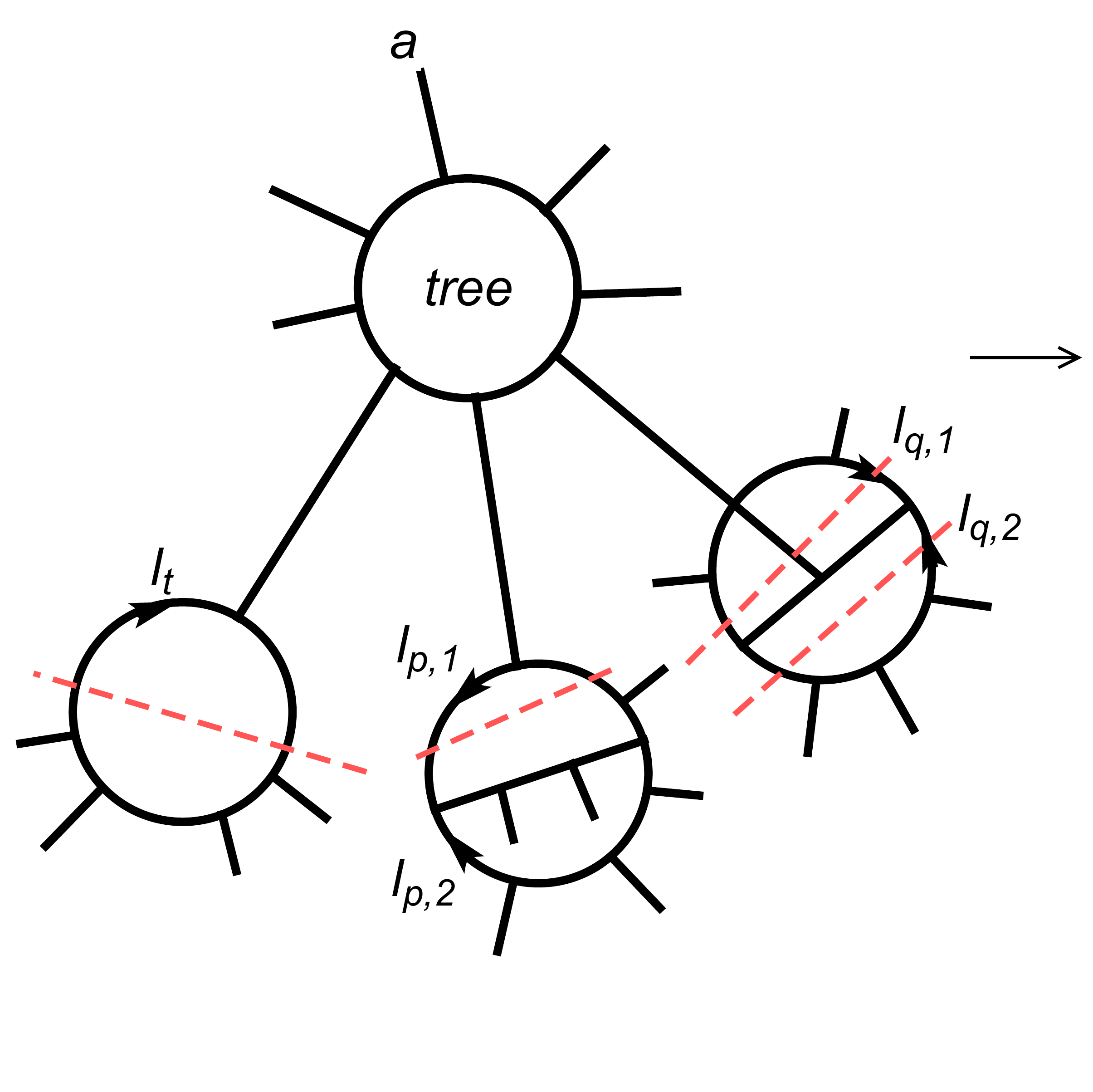}
        \includegraphics[scale=0.25]{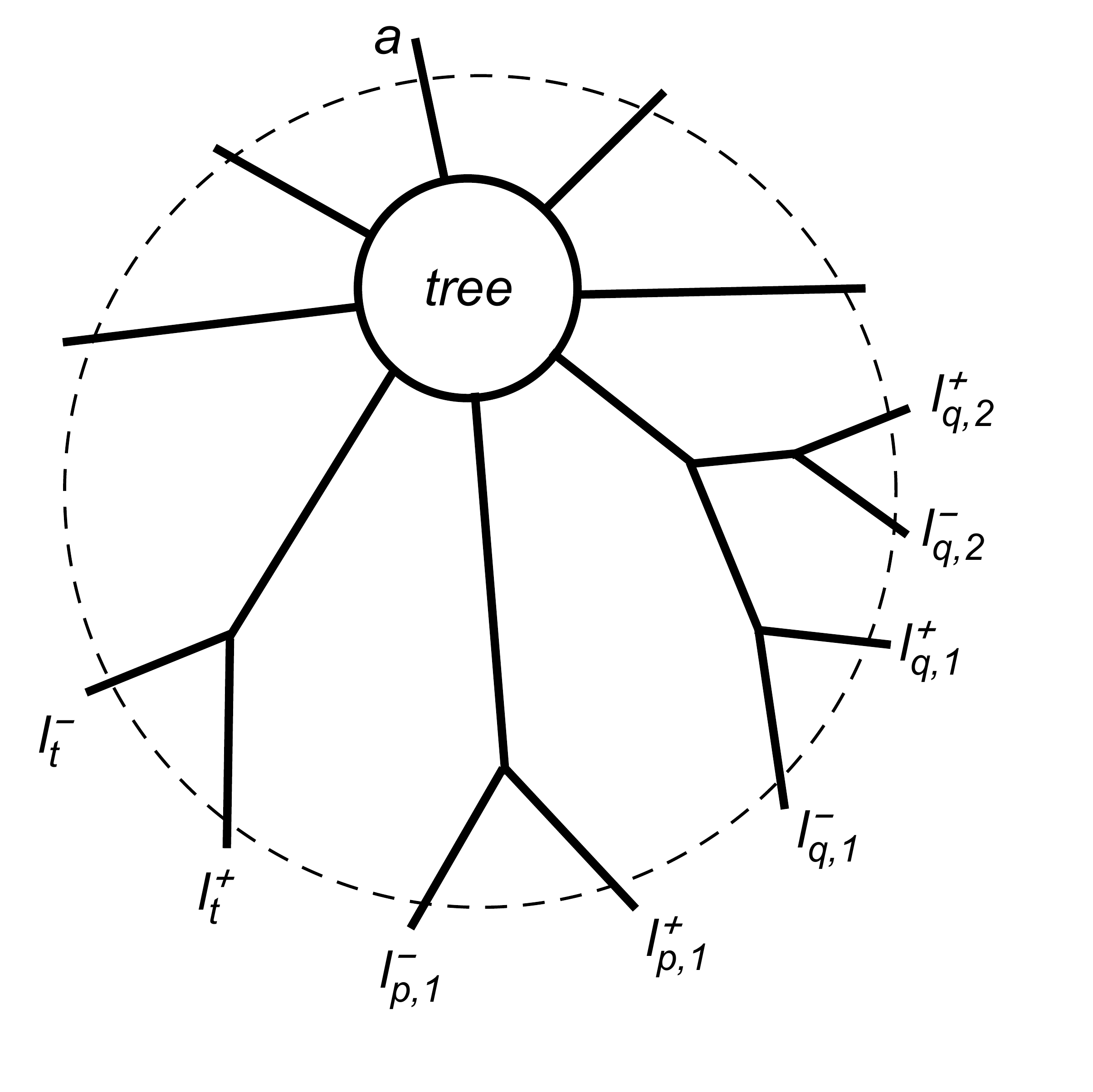}
  \caption{The definition of $F_2(a,\tilde{F})$. Each one-loop attached is replaced by a pair of loop momenta. Each two-loop is replaced by a pair of loop momenta if the leg is attached to the side and by two pairs of loop momenta if the leg is attached to the middle.}\label{2la}
\end{figure}

The key point lies in defining the {\it partially cut} version of $\tilde{F}(G)$, as a generalization of the one-loop partially cut Feynman diagram for $a$, denoted as $F_2(a,\tilde{F})$. By carefully studying examples, we propose a graphic definition of $F_2(a,\tilde{F})$ in figure \ref{2la}.

After figuring out $F_2(a,\tilde{F})$, we are able to generalize the ordered edge set and the gluing operation to two loops:
\be
&\,&{\rm OE_2}(a, G)=\textrm{${\rm Edge}(a,G)$ sorted which preserves the ordering $\alpha( F_2(a,\tilde{F}(G)))$,}\nonumber\\
&\,&(G_1,G_2)_a=\frac{(\prod_{i \in \alpha} \sigma_{a i})(\prod_{j \in \beta } \sigma_{a j})}{\sigma_{\alpha_1 \beta_1}\sigma_{\alpha_2 \beta_3}\sigma_{\alpha_3 \beta_2}\sigma_{\alpha_4 \beta_4}}G_1 G_2,\;\alpha={\rm OE_2}(a,G_1),\;\beta={\rm OE_2}(a,G_2).
\en
\newline
Since the definition of ${\rm OE_2}(a,G)$ contains the previous one-loop and tree-level cases, we will redefine ${\rm OE}(a,G):={\rm OE_2}(a,G)$ in practice. The gluing rules will be illustrated by examples in appendix \ref{GluingL}. 
\section{$\L-$Rules}\label{lambdarules}

In order to check our conjectures given in sections \ref{building-blocks} and \ref{gluesection}, we are going to consider some non-trivial examples. Nevertheless, since there is no algorithm to compute CHY integrals with four off-shell particles (two-loop computations), we will make a modification to the $\L-$algorithm in appendix \ref{L-algorithm-2L}. And the most important rules are summarized in this section.

In the CHY approach at two loops, four new punctures emerge and we denote their coordinates, in the double cover language, and momenta as
\begin{align}
\{(\s_{\ell_1^+},y_{\ell_1^+}),(\s_{\ell_1^-},y_{\ell_1^-}),(\s_{\ell_2^+},y_{\ell_2^+}),(\s_{\ell_2^-},y_{\ell_2^-})\}\quad \longrightarrow\quad \{\ell^\mu_1,-\ell^\mu_1,\ell^\mu_2,-\ell^\mu_2  \}.
\end{align}
Using the $\L-$prescription, we fix three of them $\{(\s_{\ell_1^-},y_{\ell_1^-}),(\s_{\ell_2^+},y_{\ell_2^+}),(\s_{\ell_2^-},y_{\ell_2^-})\}$ by ${\rm PSL}(2,\mathbb{C})$ symmetry, and the other one  $\{(\s_{\ell_1^+},y_{\ell_1^+})\}$ by scaling symmetry. Therefore we must know the behavior of the scattering equation $E_{\ell_1}$ so as to apply the $\L-$algorithm. This study is realized in detail in appendix \ref{Ltwoloop} and here we just summarize the results.
\begin{itemize}

\item {\bf $\s_{\ell_1}$ and $\s_{\ell_1^-}$ on the same sheet.}\\
Without loss of generality, let us consider the punctures, $\{ \s_1,\ldots, \s_{n_u},\s_{\ell_1^+},\s_{\ell^-_1}\}$ on the same sheet, so
\begin{equation}\label{ruleI}
\frac{1}{E_{\ell_1}}~~\rightarrow~~\frac{1}{k_{12\cdots n_u}},\qquad\textbf{ Rule\,I}.
\end{equation}

\item {\bf $\s_{\ell_1^+}$ and $\s_{\ell_2^+}$ on the same sheet.}\\
The next case is to consider the punctures, $\{ \s_1,\ldots, \s_{n_u},\s_{\ell_1^+},\s_{\ell_2^+}\}$ on the same sheet, where $E_{\ell_1}$ turns into
\begin{equation}\label{ruleII}
\frac{1}{E_{\ell_1}}~~\rightarrow~~\frac{1}{\frac{1}{2}(\ell_1 + \ell_2+k_1+\cdots k_n )^2},\qquad\textbf{ Rule\,II},
\end{equation}

\item {\bf $\s_{\ell_1^+}$ and $\s_{\ell_2^-}$ on the same sheet.}\\
Finally, consider the punctures $\{ \s_1,\ldots, \s_{n_u},\s_{\ell_1^+},\s_{\ell^-_2}\}$ on the same sheet.
Thus we have
\begin{equation}\label{ruleIII}
\frac{1}{E_{\ell_1}}~~\rightarrow~~\frac{1}{-\frac{1}{2}(\ell_{1}^2 + \ell_{2}^2) -\ell_1\cdot\ell_2+(\ell_1-\ell_2)\cdot(k_1+\cdots +k_{n_u})+k_{12\cdots n_u}},\qquad\textbf{ Rule\,III}.
\end{equation}
\end{itemize}
In the three cases above, after implementing the first $\Lambda$-rule, the $\L-$algorithm is performed in its usual way. 
In addition, it is important to remark that all computations have been performed when choosing the constant $\alpha=1$ which was introduced in \cite{Geyer:2016wjx}.

\subsection*{Notation}

\begin{figure}[!h]
	\centering
	\includegraphics[width=4.6in]{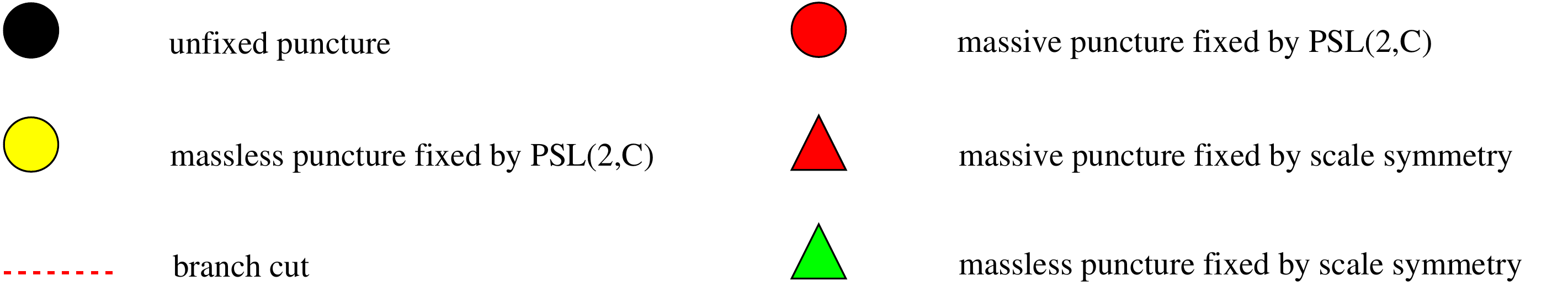}
	\caption{Color code for CHY graphs.}\label{color_cod}
\end{figure}

Since all computations will be performed using the $\L$-algorithm \cite{Gomez:2016bmv}, which is a pictorial technique, we introduce the color code given in figure \ref{color_cod} that will be used in the remaining of the paper.

\noindent
In addition, it is useful to introduce the following notation:
\begin{align}
&k_{a_1\ldots a_m}:=\sum_{a_i<a_j}^m k_{a_i}\cdot k_{a_j},\\
& [a_1,a_2,\ldots, a_m]:=k_{a_1}+k_{a_2}+\cdots + k_{a_m}.
\end{align}

\section{Examples in $\Phi^3$ Theory}\label{examplesphi3}

In this section we give three non-trivial simple examples, in order to check our conjectures and to illustrate how to use the $\L-${\bf Rules} given in section \ref{lambdarules} (for more details see appendix \ref{Ltwoloop}). We begin with a Feynman diagram with two loops separated (one-particle reducible). To construct the corresponding CHY graph, we glue two one-loop building blocks with the ones at tree level, using the technique that was developed in section \ref{gluesection}.

\subsection{One-particle Reducible Diagram}\label{non1PI}
\begin{figure}[!h]
	\centering
	\includegraphics[scale=0.5]{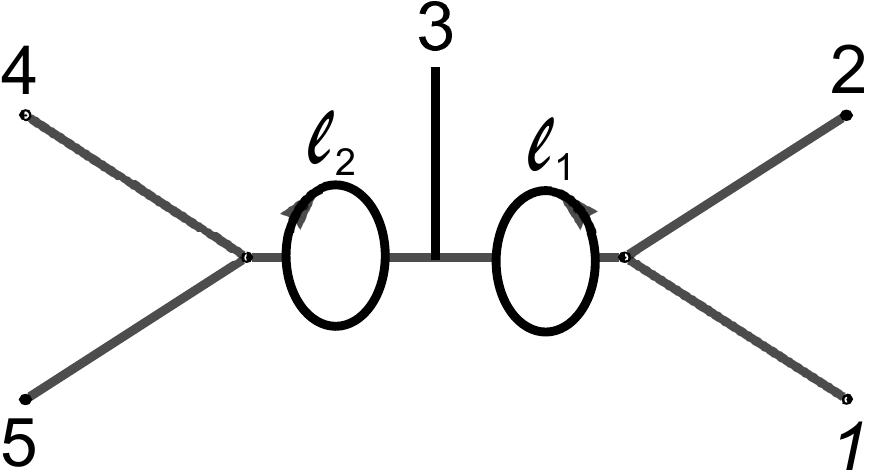}
	\caption{Two-loop one-particle reducible diagram.}\label{nononePI}
\end{figure}
\begin{figure}
  \centering
        \includegraphics[scale=0.4]{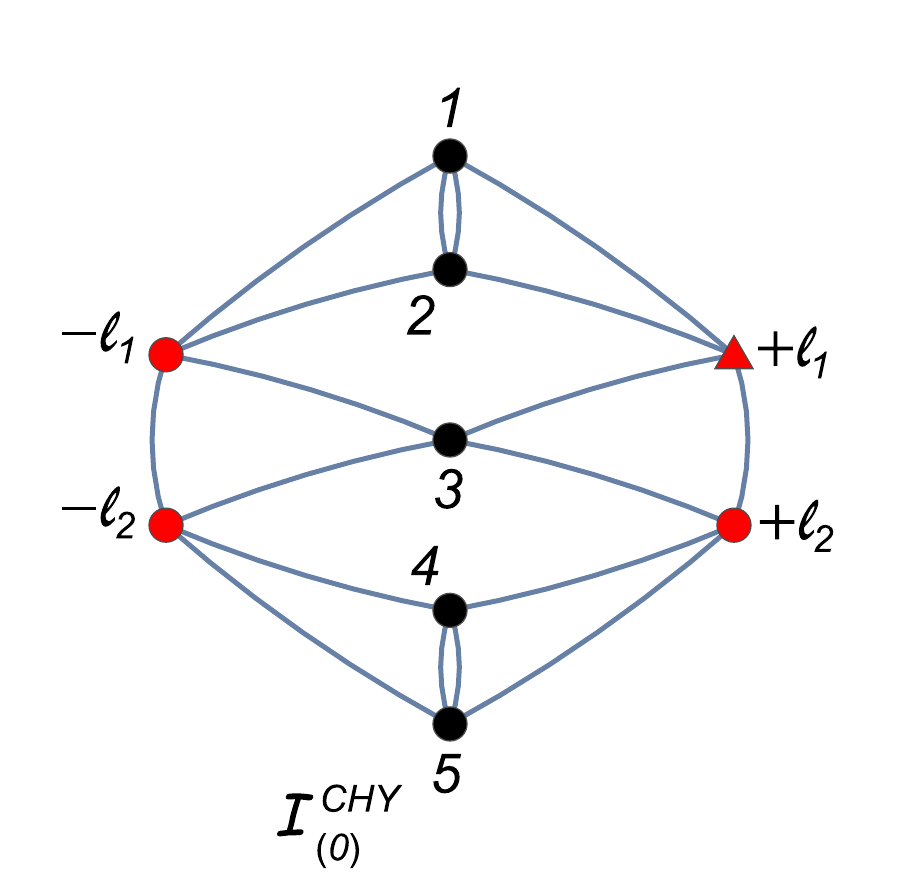}
  \caption{CHY graphs at two loops (one-particle reducible).}\label{chy_5points}
\end{figure}

Let us consider the $\Phi^3$ diagram given in figure \ref{nononePI}.
The loop integrand for it reads\footnote{Note that the $2^4$ factor comes from the propagator $s_{ab} := (k_a+k_b)^2 = 2 k_a\cdot k_b = 2k_{ab}$. }
\begin{equation}
{\cal I}_{\rm FEY}=\frac{1}{2^4\,k_{12}^2\,\ell_1^2\,(\ell_1-k_1-k_2)^2\,k_{45}^2\,\ell_2^2\, (\ell_2-k_4-k_5)^2}.
\end{equation}
Using the partial fraction identity given in \eqref{partialfrac} and shifting the loop momenta $\ell_1$ and $\ell_2$, the Feynman integrand  becomes 
\begin{align}\label{ISintegrand}
{\cal I}_{\rm FEY}\Big|_{\rm p.f \atop S}=\frac{1}{2^6\,k_{12}^2\,k_{45}^2\,\ell_1^2\,\ell_2^2}
&\left[\frac{1}{-\ell_1\cdot(k_1+k_2)+k_{12}}+ \frac{1}{\ell_1\cdot(k_1+k_2)+k_{12} } \right]\\
\times&
 \left[\frac{1}{-\ell_2\cdot(k_4+k_5)+k_{45}}+ \frac{1}{\ell_2\cdot(k_4+k_5)+k_{45} } \right].\nonumber
\end{align}
On the other hand, following the method developed in section \ref{building-blocks}  and   \ref{gluesection},
we can write the CHY integrand corresponding to the Feynman diagram represented in figure \ref{nononePI}.  The gluing process for this Feynman diagram is performed in detail in appendix \ref{glueN1PI} and the result is the CHY integrand given by the expression
\begin{align}
{\cal I}_{\rm CHY} &=\frac{1}{\ell_1^2\,\, \ell_2^2}\int d\mu^{\rm 2-loop}\, {\bf I}_{\rm CHY},  \\
{\bf I}_{\rm CHY} &= \frac{1}{(\ell_1^+,\ell_2^+,\ell_2^-,\ell_1^-)(\ell_1^+,\ell_1^-)(\ell_2^+,\ell_2^-)}\left[\frac{\o^1_{1}\o^1_{2}}{(12)}\times \o^1_{3}\o^2_{3}  \times \frac{\o^2_{4}\o^2_{5}}{(45)}\right].\nonumber
\end{align}
In addition,  the graphic representation of ${\bf I}_{\rm CHY}$ can be seen in figure \ref{chy_5points}.

\begin{figure}
  \centering
        \includegraphics[scale=0.5]{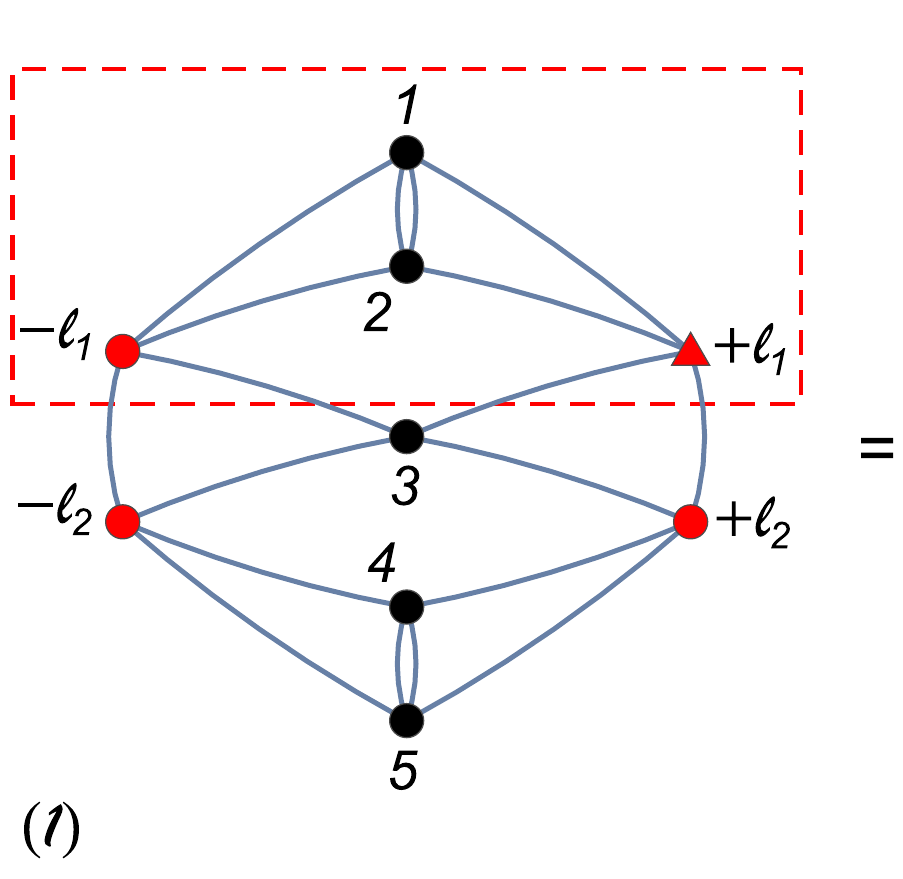}  
        \includegraphics[scale=0.5]{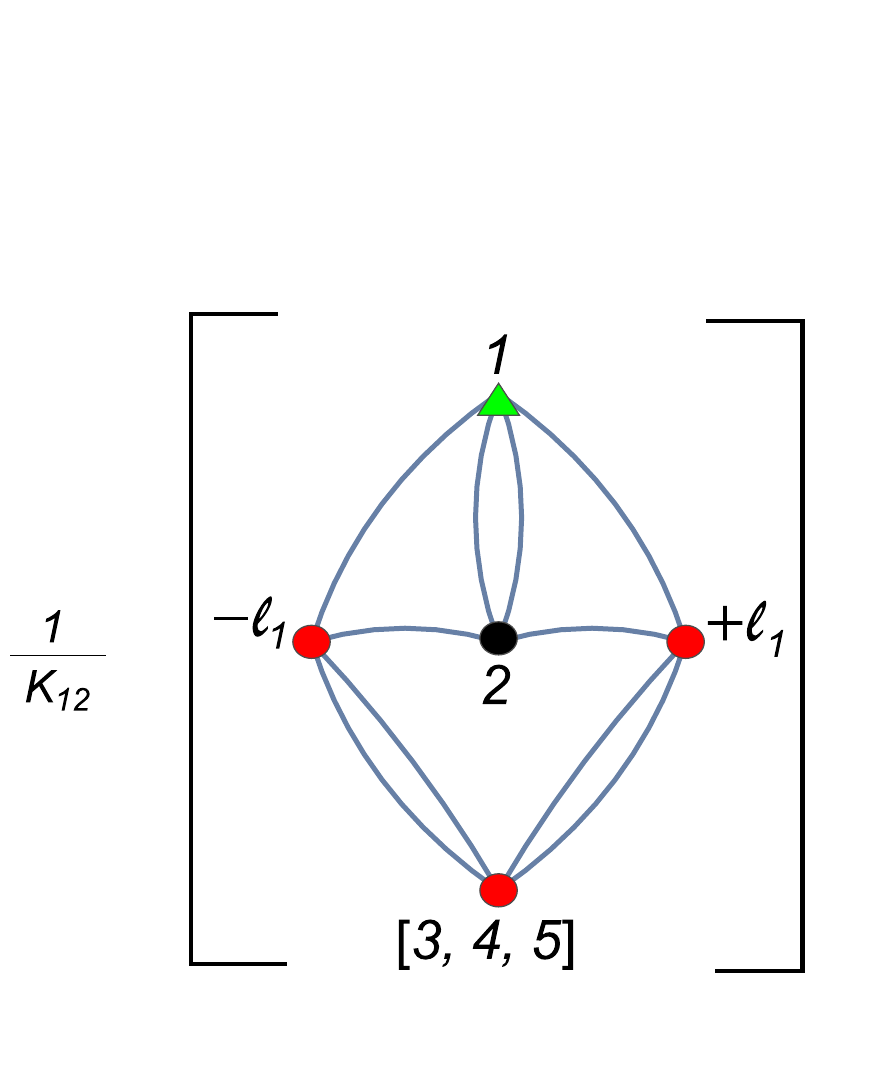}
        \includegraphics[scale=0.5]{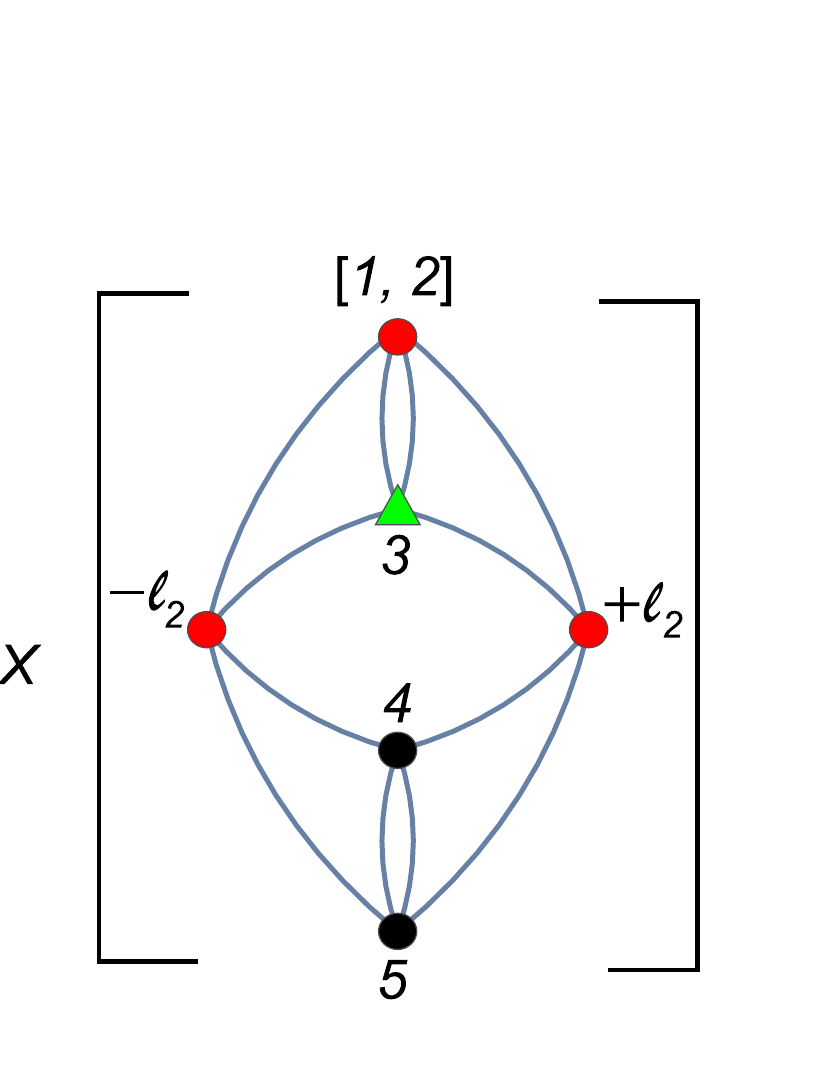}\\
        \includegraphics[scale=0.5]{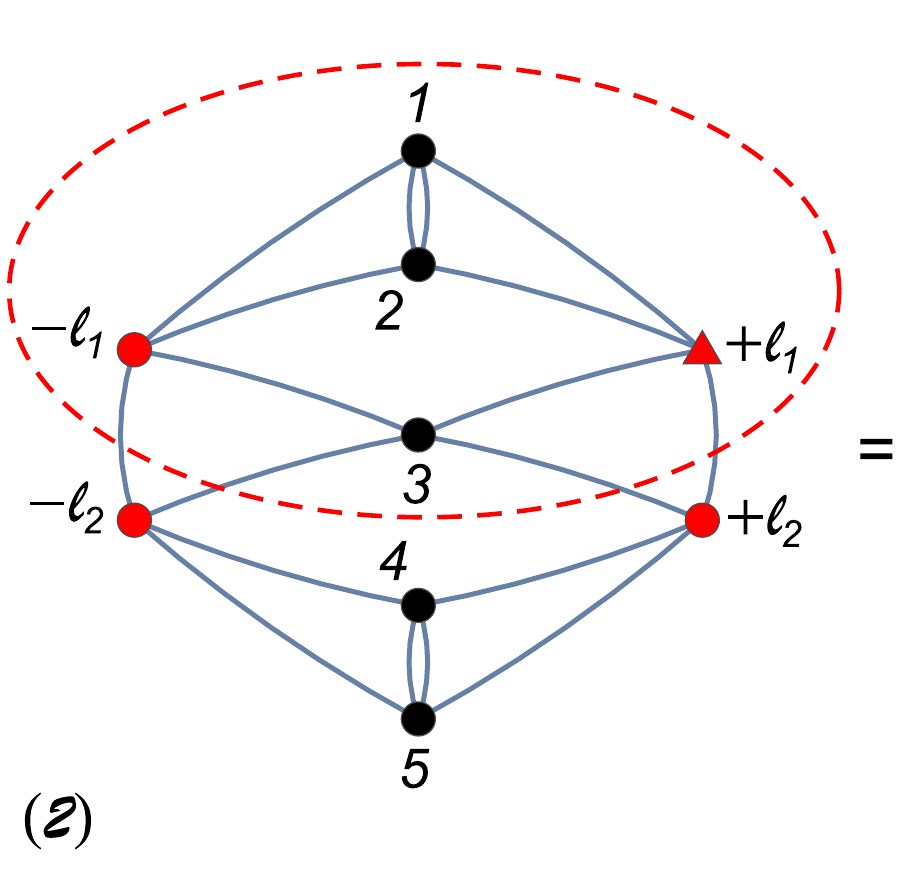}  
        \includegraphics[scale=0.5]{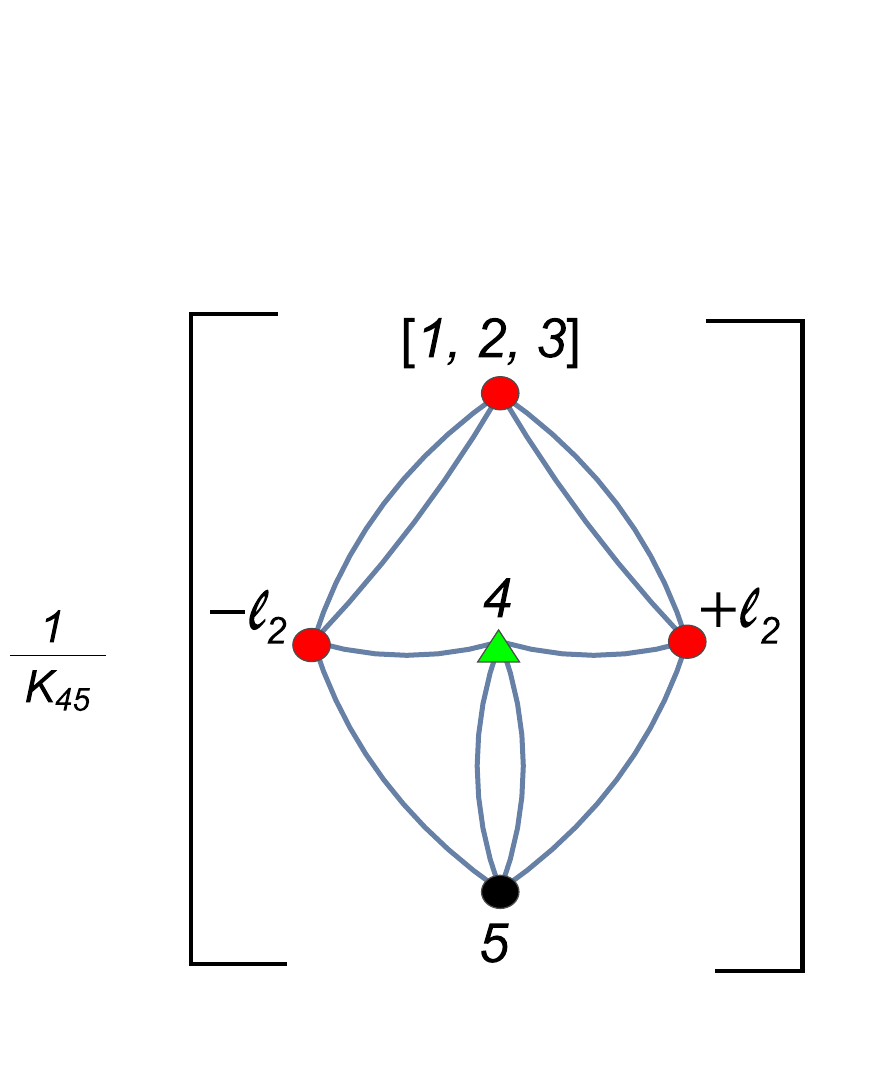}
        \includegraphics[scale=0.5]{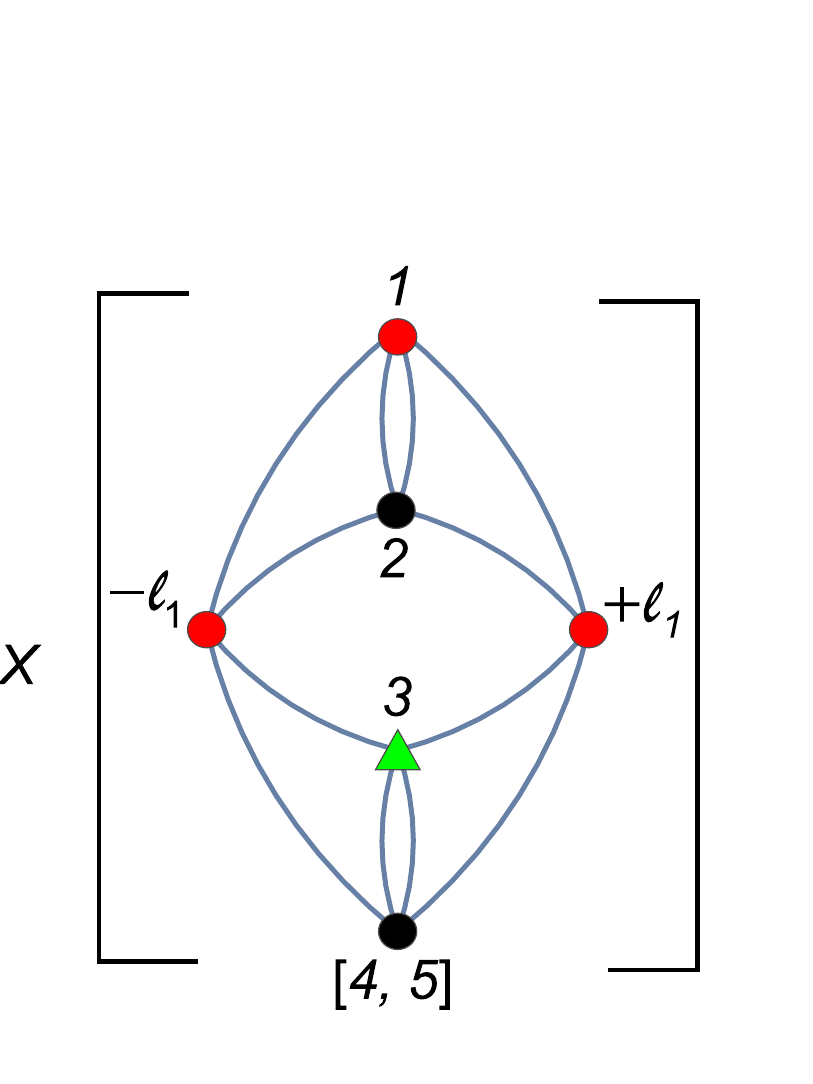}
  \caption{ All possible cuts allowed for a CHY graph at two loops.}\label{chy_cuts}
\end{figure}

In order to compute $\int d\mu^{\rm 2-loop}\, {\bf I}_{\rm CHY}$, we are going to apply the  $\L-$algorithm. From this algorithm, it is simple to note that in figure \ref{chy_5points}  there are only two allowable configurations (cuts)\footnote{For more details about allowable configuration see \cite{Gomez:2016bmv}.}  on the CHY graph,  which are given in figure \ref{chy_cuts}. 

Using the {\bf Rule I} found in  \eqref{ruleI},  we obtain two CHY subgraphs for each cut, as it is shown in figure \ref{chy_cuts}.  These subgraphs can be easily computed applying the standard $\L-$algorithm \cite{Gomez:2016bmv}. Therefore, all the non-zero configurations allowed in figure \ref{chy_cuts} have the following result
\begin{align}
({\bf 1})=\frac{1}{k_{12}^2\, k_{45}}\times\frac{1}{ k_3\cdot(k_1+k_2)}& \times \left[ \frac{1}{\ell_1\cdot(k_1+k_2)+k_{12}} + \frac{1}{-\ell_1\cdot(k_1+k_2)+k_{12}}  \right]\label{Cut1}\\
 &\times 
 \left[ \frac{1}{\ell_2\cdot(k_4+k_5)+k_{45}} + \frac{1}{-\ell_2\cdot(k_4+k_5)+k_{45}}  \right],\nonumber\\
 ({\bf 2})=\frac{1}{ k_{45}^2\,k_{12}}\times\frac{1}{ k_3\cdot(k_4+k_5)}& \times \left[ \frac{1}{\ell_1\cdot(k_1+k_2)+k_{12}} + \frac{1}{-\ell_1\cdot(k_1+k_2)+k_{12}}  \right]\label{Cut2}\\
 &\times 
 \left[ \frac{1}{\ell_2\cdot(k_4+k_5)+k_{45}} + \frac{1}{-\ell_2\cdot(k_4+k_5)+k_{45}}  \right].\nonumber
\end{align}
Clearly, adding \eqref{Cut1} with \eqref{Cut2} leads to the agreement with ${\cal I}^{\rm S}_{\rm FEY}$, i.e.
\begin{equation}
\frac{{\cal I}_{\rm CHY} }{2^6}=\frac{1}{2^6\,\, \ell_1^2 \,\,\ell_2^2}\times \left[ ({\bf 1})+({\bf 2})\right]={\cal I}_{\rm FEY}\Big|_{\rm p.f \atop S}.
\end{equation}
This computation has also been verified numerically.

\subsection{One-particle Irreducible Diagram}

In this section we consider two one-particle irreducible (1PI) diagrams, for planar and nonplanar cases.

\subsubsection{Four-particle Planar Diagram}\label{4pointsplanar}

\begin{figure}
 \centering
       \includegraphics[scale=0.5]{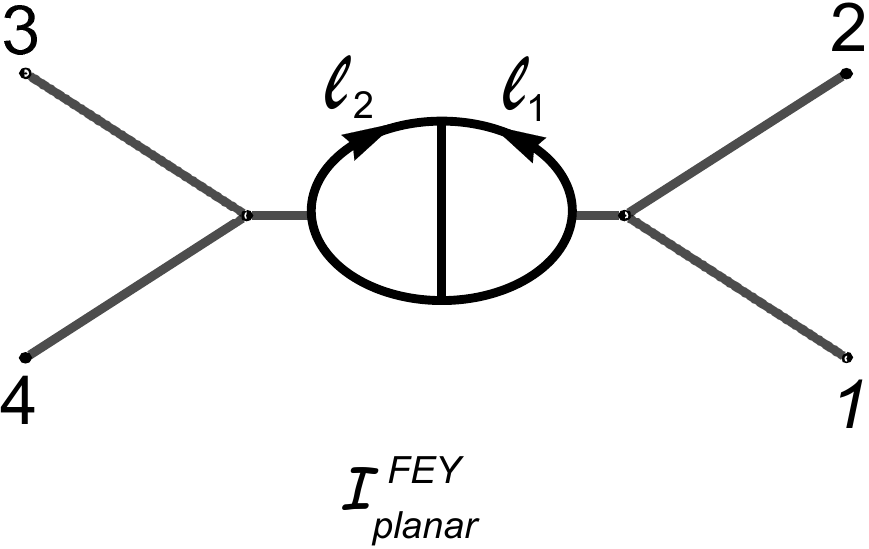}
  \caption{Two-loop 1PI planar Feynman diagram.}\label{2L_4p}
\end{figure}

Let us consider the simple Feynman diagram given in figure \ref{2L_4p}.
From this diagram it is simple to read the loop integrand:
\begin{align}\label{I2points}
{\cal I}_{\rm FEY}^{\rm planar}=&\frac{1}{2^2\,\ell_1^2\,\ell_2^2\,k_{12}\,k_{34}\,(\ell_1+\ell_2)^2\,(\ell_1-k_1-k_2)^2\,(\ell_2-k_3-k_4)^2}.
\end{align}
After using the partial fraction identity for the factors\footnote{Recall that the loop integral measure and the integration contour are invariant under translation.}
\begin{align*}
\frac{1}{\ell_1^2\,(\ell_1-k_1-k_2)^2}\times \frac{1}{\ell_2\,(\ell_2-k_3-k_4)^2},
\end{align*}
and shifting the loop momenta $\{\ell_1, \ell_2 \}$ to obtain the global factor ${1\over \ell^2_1\,\ell_2^2}$, it is straightforward to check that the Feynman integrand in \eqref{I2points} becomes
\begin{align}\label{Fey_4points}
&{\cal I}_{\rm FEY}^{\rm planar}\Big|_{\rm p.f \atop S}=\frac{1}{2^4\,\ell_1^2 \ell_2^2 k_{12} k_{34}}\left[ 
\frac{1}{(\ell_1+\ell_2)^2 \,(-\ell_1\cdot (k_1 +k_2)+k_{12}) \,\, (-\ell_2\cdot (k_3+k_4)+k_{34})}
 \right.  \nonumber\\
&
\left.
+
\frac{1}{(\ell_1+\ell_2+k_1+k_2)^2 \,(\ell_1\cdot (k_1+k_2)+k_{12}) \,\, (-\ell_2\cdot( k_3+k_4)+k_{34})}
+
\left(
\begin{matrix}
\ell_1\rightarrow -\ell_1\\
\ell_2\rightarrow -\ell_2
\end{matrix}
\right)
\right].
\end{align}

\begin{figure}
 \centering
       \includegraphics[scale=0.4]{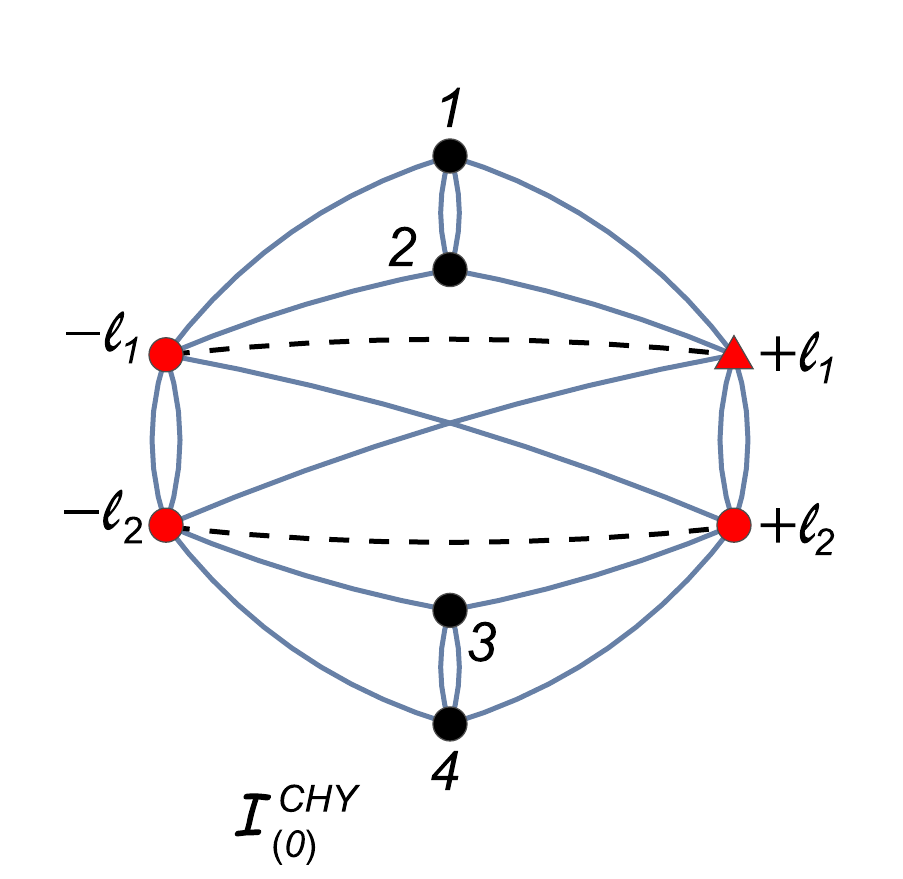}
        \includegraphics[scale=0.4]{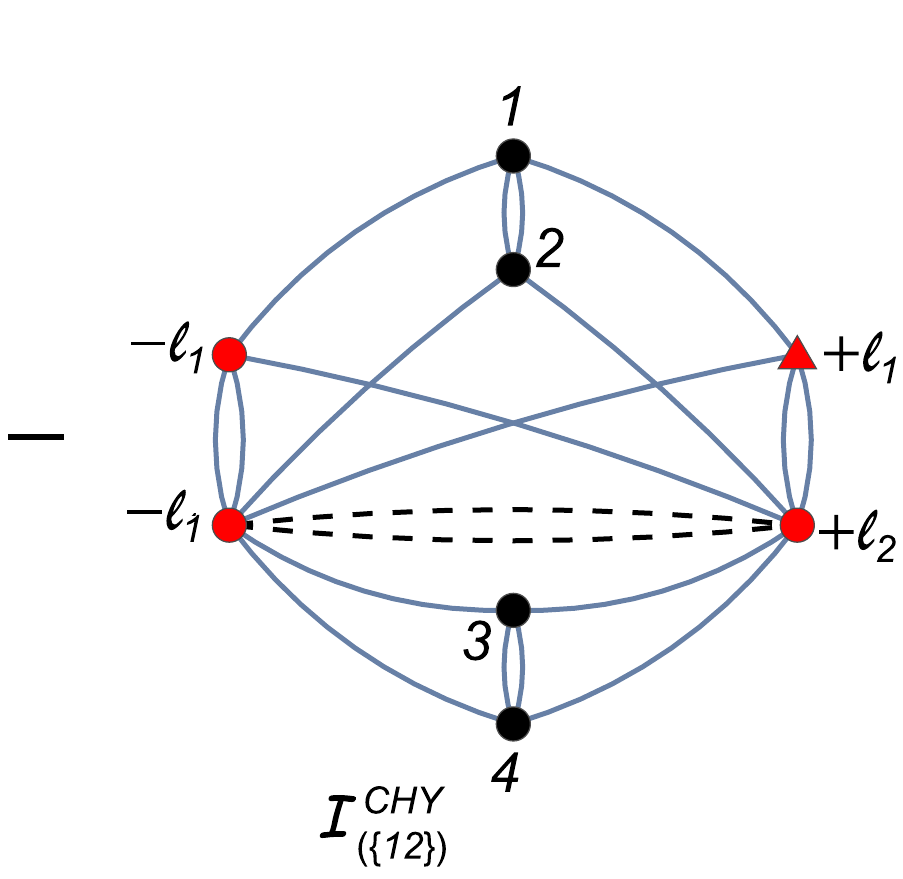}
         \includegraphics[scale=0.4]{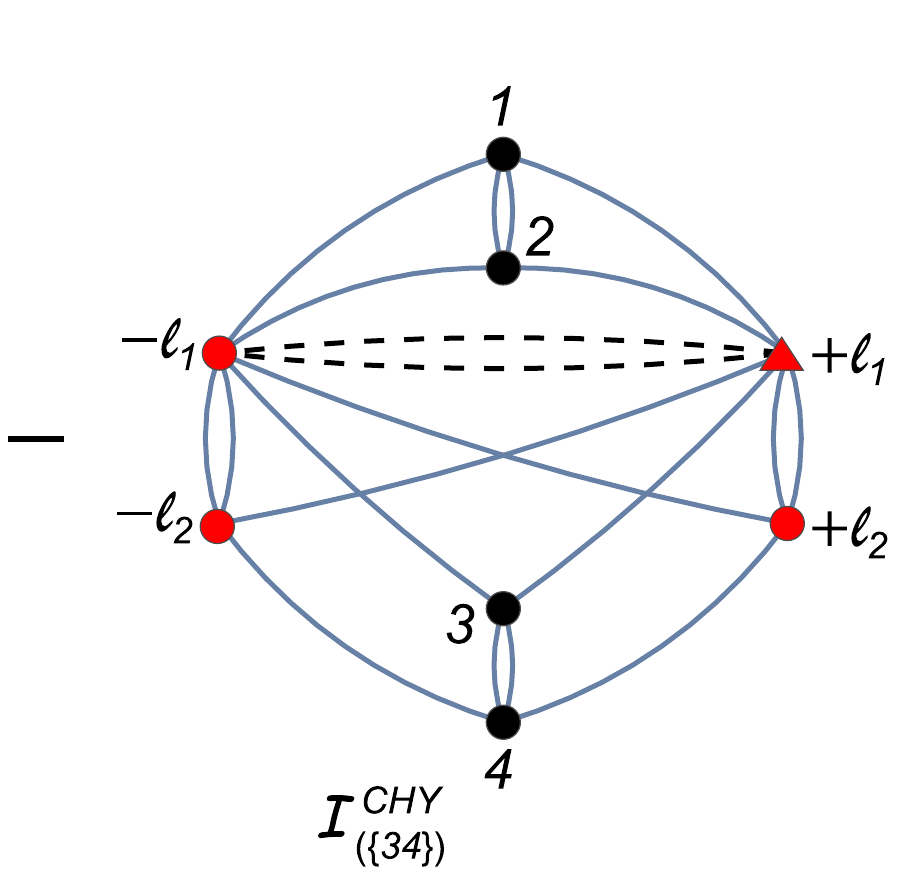}            
           \includegraphics[scale=0.4]{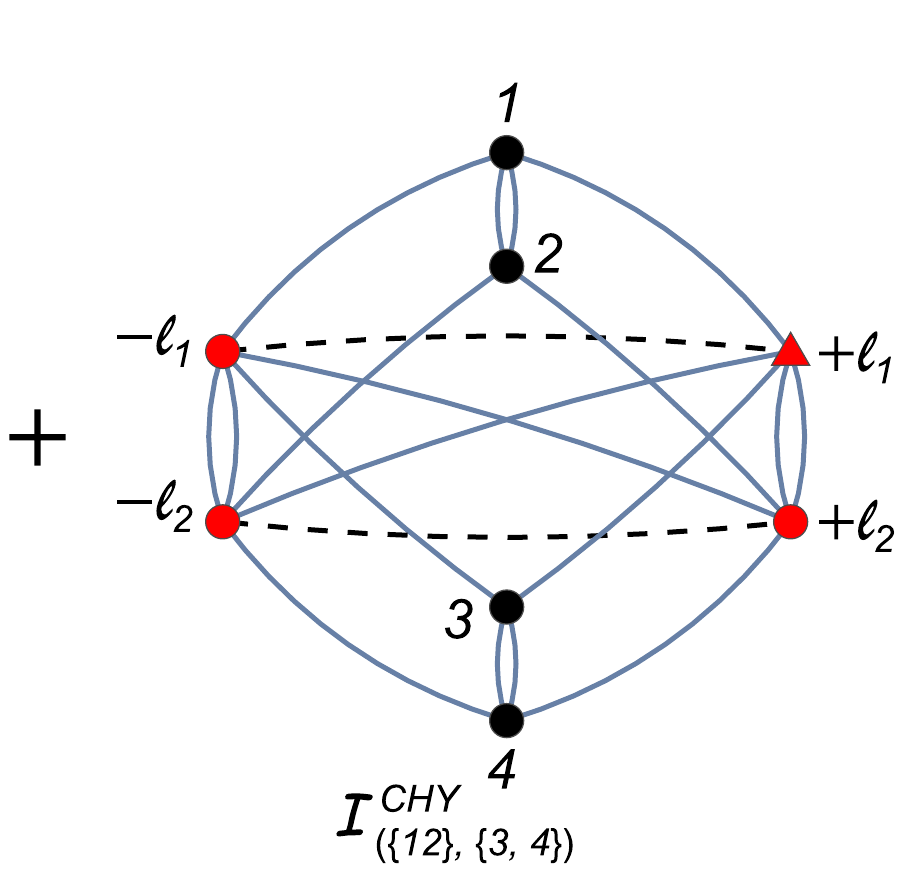}
  \caption{CHY graph for planar two-loop Feynman diagram.}\label{chy_4p_repre}
\end{figure}

In order to find the CHY integrand which corresponds to the Feynman diagram in figure \ref{2L_4p}, we should consider the planar building block given in section \ref{form} and the gluing technique developed in section \ref{gluesection}.  In appendix \ref{glue_PLANAR}, we will apply the gluing process, and the CHY integrand obtained to reproduce the Feynman diagram result is
\begin{align}\label{chy_planarI}
{\cal I}_{\rm CHY}^{\rm planar}&= \frac{1}{\ell_1^2\,\,\ell_2^2}\int\,d\mu^{\rm 2-loop}\,\, {\bf I}^{\rm planar}_{\rm CHY}, \\
{\bf I}_{\rm CHY}^{\rm planar}&= {\rm \ss^{planar}}\left[\frac{\o^1_{1}(\o^1_{2}-\o^2_{2})}{(12)}   \times  \frac{(\o^2_{3}-\o^1_{3})\o^2_{4}}{(34)}  \right].\nonumber
\end{align}
Clearly, the ${\bf I}^{\rm planar}_{\rm CHY}$ integrand is a linear combination of four CHY graphs drawn in figure \ref{chy_4p_repre}. 
 
We are going to show that by applying the $\L-$algorithm on each of the CHY graphs given in figure \ref{chy_4p_repre}to compute $\int\,d\mu^{\rm 2-loop}\,\, {\bf I}^{\rm planar}_{\rm CHY}$ and combining them to get ${\cal I}^{\rm CHY}_{(0)}-{\cal I}^{\rm CHY}_{(\{12\})} -{\cal I}^{\rm CHY}_{(\{34\})}+{\cal I}^{\rm CHY}_{(\{12\},\{3,4\})}$, we will be able to reproduce the Feynman integrand in \eqref{Fey_4points}.

\begin{figure}
 \centering
       \includegraphics[scale=0.4]{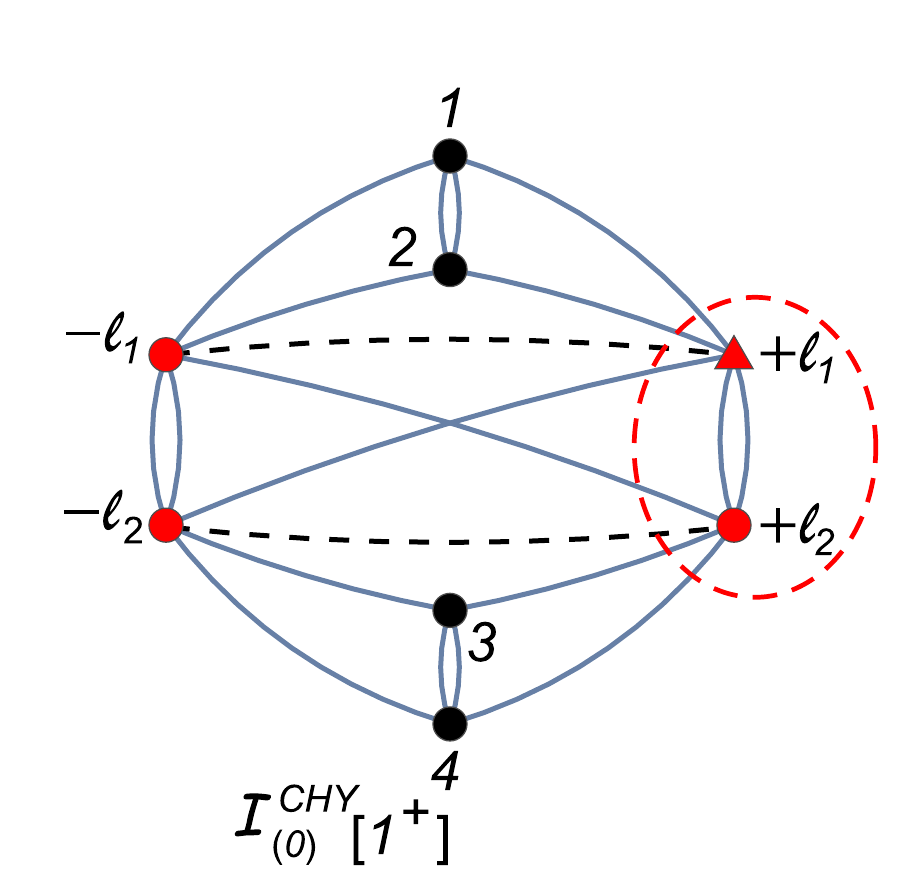}
        \includegraphics[scale=0.4]{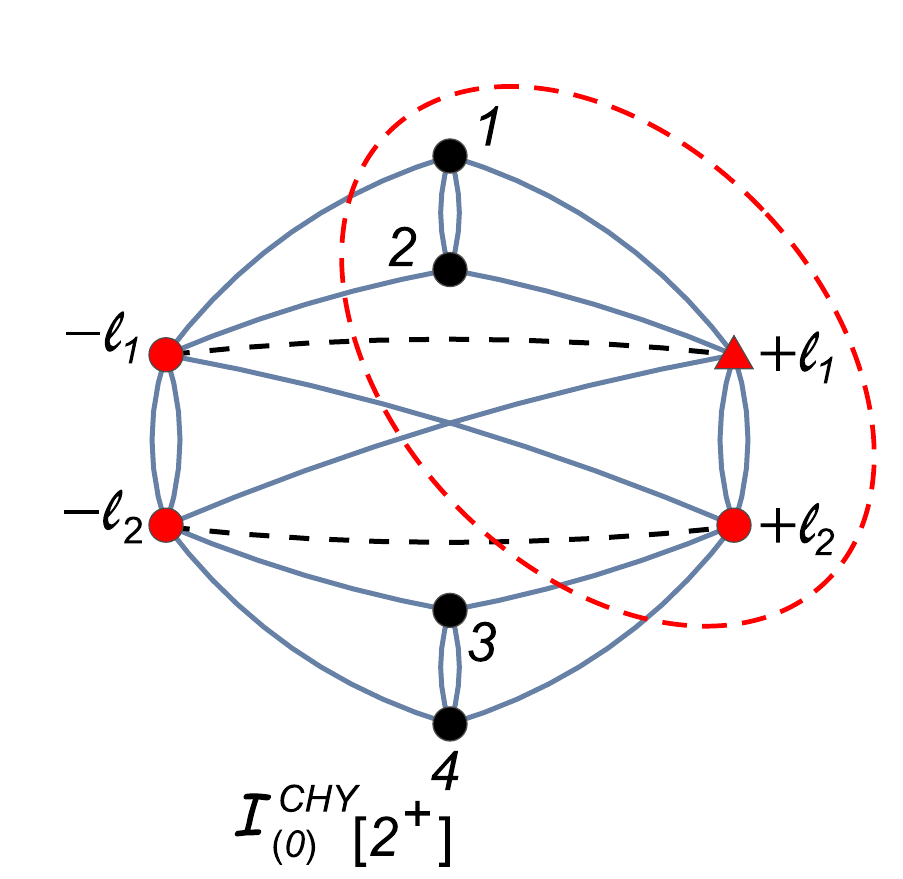}
         \includegraphics[scale=0.4]{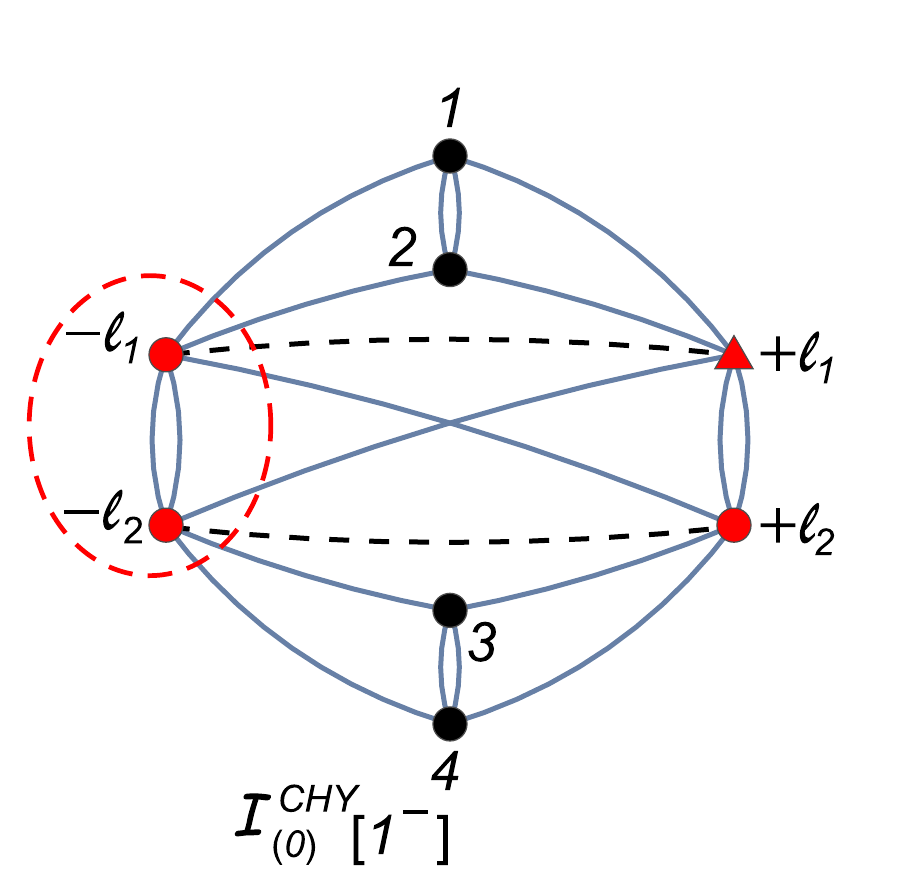}            
           \includegraphics[scale=0.4]{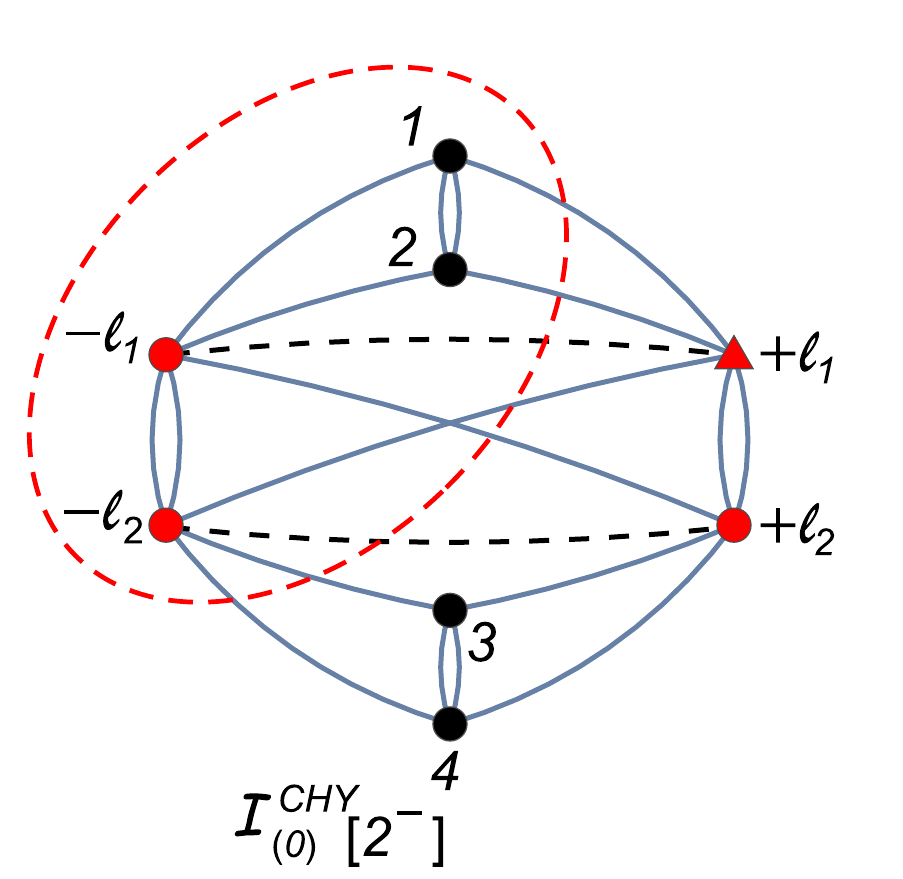}\\
                  \includegraphics[scale=0.4]{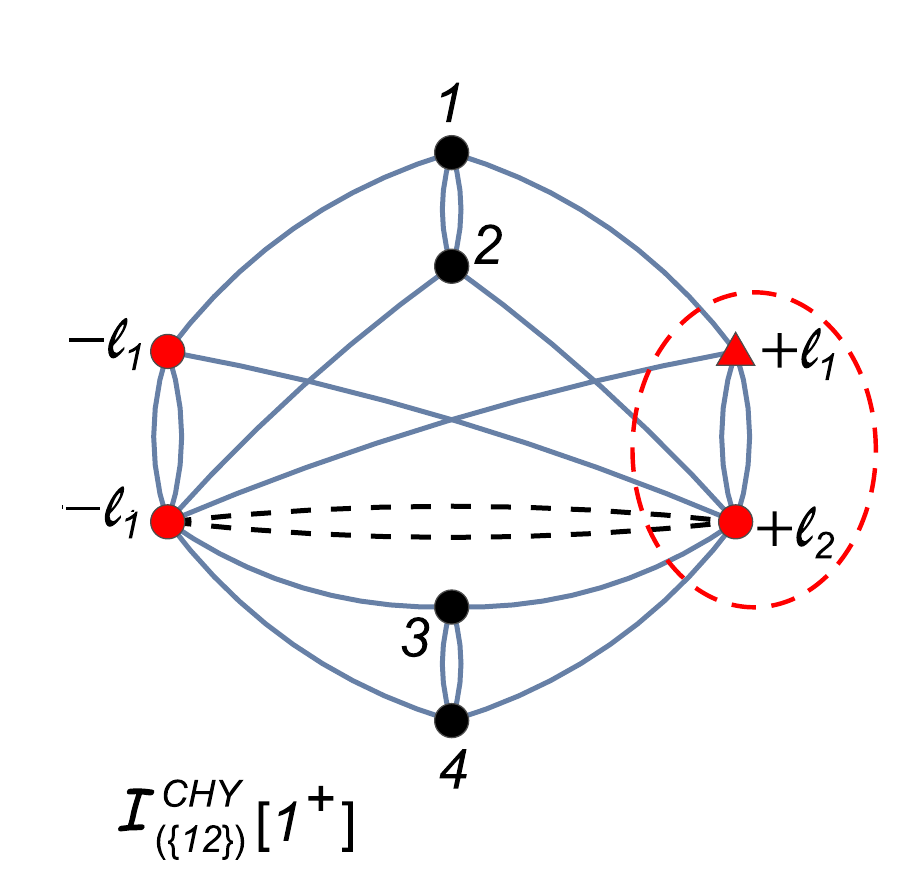}
                    \includegraphics[scale=0.4]{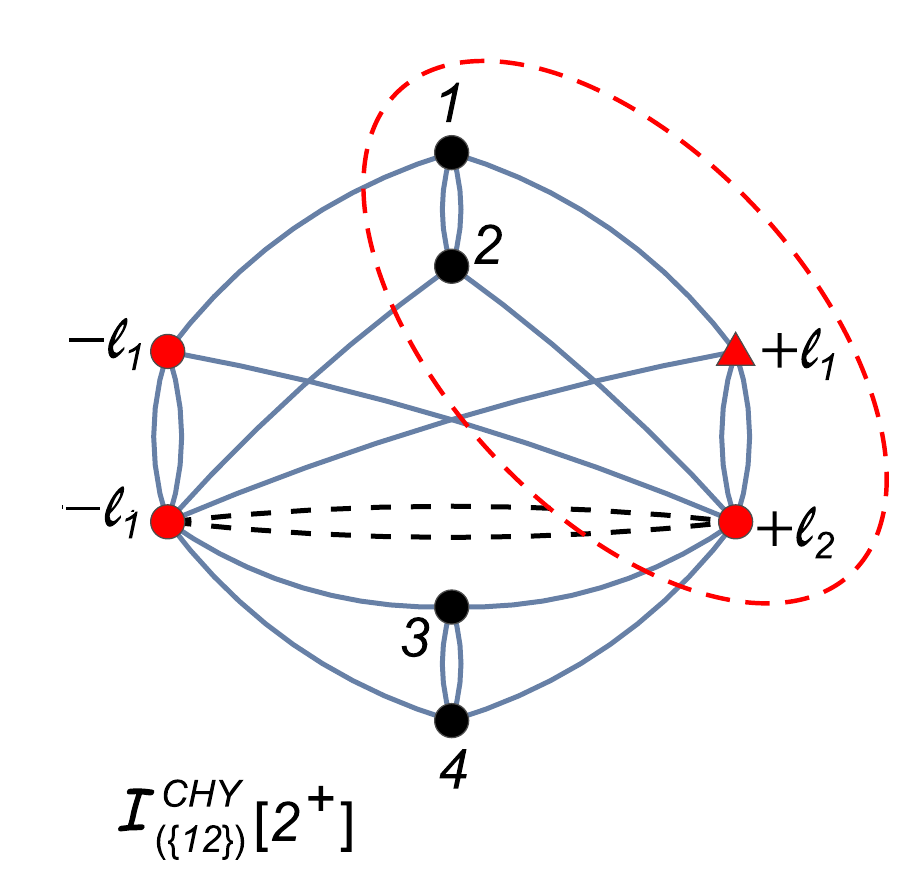}
                      \includegraphics[scale=0.4]{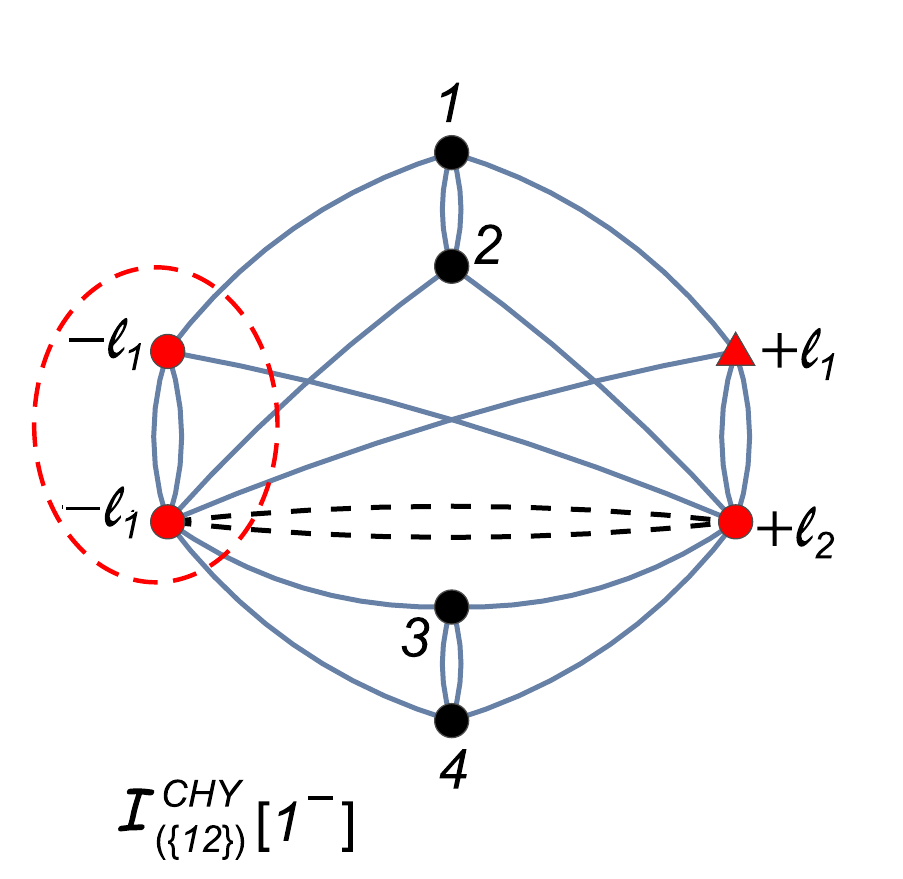}            
                        \includegraphics[scale=0.4]{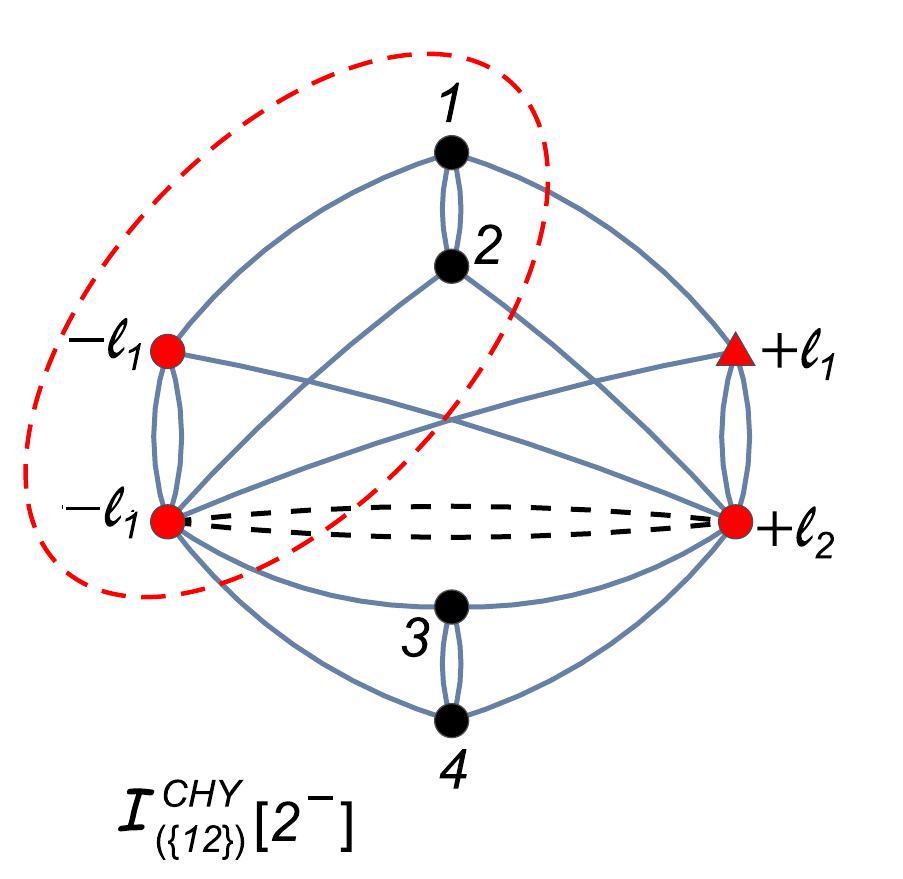}\\
                          \includegraphics[scale=0.4]{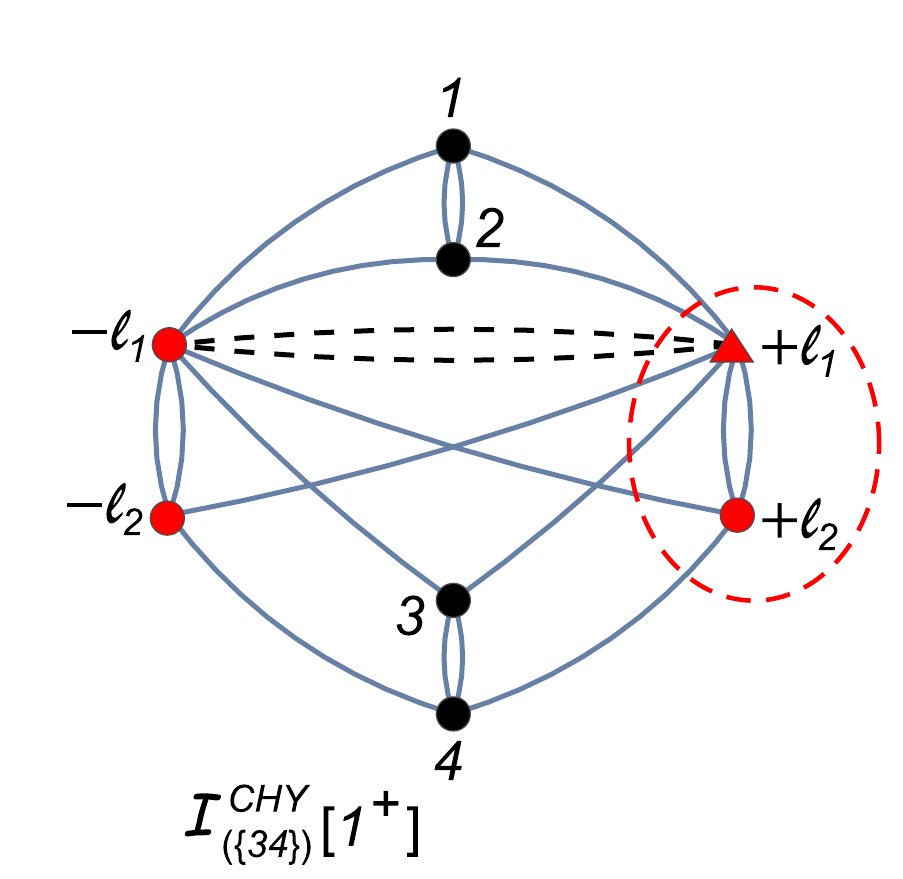}
                            \includegraphics[scale=0.4]{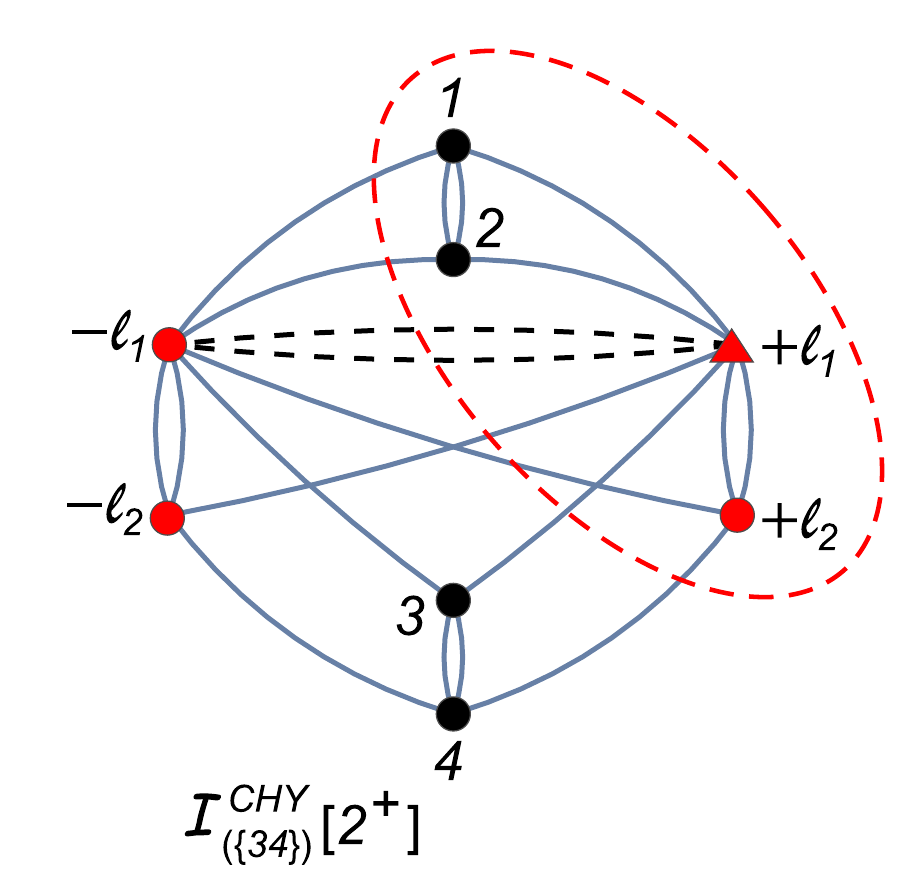}
                              \includegraphics[scale=0.4]{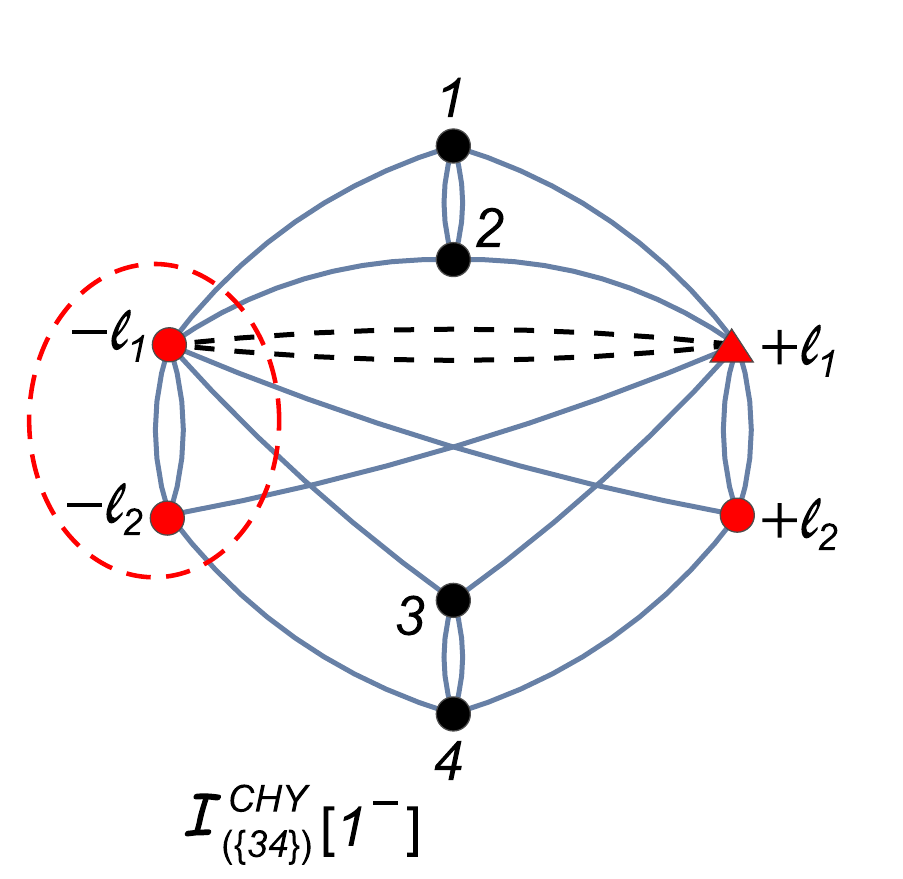}            
                                \includegraphics[scale=0.4]{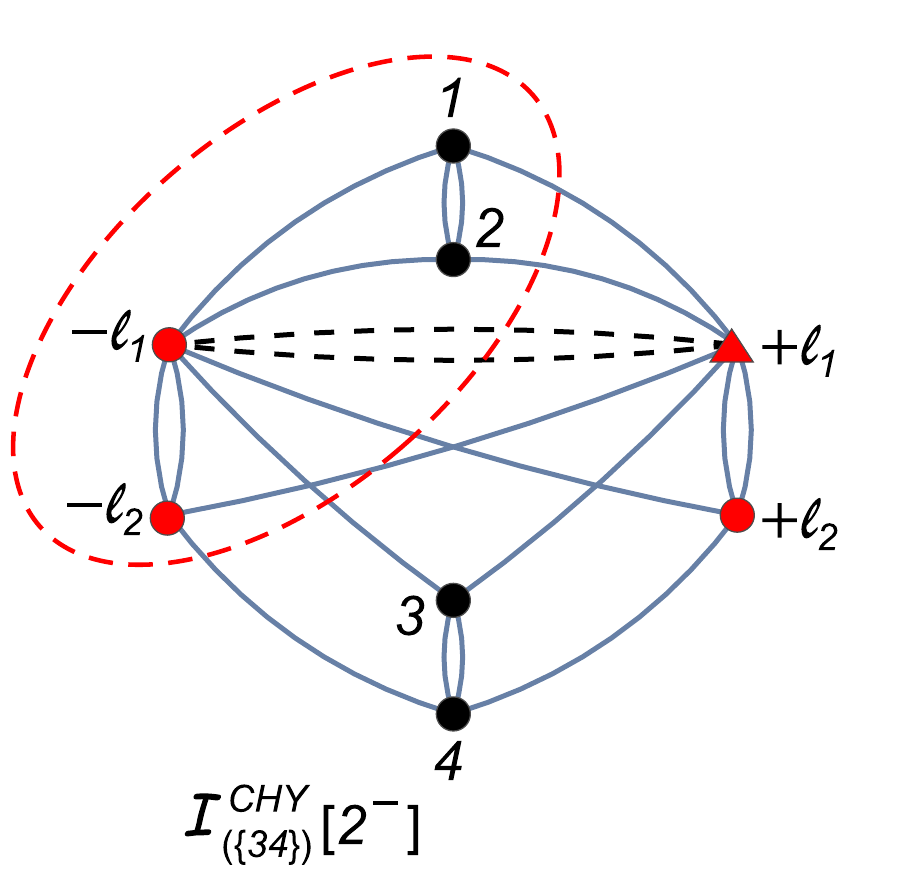}\\
                                 \includegraphics[scale=0.4]{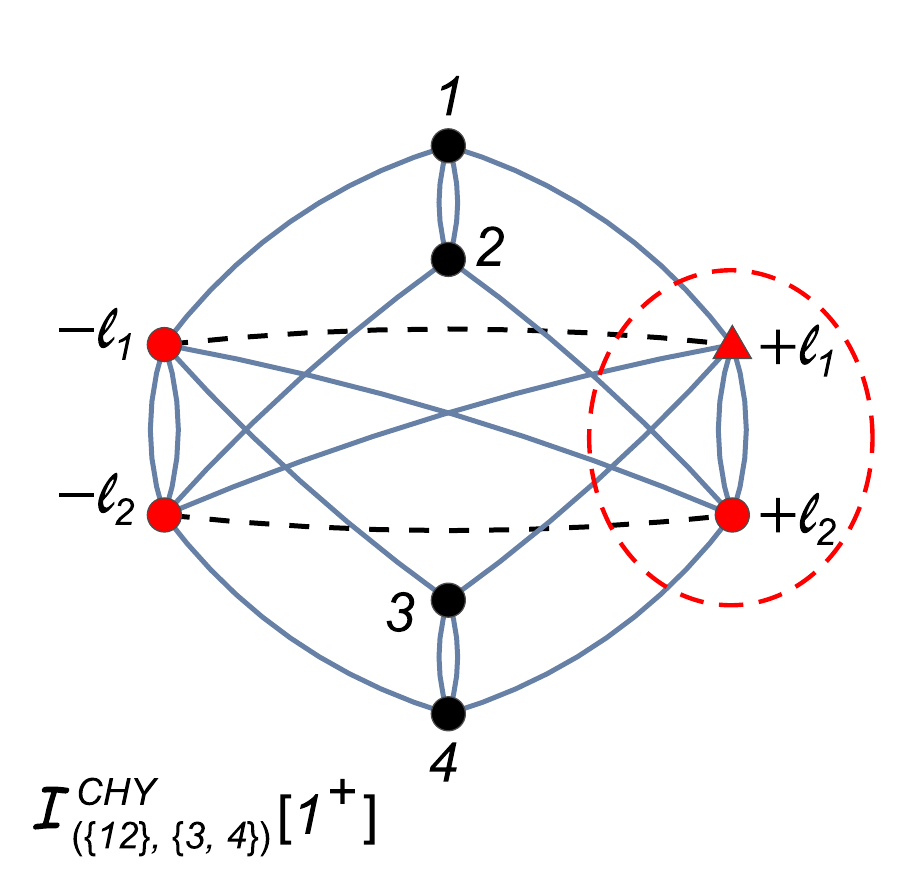}
                                   \includegraphics[scale=0.4]{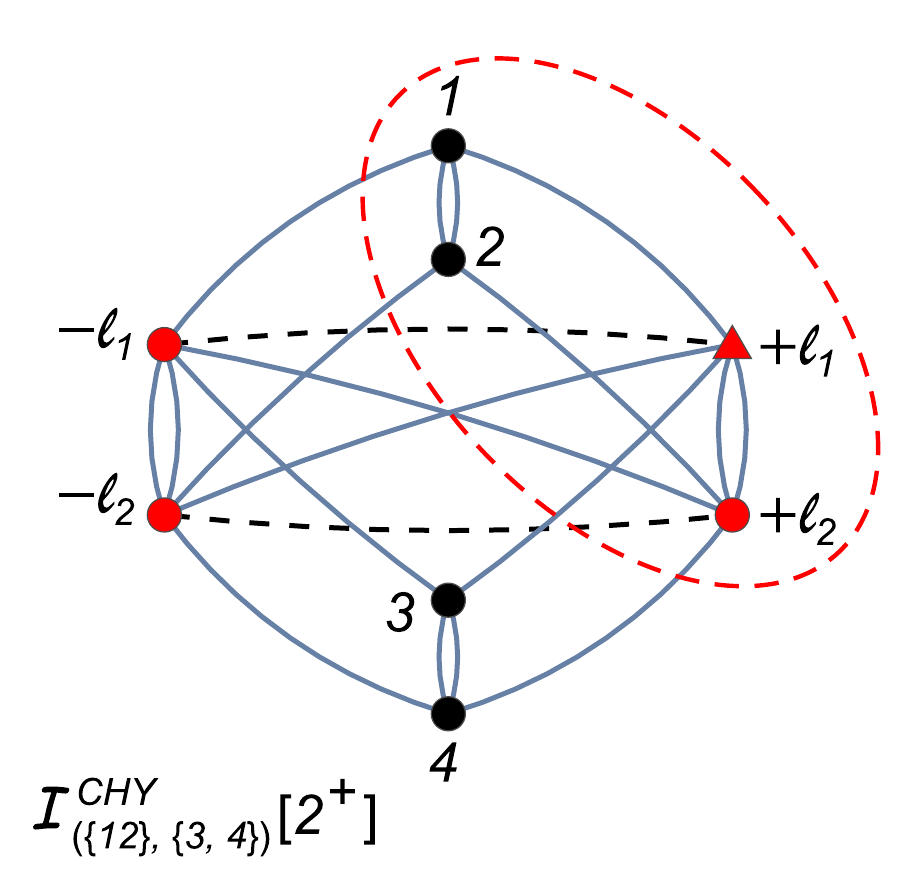}
                                     \includegraphics[scale=0.4]{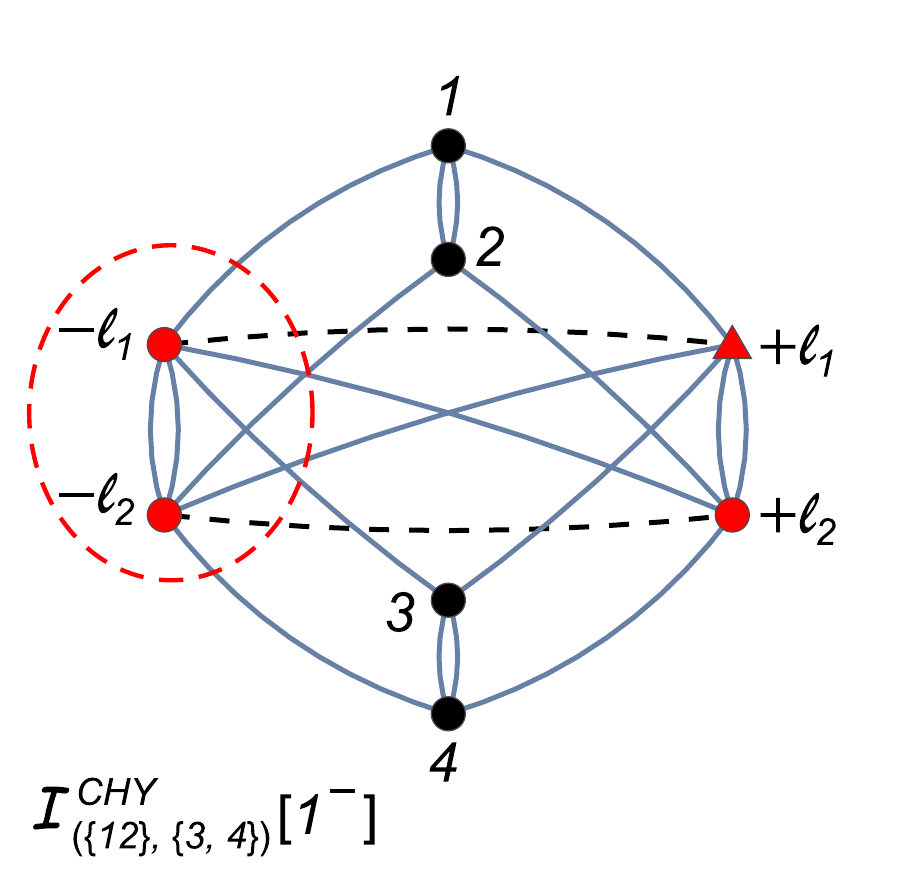}            
                                       \includegraphics[scale=0.4]{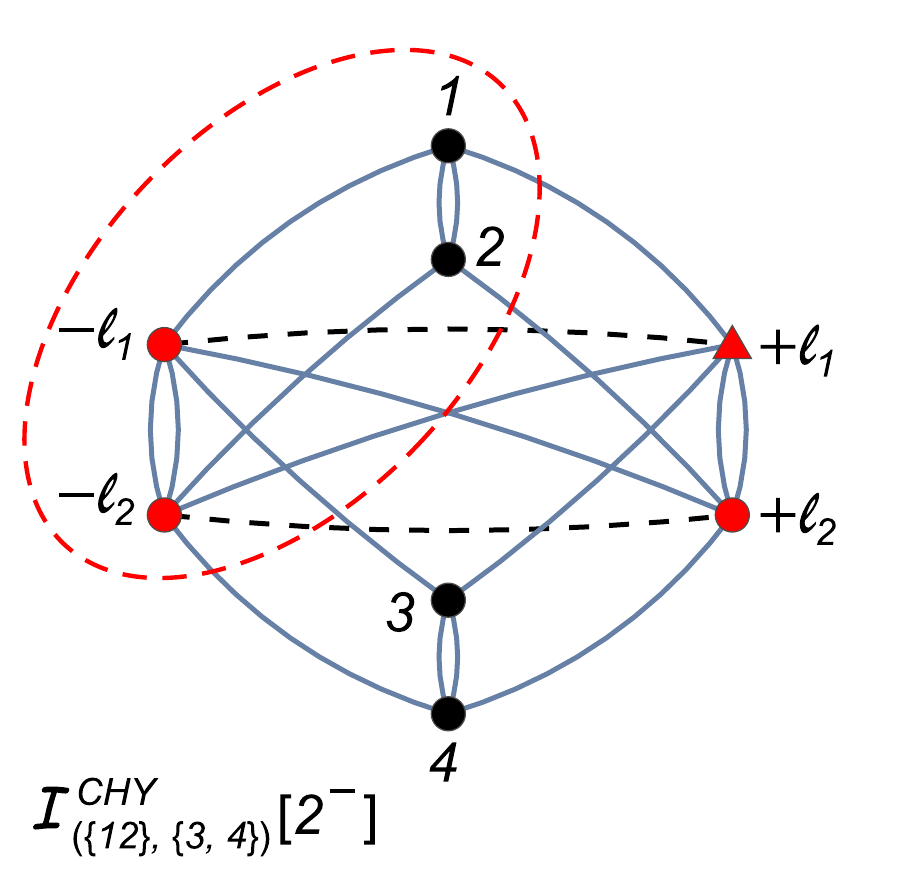}
  \caption{All possible non-zero cuts.}\label{configurations}
\end{figure}

From the $\L-$algorithm \cite{Gomez:2016bmv}, it is simple to see that there are only four non-zero cuts for each CHY graph in figure \ref{chy_4p_repre}. Those non-zero cuts are sketched in figure \ref{configurations}.
The configurations $\{ {\cal I}^{\rm CHY}_{(0)}[2^{\pm}],{\cal I}^{\rm CHY}_{(\{12\})}[2^{\pm}], {\cal I}^{\rm CHY}_{(\{34\})}[2^{\pm}], {\cal I}^{\rm CHY}_{(\{12\},\{34\})}[2^{\pm}]\}$ are easily computed using the {\bf Rule II} in \eqref{ruleII}  and the standard $\L-$algorithm. The result is simply
\begin{align}
&{\cal I}^{\rm CHY}_{(0)}[2^{+}]-{\cal I}^{\rm CHY}_{(\{12\})}[2^{+}] -{\cal I}^{\rm CHY}_{(\{34\})}[2^{+}]+ {\cal I}^{\rm CHY}_{(\{12\},\{34\})}[2^{+}]\nonumber\\
&=\frac{1}{k_{12}\,k_{34}\,{1\over 2}(\ell_1+\ell_2+k_1+k_2)^2}\frac{1}{(\ell_1\cdot (k_1+k_2)+k_{12}) \,\, (-\ell_2\cdot (k_3+k_4)+k_{34})}, \label{2p_cut2+}\\
\nonumber \\
&{\cal I}^{\rm CHY}_{(0)}[2^{-}]-{\cal I}^{\rm CHY}_{(\{12\})}[2^{-}] -{\cal I}^{\rm CHY}_{(\{34\})}[2^{-}]+ {\cal I}^{\rm CHY}_{(\{12\},\{34\})}[2^{-}]\nonumber\\
&=\frac{1}{k_{12}\,k_{34}\,{1\over 2}(\ell_1+\ell_2 - k_1 - k_2)^2}\frac{1}{(-\ell_1\cdot (k_1+k_2)+k_{12}) \,\, ( \ell_2\cdot (k_3+k_4)+k_{34})}. \label{2p_cut2-} 
\end{align}

\begin{figure}
 \centering
            \includegraphics[scale=0.4]{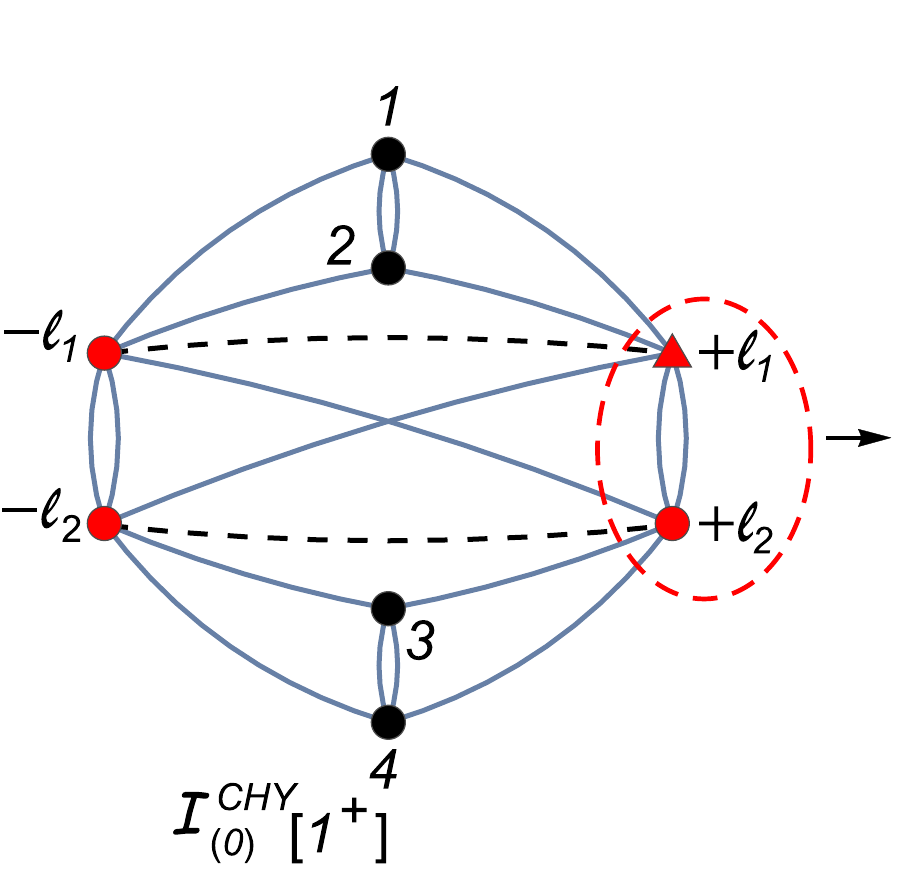}
              \includegraphics[scale=0.4]{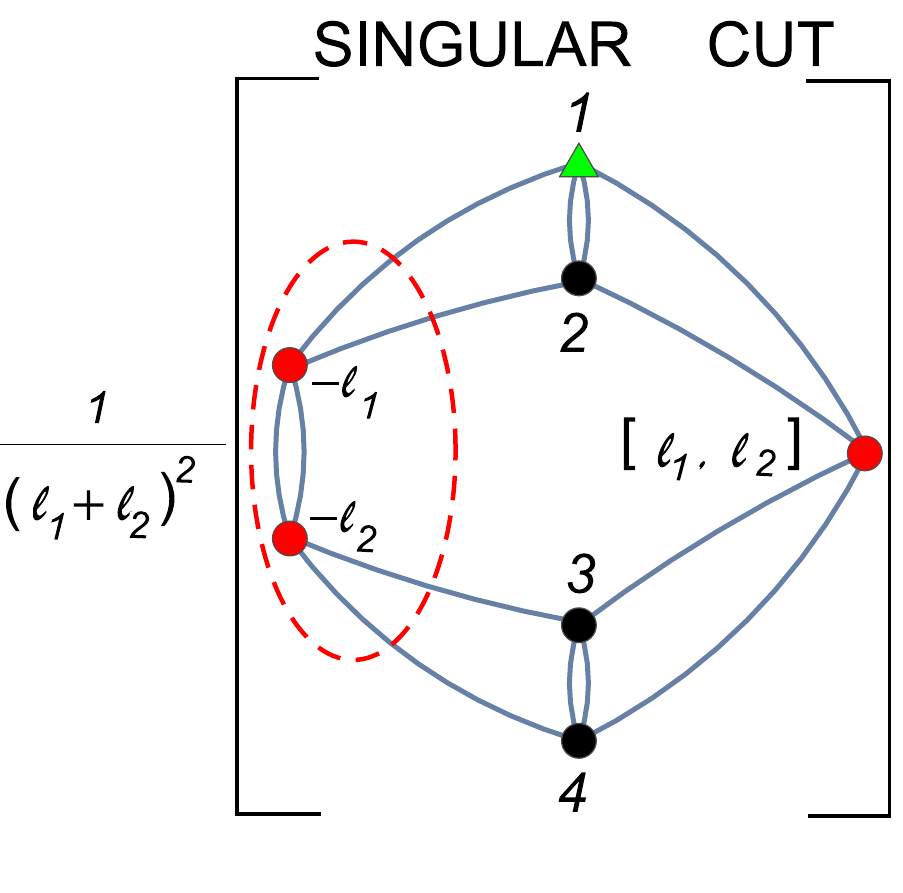} \quad
               \includegraphics[scale=0.4]{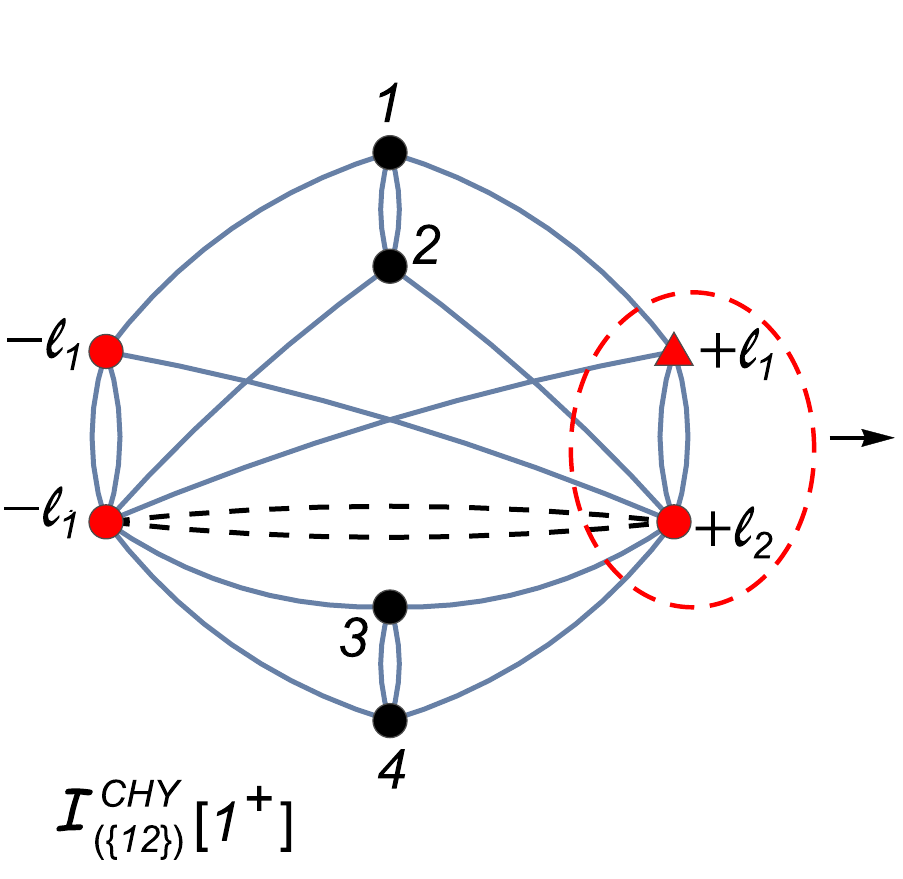}
                  \includegraphics[scale=0.4]{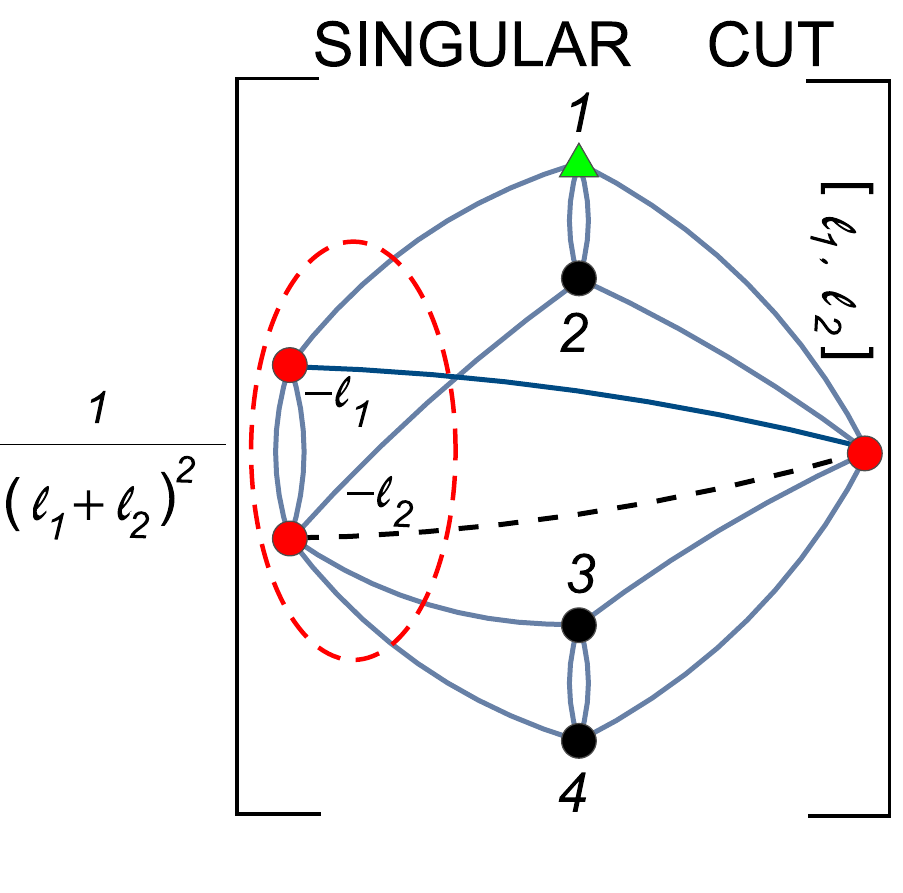}\\
                     \includegraphics[scale=0.4]{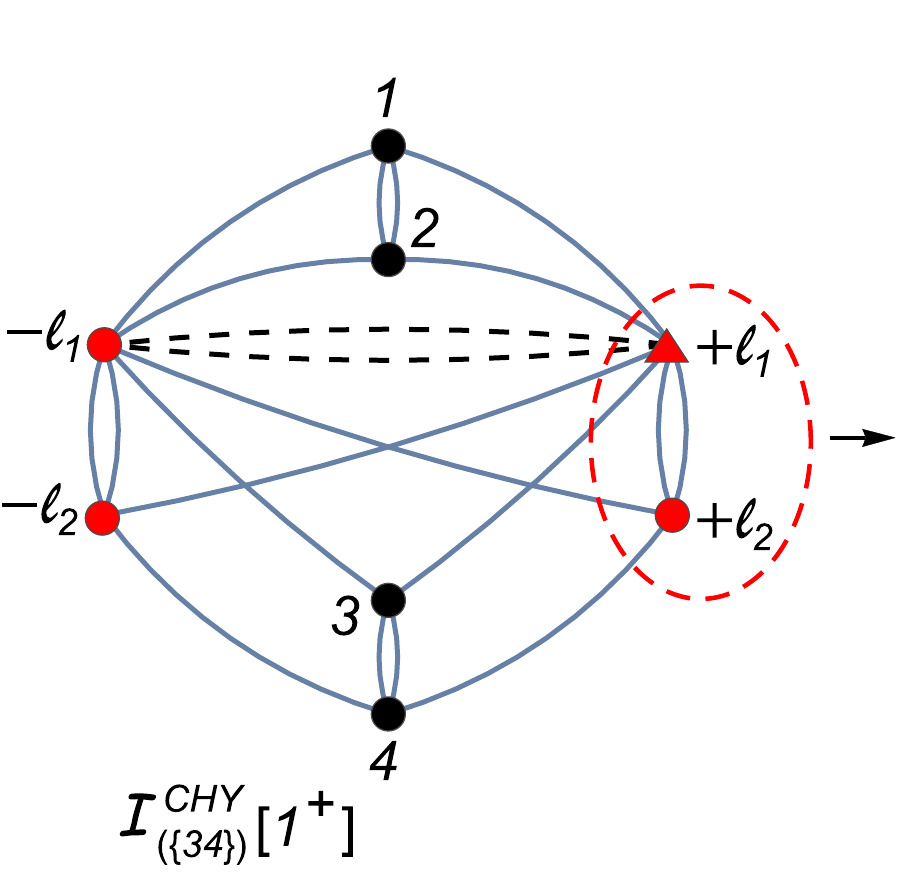}
                       \includegraphics[scale=0.4]{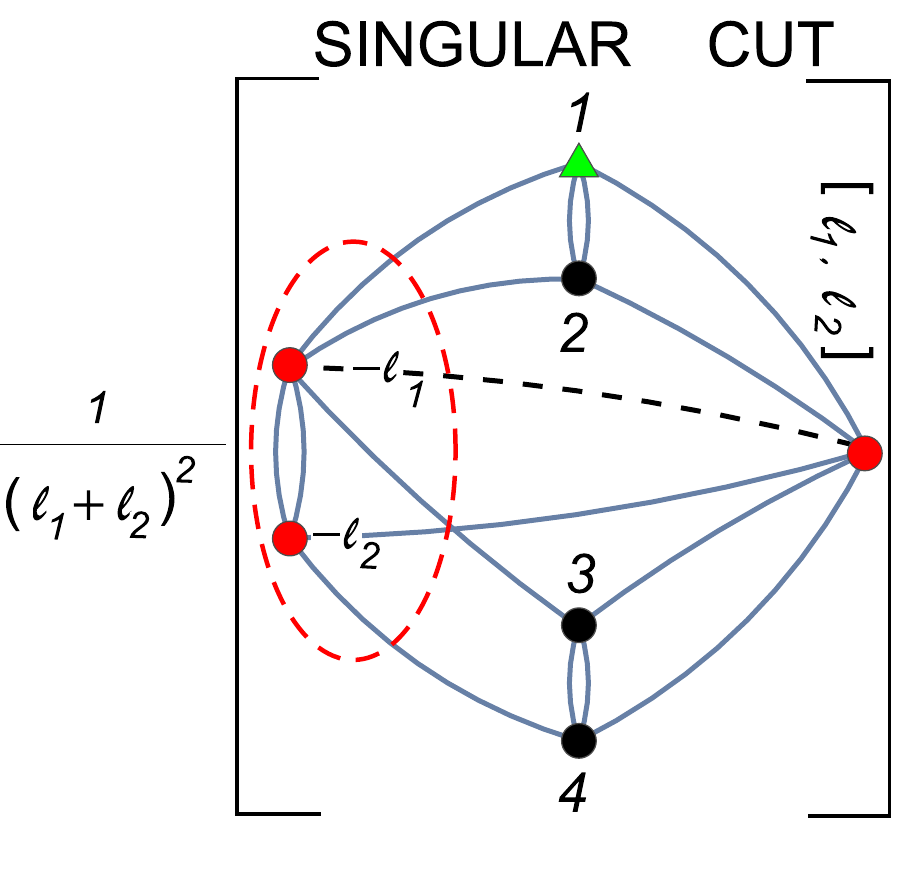} \quad
                         \includegraphics[scale=0.4]{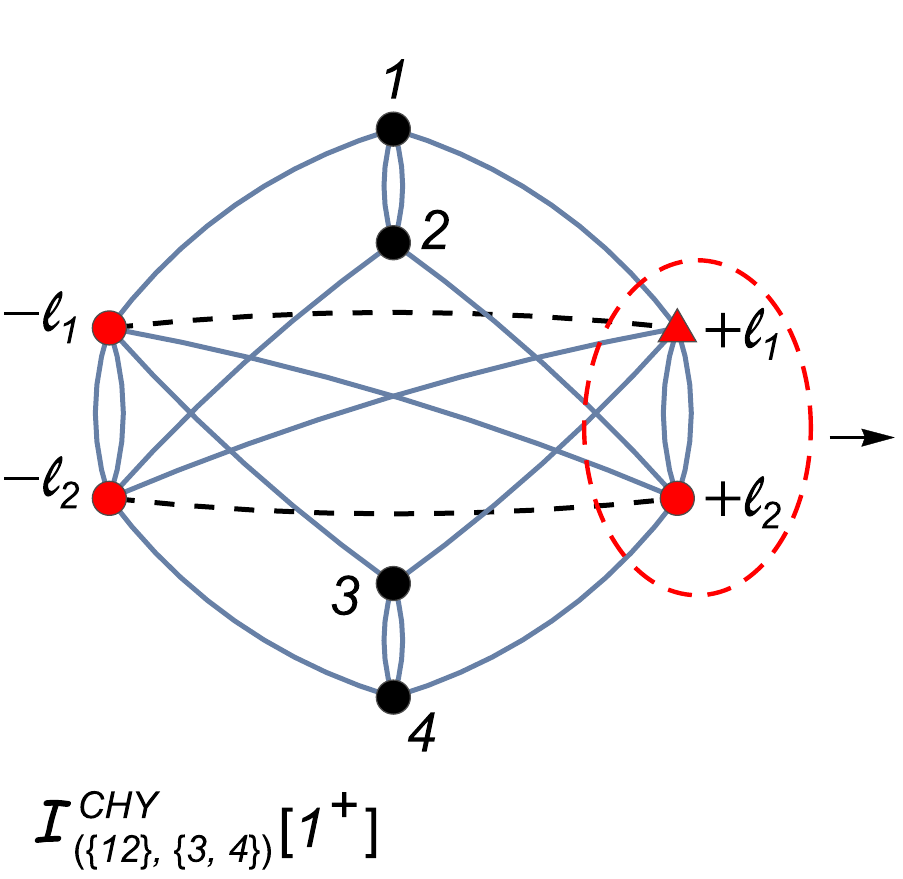}
                           \includegraphics[scale=0.4]{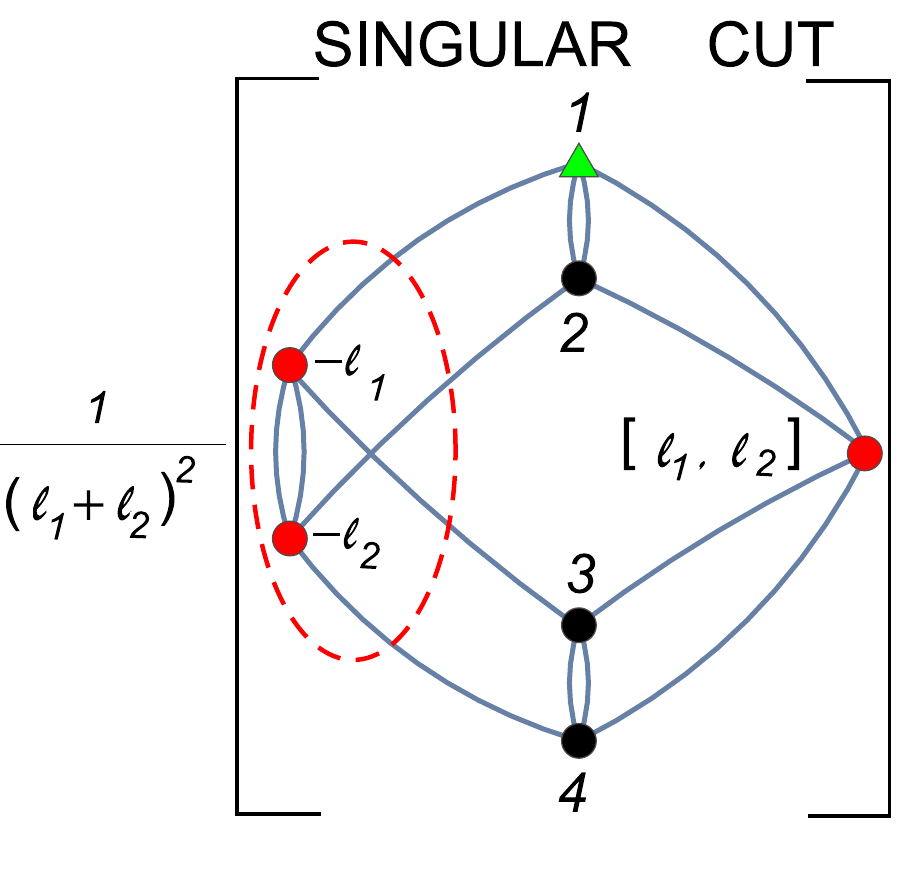}        
  \caption{Possibles singular cuts (by momentum conservation).}\label{singular}
\end{figure}

On the other hand, the configurations, $\{ {\cal I}^{\rm CHY}_{(0)}[1^{\pm}],{\cal I}^{\rm CHY}_{(\{12\})}[1^{\pm}], {\cal I}^{\rm CHY}_{(\{34\})}[1^{\pm}], {\cal I}^{\rm CHY}_{(\{12\},\{34\})}[1^{\pm}]\}$, must be carefully computed. Let us consider the $\{ {\cal I}^{\rm CHY}_{(0)}[1^{+}],{\cal I}^{\rm CHY}_{(\{12\})}[1^{+}], {\cal I}^{\rm CHY}_{(\{34\})}[1^{+}], {\cal I}^{\rm CHY}_{(\{12\},\{34\})}[1^{+}]\}$ cuts. After using the {\bf Rule II},  the CHY subgraphs obtained are singular by momentum conservation,  as it is shown in figure \ref{singular}. 
From the double cover point of view, this singularity means that when four of the six branch points of a hyperelliptic curve of genus $2$ collapse at a point, the Riemann surface becomes degenerate as if the two A-cycles were pinched at a point. This kind of singularity must cancel out.

In addition, it is also clear  that the singular cuts from ${\cal I}^{\rm CHY}_{(0)}[1^+]$ and ${\cal I}^{\rm CHY}_{(\{12\},\{34\})}[1^+]$  are exactly the same as the ones obtained from\footnote{All singular cuts have the same contribution.}  $ {\cal I}^{\rm CHY}_{(\{12\})}[1^+]$ and ${\cal I}^{\rm CHY}_{(\{34\})}[1^+]$, as it can be seen in  figure \ref{chy_4p_repre}. Therefore, considering the linear combination in figure \ref{chy_4p_repre}, the singularity cancels out, as it was required.

Consequently, we can finally write the contributions from the cuts $[1^+]$ and $[1^-]$ as
\begin{align}
&{\cal I}^{\rm CHY}_{(0)}[1^+]-{\cal I}^{\rm CHY}_{(\{12\})}[1^+]-{\cal I}^{\rm CHY}_{(\{34\})}[1^+]+{\cal I}^{\rm CHY}_{(\{12\},\{34\})}[1^+] \nonumber\\
&=
\frac{1}{k_{12}\,k_{34}\,{1\over 2}(\ell_1+\ell_2)^2}\frac{1}{(-\ell_1\cdot (k_1+k_2)+k_{12}) \,\, (-\ell_2\cdot (k_3+k_4)+k_{34})} ,\label{2p_cut1+}\\
&\nonumber\\
&{\cal I}^{\rm CHY}_{(0)}[1^-]-{\cal I}^{\rm CHY}_{(\{12\})}[1^-]-{\cal I}^{\rm CHY}_{(\{34\})}[1^-]+{\cal I}^{\rm CHY}_{(\{12\},\{34\})}[1^-] \nonumber\\
&=
\frac{1}{k_{12}\,k_{34}\,{1\over 2}(\ell_1+\ell_2)^2}\frac{1}{(\ell_1\cdot (k_1+k_2)+k_{12}) \,\, (\ell_2\cdot (k_3+k_4)+k_{34})}.\label{2p_cut1-}
\end{align}
Summing the contributions from \eqref{2p_cut2+}, \eqref{2p_cut2-}, \eqref{2p_cut1+} and \eqref{2p_cut1-}, it is straightforward to check that the CHY integrand in \eqref{chy_planarI} is in exact agreement with the Feynman integrand found in \eqref{Fey_4points}, i.e.
\begin{equation}
\frac{{\cal I}_{\rm CHY}^{\rm planar}}{2^5} =
\frac{{\cal I}^{\rm CHY}_{(0)}-{\cal I}^{\rm CHY}_{(\{12\})}-{\cal I}^{\rm CHY}_{(\{34\})}+{\cal I}^{\rm CHY}_{(\{12\},\{34\})}}{2^5\,\,\ell_1^2\,\ell_2^2}  = {\cal I}_{\rm FEY}^{\rm planar}\Big|_{\rm p.f \atop S}.
\end{equation}

\subsubsection{Four-particle Non-planar Diagram}\label{nonplanar3}

\begin{figure}
 \centering
       \includegraphics[scale=0.5]{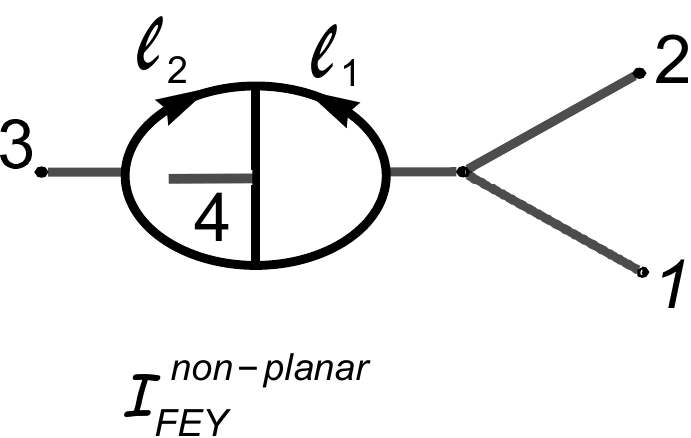}
  \caption{Four-particle non-planar Feynman diagram at two loops.}\label{Fey_NP}
\end{figure}

\begin{figure}
 \centering
       \includegraphics[scale=0.4]{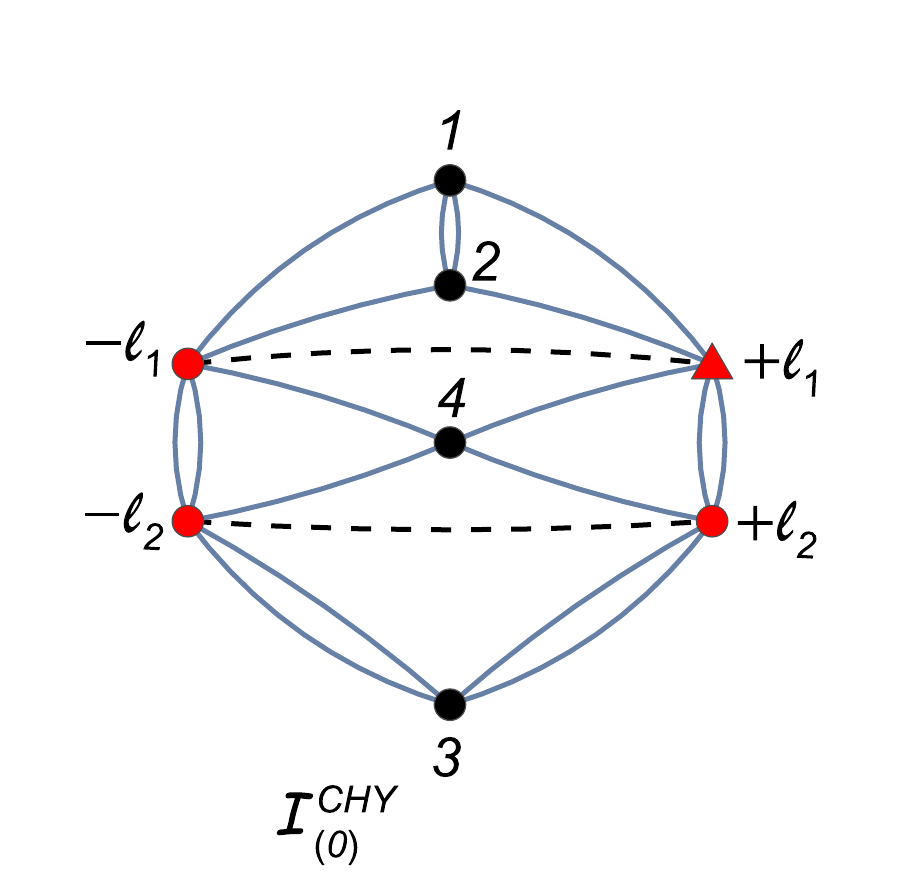}
            \includegraphics[scale=0.4]{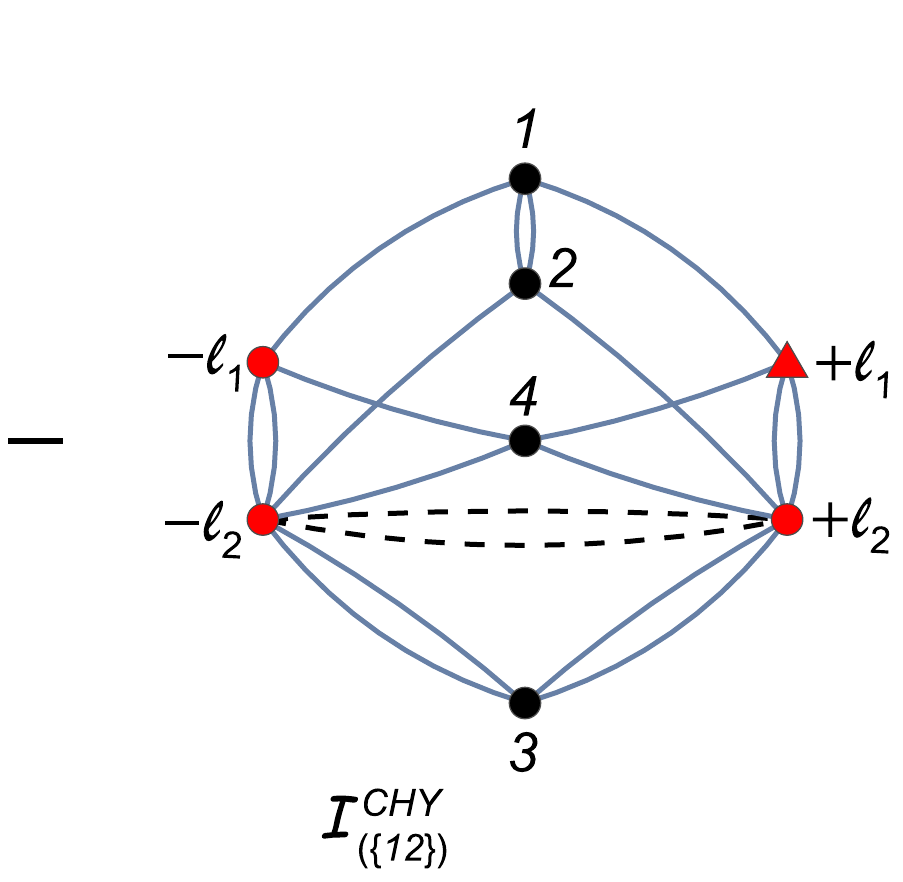}
                \includegraphics[scale=0.4]{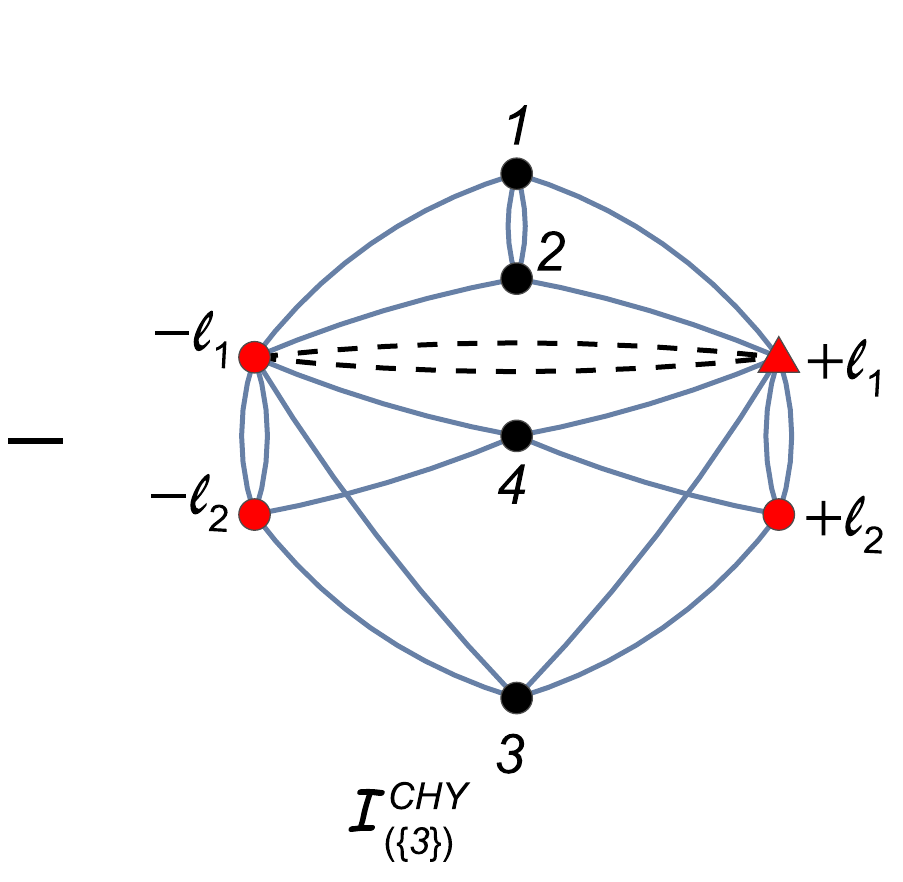}
                   \includegraphics[scale=0.4]{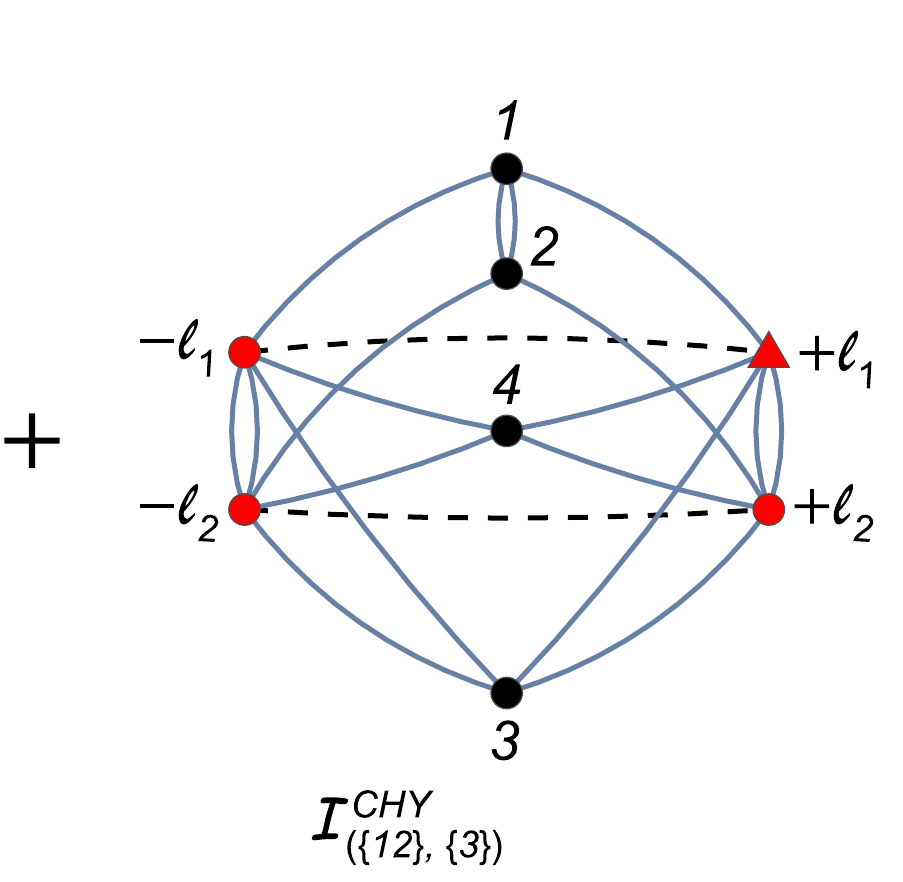}
  \caption{The CHY graphs for non-planar Feynman diagram.}\label{CHY_NP}
\end{figure}

In this section we consider a non-planar Feynman diagram at two loops. In order to give a simple but non-trivial example we focus on the diagram given in figure \ref{Fey_NP}. 
From this diagram one can easily read off its Feynman integrand, which is 
\begin{equation}
{\cal I}_{\rm FEY}^{\rm non-planar}=\frac{1}{2\,\ell_1^2\,\ell_2^2\,k_{12}\,(\ell_1+\ell_2)^2\,(\ell_1+\ell_2+k_4)^2\,(\ell_1-k_1-k_2)^2\,(\ell_2-k_3)^2}.
\end{equation}
Using the partial fraction identity in the factors\footnote{Let us recall the loop integration measure and its integration contour are invariants under these transformations.}
\begin{align*}
\frac{1}{\ell_1^2\,(\ell_1-k_1-k_2)^2}\times \frac{1}{\ell_2^2\,(\ell_2-k_3)^2} 
\end{align*}
and shifting the loop momenta $\{\ell_1, \ell_2 \}$  to obtain the global factor ${1\over \ell^2_1\,\ell_2^2}$, one can check that the Feynman integrand in \eqref{I2points} becomes
\begin{align}\label{Fey_4points_np}
&{\cal I}_{\rm FEY}^{\rm non-planar}\Big|_{\rm p.f \atop S}=\frac{1}{2^3\,k_{12}\,\,\ell_1^2\, \ell_2^2}\left[ 
\frac{1}{(\ell_1+\ell_2)^2\,(\ell_1+\ell_2+k_4)^2 \,(-\ell_1\cdot (k_1 +k_2)+k_{12}) \,\, (-\ell_2\cdot k_3)}
 \right.  \nonumber\\
&
\left.
+
\frac{1}{(\ell_1+\ell_2+k_1+k_2)^2 (\ell_1+\ell_2+k_1+k_2+k_4)^2(\ell_1\cdot (k_1+k_2)+k_{12}) (-\ell_2\cdot  k_3)}
+
\left(
\begin{matrix}
\ell_1\rightarrow -\ell_1\\
\ell_2\rightarrow -\ell_2
\end{matrix}
\right)
\right].
\end{align}
From the building block given in section \ref{building-blocks} for the non-planar case and the gluing technique developed in section \ref{gluesection}, the CHY integrand corresponds to the Feynman integrand in \eqref{Fey_4points_np} should be\footnote{For more details about the gluing process, see the examples in appendix \ref{GluingL}.}
\begin{align}\label{chy_NplanarI}
{\cal I}_{\rm CHY}^{\rm non-planar} &= \frac{1}{\ell_1^2\,\, \ell_2^2}\int\, d\mu^{\rm 2-loop} \,{\bf I}_{\rm CHY}^{\rm non-planar}, \\
{\bf I}_{\rm CHY}^{\rm non-planar} &= {\rm \ss^{non-planar}}\left[\frac{\o^1_{1}(\o^1_{2}-\o^2_{2})}{(12)}   \times  \q^3_{4}\times \q^2_{3} \right].\nonumber
\end{align}
Clearly, the ${\bf I}_{\rm CHY}^{\rm non-planar}$ integrand is a linear combination of four CHY graphs, as it is shown in  figure \ref{CHY_NP}. 

\begin{figure}
 \centering
       \includegraphics[scale=0.4]{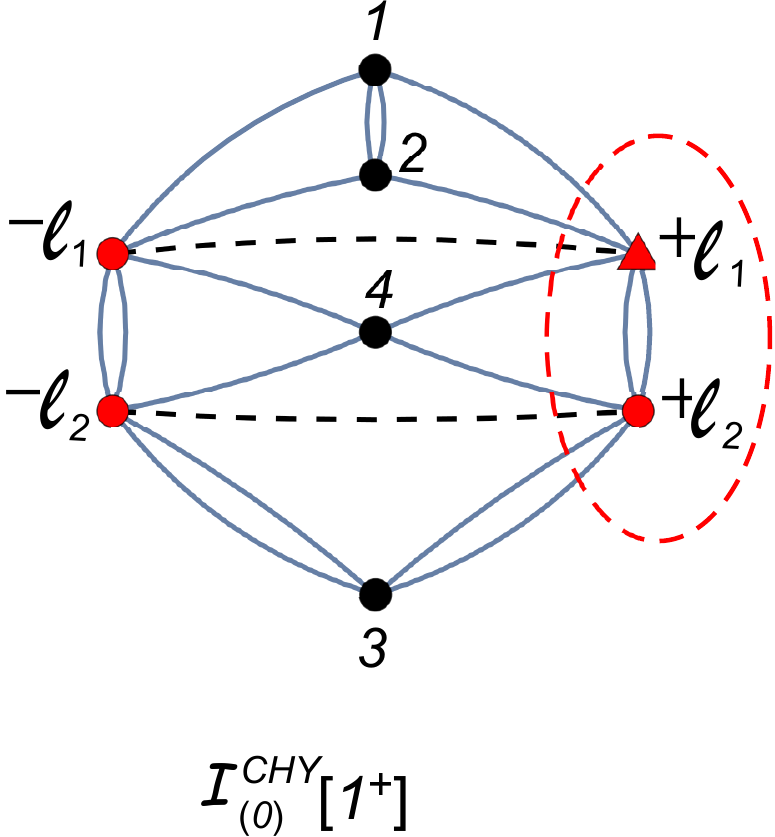}~\,
            \includegraphics[scale=0.4]{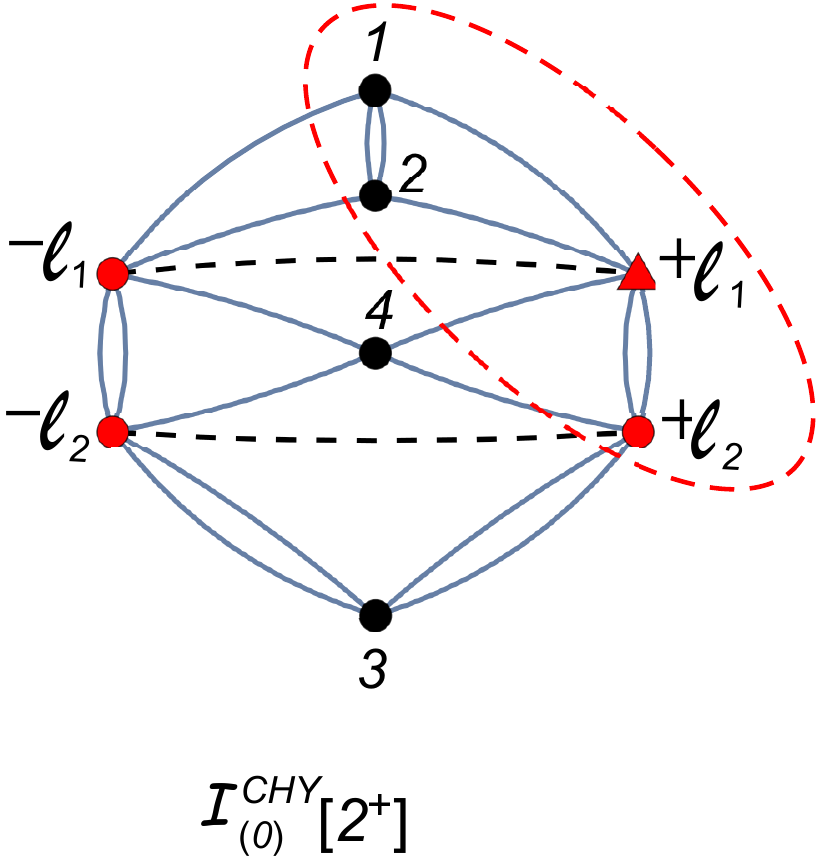}~\,
               \includegraphics[scale=0.4]{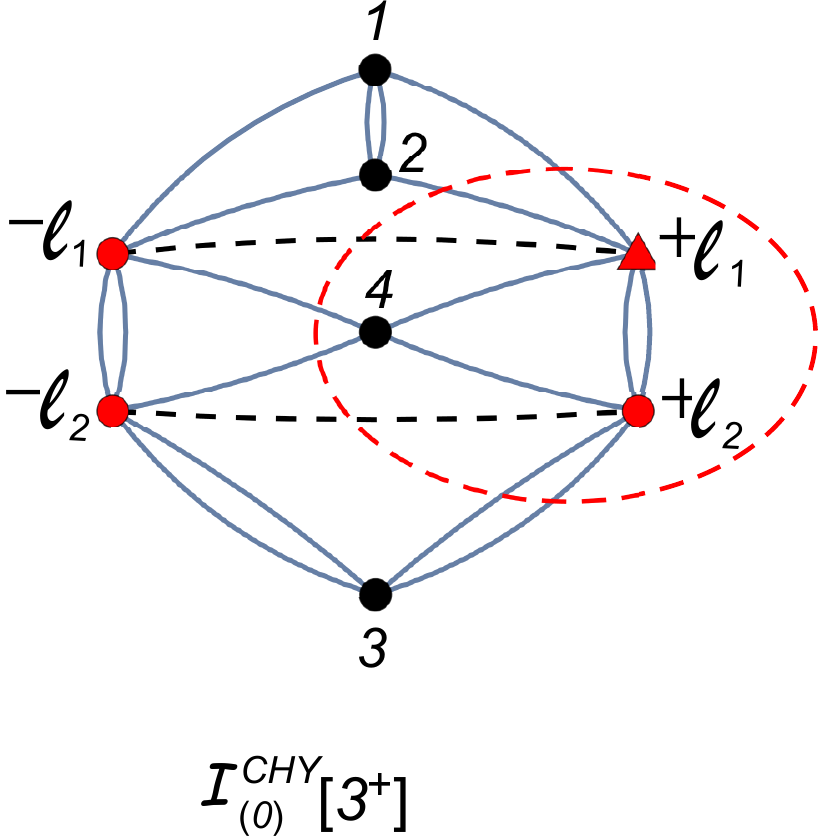}~\,
                   \includegraphics[scale=0.4]{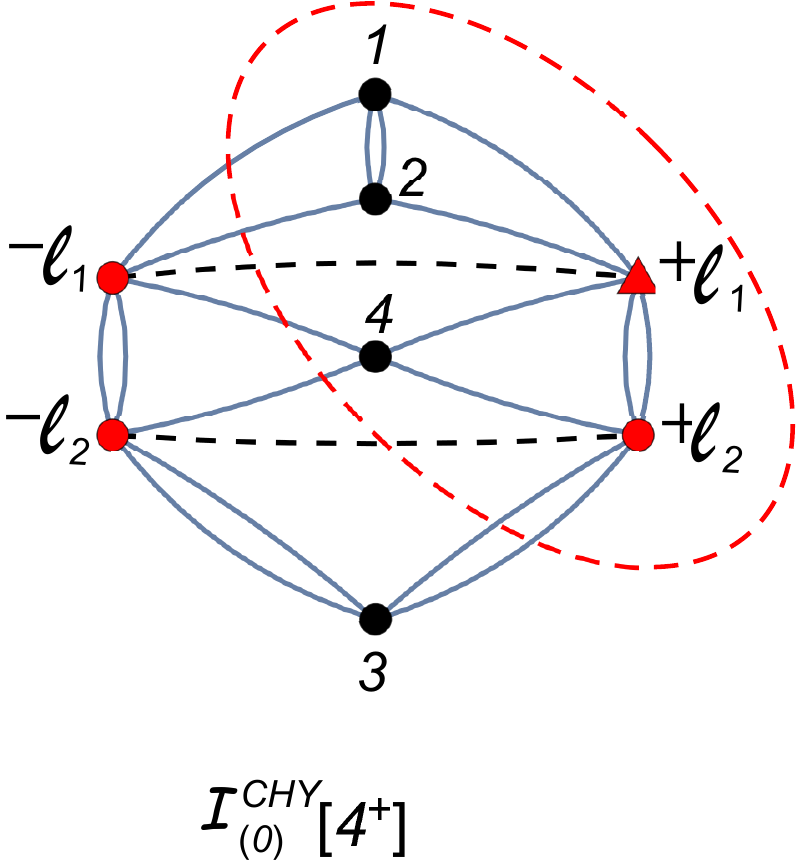}
  \caption{All possible non-zero cuts (up to transformation $\ell_1\,\leftrightarrow \,-\ell_1 $ and $\ell_2\,\leftrightarrow \,-\ell_2 $)  for ${\cal I}^{\rm CHY}_{(0)}$.}\label{CHY_NP_cuts}
\end{figure}

By applying the $\L-$algorithm to compute $\int\, d\mu^{\rm 2-loop} \,{\bf I}_{\rm CHY}^{\rm non-planar}$, it is clear that there are just eight non-zero cuts for each CHY graph given in figure \ref{CHY_NP}. In figure \ref{CHY_NP_cuts},  for the graph ${\cal I}^{\rm CHY}_{(0)}$,  we have drawn four of the eight possible non-zero cuts.  Nevertheless,  as it is simple to notice, the others four cuts, namely $\{[1^-],[2^-],[3^-],[4^-]\}$,  can be obtained by the transformation  $\ell_1\leftrightarrow-\ell_1,\, \ell_2\leftrightarrow-\ell_2$.  In addition, the cuts for the other CHY graphs, $\{{\cal I}^{\rm CHY}_{(\{ 12 \} )},{\cal I}^{\rm CHY}_{(\{ 3 \})},{\cal I}^{\rm CHY}_{(\{12\}, \{ 3\})}  \}$, are exactly the same as ones given for ${\cal I}^{\rm CHY}_{(0)}  $.

As it was explained in section \ref{4pointsplanar}, the singular cut is canceled out and we do not need to consider it.  Therefore, using the $\L-$algorithm and the rules found in section \ref{L-algorithm-2L}, it is straightforward to compute each cut, which gives the results:
\begin{align}
&{\cal I}^{\rm CHY}_{(0)}[1^+]-{\cal I}^{\rm CHY}_{(\{12\})}[1^+]-{\cal I}^{\rm CHY}_{(\{3\})}[1^+]+{\cal I}^{\rm CHY}_{(\{12\}\{3\})}[1^+] \label{NP_cut1+}\\
&=
\frac{1}{k_{12}\,{1\over 2}(\ell_1+\ell_2)^2}\frac{1}{(-\ell_1\cdot (k_1+k_2)+k_{12}) \,\, (-\ell_2\cdot k_3)(k_4\cdot (\ell_1+\ell_2))} ,\nonumber\\
&\nonumber\\
&{\cal I}^{\rm CHY}_{(0)}[2^+]-{\cal I}^{\rm CHY}_{(\{12\})}[2^+]-{\cal I}^{\rm CHY}_{(\{3\})}[2^+]+{\cal I}^{\rm CHY}_{(\{12\}\{3\})}[2^+] \label{NP_cut2+}\\
&=
\frac{1}{k_{12}\,{1\over 2}(\ell_1+\ell_2+k_1+k_2)^2}\frac{1}{(\ell_1\cdot (k_1+k_2)+k_{12}) \,\, (-\ell_2\cdot k_3)(k_4\cdot(\ell_1+\ell_2+k_1+k_2))} ,\nonumber\\
&\nonumber\\
&{\cal I}^{\rm CHY}_{(0)}[3^+]-{\cal I}^{\rm CHY}_{(\{12\})}[3^+]-{\cal I}^{\rm CHY}_{(\{3\})}[3^+]+{\cal I}^{\rm CHY}_{(\{12\}\{3\})}[3^+] \label{NP_cut3+}\\
&=
\frac{1}{k_{12}\,{1\over 2}(\ell_1+\ell_2+k_4)^2}\frac{1}{(-\ell_1\cdot (k_1+k_2)+k_{12}) \,\, (-\ell_2\cdot k_3)(-k_4\cdot(\ell_1+\ell_2))} ,\nonumber\\
&\nonumber\\
&{\cal I}^{\rm CHY}_{(0)}[4^+]-{\cal I}^{\rm CHY}_{(\{12\})}[4^+]-{\cal I}^{\rm CHY}_{(\{3\})}[4^+]+{\cal I}^{\rm CHY}_{(\{12\}\{3\})}[4^+] \label{NP_cut4+}\\
&=
\frac{1}{k_{12}\,{1\over 2}(\ell_1+\ell_2+k_1+k_2+k_4)^2}\frac{1}{(\ell_1\cdot (k_1+k_2)+k_{12}) \,\, (-\ell_2\cdot k_3)(-k_4\cdot(\ell_1+\ell_2+k_1+k_2))} .\nonumber
\end{align}
Let us recall that to obtain the result for the other four cuts, one just makes the transformation, $\ell_1\,\leftrightarrow \,-\ell_1 $ and $\ell_2\,\leftrightarrow \,-\ell_2 $. Summing over all contributions  given in \eqref{NP_cut1+}, \eqref{NP_cut2+}, \eqref{NP_cut3+}, \eqref{NP_cut4+} and the other four cuts obtained by, $\ell_1\,\leftrightarrow \,-\ell_1 $ and $\ell_2\,\leftrightarrow \,-\ell_2 $, it is straightforward to check
\begin{equation}
\frac{{\cal I}_{\rm CHY}^{\rm non-planar}}{2^4} =
\frac{{\cal I}^{\rm CHY}_{(0)}-{\cal I}^{\rm CHY}_{(\{12\})}-{\cal I}^{\rm CHY}_{(\{3\})}+{\cal I}^{\rm CHY}_{(\{12\},\{3\})}}{2^4\,\,\ell_1^2\,\ell_2^2}  = {\cal I}_{\rm FEY}^{\rm non-planar}\Big|_{\rm p.f \atop S}.
\end{equation}

In this section we have computed explicitly some non-trivial examples and verified our conjecture given in section \ref{building-blocks}, with the help of the new two-loop $\L$ rules. We have also verified more complicated examples up to seven external particles at two loops. In addition, it is useful to remember that all computations were checked numerically.

\section{Conclusions}\label{conclusions}

In this work we have introduced a way of computing the CHY integrands corresponding to given Feynman diagrams up to two loops. Starting from the holomorphic forms on the Riemann surfaces, we have defined appropriate quadratic
differentials that serve as building blocks for constructing the CHY integrands. Together with the gluing rules, they allow for the reconstruction of arbitrary Feynman diagrams in the CHY language.

We have used the two-loop scattering equations defined in \cite{Geyer:2016wjx} to generalize the $\Lambda$-algorithm \cite{Gomez:2016bmv,Cardona:2016bpi} to two loops. This prescription allows for easy computation of the CHY integrals using graphical rules. We have demonstrated on several examples the usefulness of this algorithm in explicit computations of CHY integrands.

Importantly, all the integrands defined in this work are free of the poles of the form $1/\sigma_{l^+ l^-}$. Because of this, all classes of solutions, i.e., the degenerate and non-degenerate ones \cite{He:2015yua,Cachazo:2015aol,Geyer:2015bja,Geyer:2016wjx}, give finite contributions and there is no need for different treatment of different solutions. Hence, these CHY integrals can be simply evaluated using the other methods or numerically. Nevertheless, in this work we have utilized the $\Lambda-$algorithm to make the calculations even simpler.

As always, there is a question of generalizing to higher-loop orders. We hope that the procedure of defining the holomorphic and quadratic
differentials, together with the physical constraints of the factorization channels, described in sections \ref{holo} and \ref{building-blocks} can pave a new way for generalizing the CHY approach to higher loops. We leave the analysis of the three-loop case as a future research direction. 

For now, we have focused on studying the structure of factorizations and scattering equations, for which the $\Phi^3$ theory is a perfect playground. Once these properties are well-understood, an interest would lie in generalizing this approach to other theories. The first theory to consider would be the bi-adjoint scalar \cite{Cachazo:2013iea}, which shares the greatest similarity with $\Phi^3$ theory while there is no symmetrization of particles.  After the bi-adjoint scalar theory is settled, one future direction would be to express more complicated amplitudes such as Yang--Mills and Einstein gravity into a basis of the bi-adjoint scalars, along the lines of \cite{Lam:2016tlk,Bjerrum-Bohr:2016axv}. It would also be  interesting to start from our two-loop $\Phi^3$ theory answers and to generalize the results of \cite{Cachazo:2015aol,Geyer:2015bja} which show that at one loop one can define compact expression for the CHY integrands for the bi-adjoint, Yang--Mills and Einstein gravity theories that preserve the double copy structure \cite{Geyer:2015jch}.

In particular, an exciting approach of Cachazo, He, and Yuan \cite{Cachazo:2015aol} treats one-loop amplitudes in four-dimensions as a dimensional reduction of the five-dimensional tree-level amplitude. It would be interesting to see whether a similar procedure can be followed in the two-loop case, this time reducing from six-dimensional amplitudes. In conjunction with the ambitwistor approach \cite{Geyer:2015bja,Geyer:2016wjx}, it could be useful in deriving compact expression for two-loop CHY integrands.

Finally, we would like to comment on the choice of building blocks we have used. Namely, at two loops two skeleton functions make appearance, depending on planarity of the diagram we wish to reproduce:
\begin{align}
{\rm \ss^{ planar}}=\frac{1}{(\ell_1^+,\ell_2^+,\ell_2^-,\ell_1^-)(\ell_2^+,\ell_1^+,\ell_2^-,\ell_1^-)} \quad{\rm and}\quad {\rm \ss^{non-planar}} =\frac{1}{(\ell_1^+,\ell_2^+,\ell_2^-,\ell_1^-)^2}.
\end{align}
However, in principle other ${\rm PSL}(2,\mathbb{C})$ combinations could have entered. What singles out these two? We would like to understand the constraints, coming from factorization properties, placed on this choice in the future work.

Similarly, other combinations of the quadratic
differentials could have been used. Let us briefly consider one choice,
\begin{align}
\q^4_a &:= (\o^1_a - \o^2_a)(\o^1_a - \o^2_a)\nonumber\\
&= \o^1_a (\o^1_a - \o^2_a) + \o^2_a (\o^2_a - \o^1_a)\nonumber\\
&= \q^1_a + \q^2_a.
\end{align}
Hence, this object sums over two possibilities of attaching the external leg with label $a$ to both left and right loops. It strongly suggests that this quadratic
differential should appear in constructing the CHY representation of the full loop integrand for $\Phi^3$ theory, as a sum over all possible Feynman diagrams. We leave this as a future research direction.

\acknowledgments

We thank F. Cachazo for many useful discussions. H.G. would like to thank to C. Cardona for discussions. H.G. is very grateful to the Perimeter Institute for hospitality during this work. This research was supported in part by Perimeter Institute for Theoretical Physics. Research at Perimeter Institute is supported by the Government of Canada through the Department of Innovation, Science and Economic Development Canada and by the Province of Ontario through the Ministry of Research, Innovation and Science. The work of  H.G.  is supported by USC grant DGI-COCEIN-No 935-621115-N22.


\appendix

\section{Gluing  Loop CHY integrands}\label{GluingL}

In this appendix we perform the gluing procedure  developed in section \ref{gluesection} for the examples given in section \ref{non1PI} and  \ref{4pointsplanar} in an explicit way.

\subsection{Gluing of One-loop Building Block}\label{glueN1PI}

\begin{figure}[!h]
 \centering
       \includegraphics[scale=0.7]{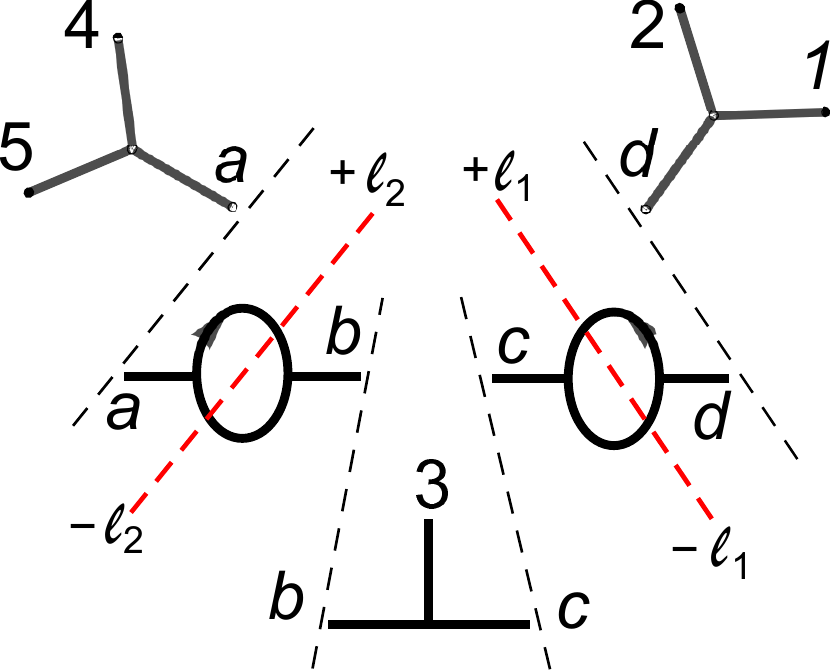}~\,
  \caption{Gluing process for the Feynman diagram in figure \ref{nononePI}.}\label{glue_n1pi}
\end{figure}

In this section we build the CHY integrand corresponding to the Feynman diagram given in figure \ref{nononePI} from the one-loop building block.  The idea is to cut the Feynman digram, as it is shown in figure \ref{glue_n1pi}, 
and to glue the building blocks following the rules obtained in section \ref{gluesection}.\\
Clearly, there are three tree-level building blocks given by
\begin{align}
{\bf I}^{\rm tree}_{\rm CHY}(a) &= \frac{1}{(\s_{45} \, \s_{5a}\,\s_{a4})^2} \label{ITa},\\
{\bf I}^{\rm tree}_{\rm CHY}(b|c) &= \frac{1}{(\s_{3b} \, \s_{bc}\,\s_{c3})^2} \label{ITbc},\\
{\bf I}^{\rm tree}_{\rm CHY}(d) &= \frac{1}{(\s_{12} \, \s_{2d}\,\s_{d1})^2} \label{ITd},
\end{align}
and two one-loop building blocks
\begin{align}
{\bf I}^{\rm one-loop}_{\rm CHY}(a|b) &= \frac{1}{(\s_{\ell_2^+\ell_2^-} \, \s_{\ell_2^-\ell_2^+})^2}\left[( \o^2_{a})^2 ( \o^2_{b})^2 \right] \nonumber\\
&= (\s_{\ell_2^+ \ell_2^-} \, \s_{\ell_2^-\ell_2^+})^2  \left[\frac{1}{( \s_{a\ell_2^+}\,\s_{\ell_2^+\ell_2^-}\s_{\ell_2^-a})^2 }  \frac{1} { ( \s_{b\ell_2^+}\,\s_{\ell_2^+\ell_2^-}\s_{\ell_2^-b})^2 } \right],\label{ILab} \\
{\bf I}^{\rm one-loop}_{\rm CHY}(c|d) &= \frac{1}{(\s_{\ell_1\ell_1^-} \, \s_{\ell_1^-\ell_1})^2}\left[( \o^1_{c})^2 ( \o^1_{d})^2 \right] \nonumber\\
&= (\s_{\ell_1^+ \ell_1^-} \, \s_{\ell_1^-\ell_1^+})^2 \left[\frac{1}{( \s_{c\ell_1^+}\,\s_{\ell_1^+\ell_1^-}\s_{\ell_1^-c})^2 }  \frac{1} { ( \s_{d\ell_1^+}\,\s_{\ell_1^+\ell_1^-}\s_{\ell_1^-d})^2 } \right]. \label{ILcd}
\end{align}
In the brackets of  \eqref{ILab} and \eqref{ILcd}, one can see two three-point tree-level CHY integrands, which are represented by the dotted red lines in figure \ref{glue_n1pi}.  Now, we are ready to perform the gluing procedure that should be carried out graph by graph. Using the rules found in section \ref{gluesection} while taking advantage of ${\rm OE}(a,{\cal I}^{\rm tree}_{\rm CHY}(a) )=\{4,4,5,5\}$ and ${\rm OE}(a,{\cal I}^{\rm one-loop}_{\rm CHY}(a|b) )=\{\ell_2^-,\ell_2^-,\ell_2^+,\ell_2^+\}$, one obtains
\begin{align}
 {\bf I}^{\rm loop-tree}_{\rm CHY}(b)  &:= \left( {\bf I}^{\rm tree}_{\rm CHY}(a)  , \, {\bf I}^{\rm one-loop}_{\rm CHY}(a|b)    \right)_a \nonumber\\
 &=
\left[\frac{\o^2_{4} \, \o^2_{5}}{(4,5) } \times  \frac{1} { ( \s_{b\ell_2^+}\,\s_{\ell_2^+\ell_2^-}\s_{\ell_2^-b})^2 } \right]\label{ILb},
\end{align}
where we have used the definition given in \eqref{chainS}. In analogy, from ${\rm OE}(d,{\cal I}^{\rm tree}_{\rm CHY}(d) )=\{1,1,2,2\}$ and ${\rm OE}(d,{\cal I}^{\rm one-loop}_{\rm CHY}(c|d) )=\{\ell_1^-,\ell_1^-,\ell_1^+,\ell_1^+\}$ we can glue  \eqref{ITd} with \eqref{ILcd},
\begin{align}
 {\bf I}^{\rm loop-tree}_{\rm CHY}(c)  &:= \left( {\bf I}^{\rm tree}_{\rm CHY}(d)  , \, {\bf I}^{\rm one-loop}_{\rm CHY}(c|d)    \right)_d \nonumber\\
 &=
\left[   \frac{1} { ( \s_{c\ell_1^+}\,\s_{\ell_1^+\ell_1^-}\s_{\ell_1^-c})^2 } \times     \frac{\o^1_{1} \, \o^1_{2}}{(1,2) }\right]\label{ILc}.
\end{align}
The next step is to glue \eqref{ITbc} with \eqref{ILb} by using ${\rm OE}(b,{\cal I}^{\rm tree}_{\rm CHY}(b|c) )=\{3,3,c,c\}$ as well as ${\rm OE}(b,{\cal I}^{\rm loop-tree}_{\rm CHY}(a|b) )=\{\ell_2^-,\ell_2^-,\ell_2^+,\ell_2^+\}$, so
\begin{align}
 {\bf I}^{\rm loop-tree^2}_{\rm CHY}(c)  &:= \left( {\bf I}^{\rm tree}_{\rm CHY}(b|c)  , \, {\bf I}^{\rm loop-tree}_{\rm CHY}(b)    \right)_b \nonumber\\
 &=\frac{1}{(\s_{\ell_2^+\ell_2^-})^3}
\left[\frac{\o^2_{4} \, \o^2_{5}}{(4,5) } \times  \frac{\o^2_{3}} {  ( \s_{c\ell_2^+}\,\s_{c\ell_2^-}\,\s_{c3}\,\s_{3c}) } \right]\label{ILTc}.
\end{align}
Finally, gluing \eqref{ILc} and \eqref{ILTc} after choosing the ordered edge sets ${\rm OE}(c,{\cal I}^{\rm loop-tree}_{\rm CHY}(c) )=\{\ell_1^-,\ell_1^-,\ell_1^+,\ell_1^+\}$ and ${\rm OE}(c,{\cal I}^{\rm loop-tree^2}_{\rm CHY}(c))=\{3,3,\ell_2^-,\ell_2^+\}$ we obtain 
\begin{align}
 {\bf I}_{\rm CHY}  &:= \left( {\bf I}^{\rm loop-tree}_{\rm CHY}(c)  , \, {\bf I}^{\rm loop-tree^2}_{\rm CHY}(c)    \right)_c \nonumber\\
 &=\frac{1}{(\s_{\ell_2^+\ell_2^-})^3\,(\s_{\ell_1^+\ell_1^-})^3\,(\s_{\ell_1^+\ell_2^+})\,(\s_{\ell_1^-\ell_2^-})}\times
\left[\frac{\o^2_{4} \, \o^2_{5}}{(4,5) } \times \o^1_{3} \o^2_{3}  \times   \frac{\o^1_{1} \, \o^1_{2}}{(1,2) }
\right]\label{ILTc}.
\end{align}

\subsection{Gluing of Two-loop Planar Building Block}\label{glue_PLANAR}

\begin{figure}
 \centering
       \includegraphics[scale=0.7]{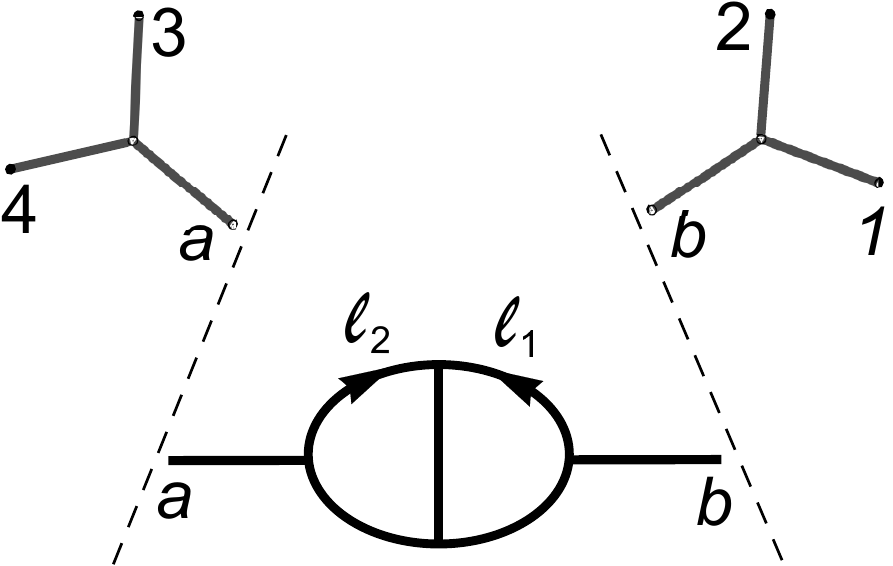}
  \caption{Gluing process for the Feynman diagram in figure \ref{2L_4p}.}\label{glue_planar}
\end{figure}

After showing how to glue one-loop CHY building block from gluing operation defined in section \ref{gluesection}, we are going to show a two-loop 1PI case by building the CHY integrand which should correspond to the two-loop planar Feynman diagram given in figure \ref{2L_4p} as an example. By cutting the Feynman digram, as it is shown in figure \ref{glue_planar}, one could find two building blocks at tree level, given by
\begin{align}
{\bf I}^{\rm tree}_{\rm CHY}(a) &= \frac{1}{(\s_{34} \, \s_{4a}\,\s_{a3})^2}\,, \label{treea}\\
{\bf I}^{\rm tree}_{\rm CHY}(b) &= \frac{1}{(\s_{12} \, \s_{2b}\,\s_{b1})^2}\,, \label{treeb}
\end{align}
and another building block at two loops
\begin{align}\label{bbtwoL}
{\bf I}_{\rm CHY}^{\rm planar}(a|b)&= {\rm \ss^{planar}}\left[ (\o^2_{a}-\o^1_{a}) \o^2_{a} \o^1_{b} (\o^1_{b}-\o^2_{b})
\right]\nonumber\\
&={\rm \ss^{planar}}\left[(\o^2_{a})^2 (\o^1_{b})^2-\o^1_{a} \o^2_{a}(\o^1_{b})^2-(\o^2_{a})^2 \o^1_{b} \o^2_{b}+\o^1_{a} \o^2_{a}\o^1_{b} \o^2_{b}\right].
\end{align}
Notice here the main difference is that we need to separate ${\cal I}_{\rm CHY}^{\rm planar}(a|b)$ into smaller pieces in order to implement our gluing operation in section \ref{2lg}. However, the prodecures are quite similar: we use the rules shown in figure \ref{fg} and \ref{2la} to obtain the ordered edge sets ${\rm OE}(a,G)$. For example, in \eqref{bbtwoL} the term with $\o^1_{a} \o^2_{a}\o^1_{b} \o^2_{b}$ has ${\rm OE}(a,{\cal I}_{\rm CHY}^{\rm planar}(a|b))=\{\ell_1^+,\ell_1^-,\ell_2^+,\ell_2^-\}$. After figuring out all the ordered edge sets we glue term by term following the rules in figure \ref{glue}.
 Gluing the CHY building blocks in \eqref{treea} and \eqref{bbtwoL}, we obtain
\begin{align}\label{glue_L-T}
 {\bf I}_{\rm CHY}^{\rm planar-tree} (b) &:=   \left( {\bf I}^{\rm tree}_{\rm CHY}(a),     {\bf I}_{\rm CHY}^{\rm planar}(a|b)  \right)_a\nonumber \\
&=
{\rm \ss^{planar}}
\times  \left[ \frac{(\o^2_{3}-\o^1_{3}) \o^2_{4}}{(3,4)}\times  \o^1_{b}(\o^1_{b}-\o^2_{b})
\right].
\end{align}
And gluing the tree-level building block in \eqref{treeb} with the CHY integrand found in \eqref{glue_L-T}
one gets the answer
\begin{align}\label{glue_final_planar}
 {\bf I}_{\rm CHY}^{\rm planar} &:=   \left({\bf I}_{\rm CHY}^{\rm planar-tree}(b),{\bf  I}^{\rm tree}_{\rm CHY}(b)  \right)_b\nonumber \\
&=
{\rm \ss^{planar}}
\times  \left[ \frac{(\o^2_{3}-\o^1_{3}) \o^2_{4}}{(3,4)}\times  \frac{\o^1_{1}(\o^1_{2}-\o^2_{2})  }{(1,2)}
\right],
\end{align}
which is the CHY integrand computed in section \ref{4pointsplanar}.

\section{Tree-level Scattering Equations}\label{Sec2}

So far, we have worked with the original embedding proposed by Cachazo, He and Yuan (CHY) in \cite{Cachazo:2013gna,Cachazo:2013hca,Cachazo:2013iaa}, namely the marked points $\{\s_i\}$ on a Riemann sphere with a single cover. Nevertheless, in order to perform analytical computations, it is well-known that the $\L-$prescription is a powerful tool. Hence, in this appendix we summarize the results of \cite{Cardona:2016bpi, Gomez:2016bmv, Cardona:2016wcr}, which are used in the calculation of the examples in section \ref{examplesphi3}.

\subsection{$\L-$Prescription}\label{Tree_prescription}

In \cite{Gomez:2016bmv}, a prescription for the computation of scattering amplitudes at tree level into the CHY framework was proposed by means of a double cover approach. The $n-$particle amplitude is given by the expression\footnote{Without loss of generality, we have fixed the $\{\s_1,\s_2,\s_3,\s_n\}$ punctures and the $\{E_{1},E_{2},E_{3}  \}$ scattering equations.}
\begin{align}\label{treeAmplitude}
&{\cal A}_n(1,2,\ldots,n)\nonumber\\
&=\frac{1}{2^2}\int_{\Gamma^{\rm t}}\left( \frac{d\L}{\L} \right) \left(\prod_{a=1}^n \frac{y_a \, d y_a}{C_a}  \right)\times \left( \prod _{i=4}^{n-1} \frac{d\s_i}{E^{\rm t}_i} \right)\times |1,2,3|\,\,\Delta_{\rm FP}(123n)\times  \frac{{\bf I}_n(\s,y)}{E^{\rm t}_n},
\end{align}
where the $\Gamma^{\rm t}$ integration contour is defined by the $2n-3$ equations
\begin{align}
&\L=0,~~~C_a:=y_a^2-(\s_a^2-\L^2)=0,~~ a=1,\ldots, n, \nonumber\\
& E^{\rm t}_i:=\half\sum_{j=1\atop j\ne i}^n {k_i\cdot k_j \over \s_{ij}}\left({y_j\over y_i}+1 \right)=0, ~~ i=4,5,\ldots, n-1, \qquad {\rm with}~~\s_{ij}:=\s_i-\s_j.\label{Lscattering}
\end{align}
The $\{E^{\rm t}_i=0\}$ corresponds to the tree-level scattering equations and the $\{C_a=0\}$  constraints define the double covered sphere.

The Faddeev--Popov determinants, $|1,2,3|$ and $\Delta_{\rm FP}(123n)$, are given by the expressions
\begin{align}
|1,2,3| &= \frac{1}{\Lambda^2} \left|
                            \begin{array}{cccc}
                              y_1\,\, &\,\,y_1(\sigma_1+y_1)\,\,  & \,\, y_1(\sigma_1-y_1) \\
                             y_2\,\, &\,\,y_2(\sigma_2+y_2)\,\,  & \,\, y_2(\sigma_2-y_2)\\  
                               y_3\,\, &\,\,y_3(\sigma_3+y_3)\,\,  & \,\, y_3(\sigma_3-y_3)\\
                            \end{array}
                          \right |, \label{FPone}\\ 
\Delta_{\rm FP}(123n) &= \frac{1}{\Lambda^2} \left|
                            \begin{array}{cccc}
                              y_1\,\, &\,\,y_1(\sigma_1+y_1)\,\,  & \,\, y_1(\sigma_1-y_1)\,\,  &\,\, \s_1 \\
                             y_2\,\, &\,\,y_2(\sigma_2+y_2)\,\,  & \,\, y_2(\sigma_2-y_2)\,\,  &\,\, \s_2 \\  
                               y_3\,\, &\,\,y_3(\sigma_3+y_3)\,\,  & \,\, y_3(\sigma_3-y_3)\,\,  &\,\, \s_3 \\
                                y_4\,\, &\,\,y_4(\sigma_4+y_4)\,\,  & \,\, y_4(\sigma_4-y_4)\,\,  &\,\, \s_4 \\
                            \end{array}
                          \right|. \label{FPtwo}
\end{align}

The ${\bf I}_n(\s,y)$ is the integrand which defines the theory and is a rational function in terms of {\bf chains}.  For the sake of completeness let us remind that we define a {\bf $k$-chain} as a  sequence of $k$-objects \cite{Cachazo:2015nwa}, in this case a $k$-chain is read as
\begin{equation}\label{chains}
\tau_{i_1:i_2} \tau_{i_2:i_3}\cdots \tau_{i_{k-1}:i_k} \tau_{i_{k}:i_1}:=(i_1: i_2:\cdots : i_k),
\end{equation}
where the $\tau_{a:b}$'s are the third-kind forms 
\begin{equation}\label{tau}
\tau_{a:b}:={1\over 2\,y_a}\left({y_a+y_b+\s_{ab}\over\s_{ab}}\right).
\end{equation}
After integration over the moduli parameter $\Lambda$, the $\tau_{a:b}$ becomes the more familiar $1/z_{ab}$ over the sphere.\footnote{In this note we will focus on computations over the punctured sphere only, and hence  the integrands and other quantities will be given in terms of the usual $z_{ab}$ only. }
Note that the chains have a maximum length, which is the total number of particles $n$. 

\subsubsection{CHY Tree-level Graph}

\begin{figure}[h]
  \centering
  \includegraphics[width=1.2in]{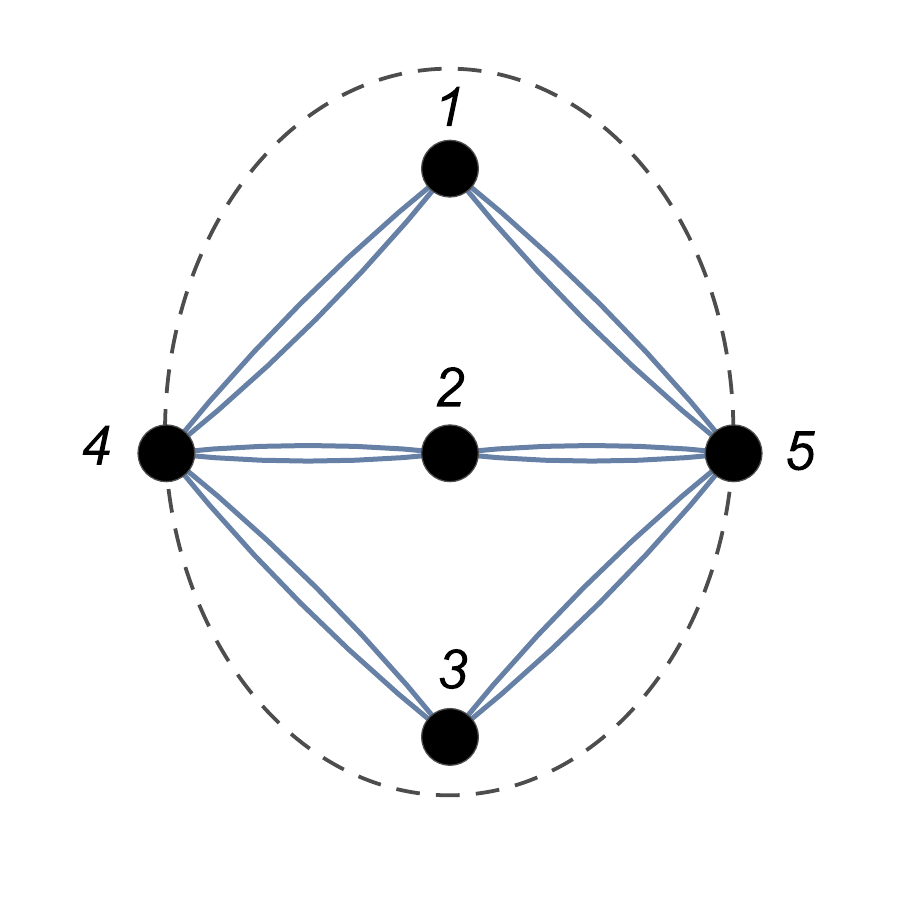}\\
  \caption{The ${\cal I}_5(1,2,3,4,5)$  {\sl regular graph}.}\label{example_CHYG}
\end{figure}

Let us recall here that each ${\bf I}_n(\s,y)$ integrand has a {\bf regular graph}\footnote{A $G$ graph  is defined by the two finite sets, $V$ and $E$.  $V$ is the vertex set and $E$ is the edge set.}  (bijective map) associated, which we denoted by $G=(V_G,E_G)$ \cite{graph1,graph2,Cachazo:2015nwa}.  The vertex set of $G$ is given by the $n$-labels (punctures) 
$$
V_G=\{1,2,\ldots,n\},
$$ 
and the edges are given by the lines and anti-lines:
\begin{align}
&\tau_{a:b}\,\leftrightarrow\,a ~\overline{~~~~~~~~~~~~~}~ b ~~~{\rm (line)}  \\
& \tau^{-1}_{a:b}\,\leftrightarrow\, a\,- \, - \,  - \, -\, b ~~~     {\rm  (anti-line)}.
 \end{align}
Since $\tau_{a:b}$ always appears into a chain,  the graph is not a directed graph,  in the same way as in \cite{Cachazo:2015nwa}. For example, let us consider the integrand
\begin{equation}\label{fexample}
{\bf I}_5(1,2,3,4,5)=\frac{(1:5:2:4) (3:4:2:5)\times (1:4:2:5) (3:5:2:4)} {(4:5)}.
\end{equation}
This integrand is represented by the  $G$ graph in figure \ref{example_CHYG}.
\\
Note that for each vertex the number of lines minus anti-lines  must always be 4,
$$
\#\, {\rm  lines} - \#\,{\rm antilines}=4,
$$
which is a consequence of the ${\rm PSL}(2,\mathbb{C})$ symmetry.

\section{$\L-$Scattering Equations at Two Loops}\label{Ltwoloop}

In a similar way as on a torus, a double torus can be represented as a double cover of a sphere with three branch cuts, i.e. an hyperelliptic curve in $\mathbb{CP}^2$. After collapsing two of three branch cuts,  four new massive particles arise with momentum $\{\ell_1^\mu,-\ell_1^\mu,\ell_2^\mu,-\ell_2^\mu\}$, respectively, and it should give a CHY graph as in figure \ref{chy_5points}. This process will not be discussed here, but we will explain later how to obtain some of these  graphs. Finally, the third branch cut is used to perform the $\L-$algorithm on this graph. In this section we focus on the $\L-$scattering equations and our starting point is the scattering equations given in \cite{Geyer:2016wjx}.

In \cite{Gomez:2016bmv}, it is simple to notice that the map from the original scattering equations \cite{Cachazo:2013iaa,Cachazo:2013gna,Cachazo:2013hca} to the $\L-$scattering equations (see \eqref{Lscattering})  is given by the replacement 
\begin{equation}
\frac{1}{\s_{ij}}~~\longrightarrow ~~\frac{1}{2\, \s_{ij}}\left( \frac{y_j}{y_i}+1\right), ~~~{\rm with} ~~~y_k^2=\s_k^2-\L^2\,,~~i,j,k=1,2,\ldots, n.  
\end{equation}

Following this idea and from the two-loop scattering equations in \cite{Geyer:2016wjx}, we propose the $\L-$scattering equations at two loops as
\begin{align}\label{scatteringtwo}
E_i:=&\sum_{j=1\atop j\ne i}^n {k_i\cdot k_j \over 2\,\s_{ij}}\left({y_j\over y_i}+1 \right)+
 {k_i\cdot k_{n+1} \over 2\, \s_{i(n+1)}}\left({y_{n+1}\over y_i}+1 \right)+
 {k_i\cdot k_{n+2} \over 2\, \s_{i(n+2)}}\left({y_{n+2}\over y_i}+1 \right)\\
 &+ 
 {k_i\cdot k_{n+3} \over 2\, \s_{i(n+3)}}\left({y_{n+3}\over y_i}+1 \right)+ 
 {k_i\cdot k_{n+4} \over 2\, \s_{i(n+4)}}\left({y_{n+4}\over y_i}+1 \right), \nonumber \\
 E_{n+1}:=&\sum_{j=1}^n {k_{n+1}\cdot k_j \over 2\,\s_{(n+1)j}}\left({y_j\over y_{n+1}}+1 \right)+
 {k_{n+1}\cdot k_{n+3}+\frac{\a}{2}(k_{n+1}^2 + k_{n+3}^2) \over 2\, \s_{(n+1)(n+3)}}\left({y_{n+3}\over y_{n+1}}+1 \right)\nonumber\\
 &+ 
{k_{n+1}\cdot k_{n+4}-\frac{\a}{2}(k_{n+1}^2 + k_{n+4}^2) \over 2\, \s_{(n+1)(n+4)}}\left({y_{n+4}\over y_{n+1}}+1 \right),  \nonumber \\
 E_{n+2}:=&\sum_{j=1}^n {k_{n+2}\cdot k_j \over 2\,\s_{(n+2)j}}\left({y_j\over y_{n+2}}+1 \right)+
 {k_{n+2}\cdot k_{n+3}-\frac{\a}{2}(k_{n+2}^2 + k_{n+3}^2) \over 2\, \s_{(n+2)(n+3)}}\left({y_{n+3}\over y_{n+2}}+1 \right)\nonumber\\
 &+ 
{k_{n+2}\cdot k_{n+4}+\frac{\a}{2}(k_{n+2}^2 + k_{n+4}^2) \over 2\, \s_{(n+2)(n+4)}}\left({y_{n+4}\over y_{n+2}}+1 \right)=0,  \nonumber \\
 E_{n+3}:=&\sum_{j=1}^n {k_{n+3}\cdot k_j \over 2\,\s_{(n+3)j}}\left({y_j\over y_{n+3}}+1 \right)+
 {k_{n+3}\cdot k_{n+1}+\frac{\a}{2}(k_{n+3}^2 + k_{n+1}^2) \over 2\, \s_{(n+3)(n+1)}}\left({y_{n+1}\over y_{n+3}}+1 \right)\nonumber\\
 &+ 
{k_{n+3}\cdot k_{n+2}-\frac{\a}{2}(k_{n+3}^2 + k_{n+2}^2) \over 2\, \s_{(n+3)(n+2)}}\left({y_{n+2}\over y_{n+3}}+1 \right),  \nonumber \\
 E_{n+4}:=&\sum_{j=1}^n {k_{n+4}\cdot k_j \over 2\,\s_{(n+4)j}}\left({y_j\over y_{n+4}}+1 \right)+
 {k_{n+4}\cdot k_{n+1}-\frac{\a}{2}(k_{n+4}^2 + k_{n+1}^2) \over 2\, \s_{(n+4)(n+1)}}\left({y_{n+1}\over y_{n+4}}+1 \right)\nonumber\\
 &+ 
{k_{n+4}\cdot k_{n+2}+\frac{\a}{2}(k_{n+4}^2 + k_{n+2}^2) \over 2\, \s_{(n+4)(n+2)}}\left({y_{n+2}\over y_{n+4}}+1 \right),  \nonumber
\end{align} 
where, without loss of generality,  we choose\footnote{For more details about $\alpha$ parameter see \cite{Geyer:2016wjx}.} $\a=1$. These scattering equations are supported on the curves, $y_A^2=\s_A^2-\L^2,~~A=1,\ldots, n+4$, and we have defined 
\begin{align}\label{Notation_ell}
\{k^\mu_{n+1},k^\mu_{n+2},k^\mu_{n+3},k^\mu_{n+4}  \}:=&\{\ell^\mu_1,-\ell^\mu_1,\ell^\mu_2,-\ell^\mu_2  \},\nonumber\\
\{\s_{(n+1)},\s_{(n+2)},\s_{(n+3)},\s_{(n+4)}\}:=& \{\s_{\ell_1},\s_{-\ell_1},\s_{\ell_2},\s_{-\ell_2}\},\\
\{y_{n+1},y_{n+2},y_{n+3},y_{n+4}\}:=& \{y_{\ell_1},y_{-\ell_1},y_{\ell_2},y_{-\ell_2}\}\nonumber.
\end{align}
It is straightforward to check that the set $\{E_i, E_{n+1},E_{n+2},E_{n+3},E_{n+4} \}$  satisfies
\begin{equation}
\sum_{A=1}^{n+4} y_A\,E_A=0,\quad \sum_{A=1}^{n+4} (\s_A+y_A)\,y_A\,E_A=0,\quad\sum_{A=1}^{n+4} (\s_A-y_A)\,y_A\,E_A=0,
\end{equation}
on the support of the momentum conservation, $\sum_{i=1}^nk_i=0$.  Therefore, these scattering equations are invariants under the operation of the global vectors 
\begin{equation}
L_0=\sum_{A=1}^{n+4}y_A\, \p_{\s_A},\quad  
L_{\pm 1}=\sum_{A=1}^{n+4}\frac{1}{\L} (\s_A\mp y_A)y_A\, \p_{\s_A}, \quad y_A^2=\s_A^2-\L^2, 
\end{equation}
which are the  generators  of the ${\rm PSL}(2,\mathbb{C})$ symmetry.

\subsection{Prescription}

The prescription to compute scattering amplitudes is totally analogous to the one given in \eqref{Tree_prescription}.  In a similar way as in \cite{Geyer:2016wjx}, we propose the scattering amplitude prescription at two loops by the expression
\begin{equation}\label{integralL}
{\rm A^{2-loop}}(1,2,\ldots, n)=\int\frac{d^D\ell_1}{\ell_1^2}\frac{d^D\ell_2}{\ell_2^2}\,{\cal A}_n^{\rm 2-loop}(1,\ldots,n|n+1,n+2|n+3,n+4),
\end{equation}
where
\begin{align}\label{prescriptionL}
&{\cal A}_n^{\rm 2-loop}(1,\ldots,n|n+1,n+2|n+3,n+4)\\
&=\frac{1}{2^2}\int_{\Gamma}\left( \frac{d\L}{\L} \right) \left(\prod_{A=1}^{n+4} \frac{y_A \, d y_A}{C_A}  \right)\times \left( \prod _{i=1}^{n} \frac{d\s_i}{E_i} \right)\times |n+2,n+3,n+4|\nonumber\\
&~~~~\times
\Delta_{\rm FP}(n+1,n+2,n+3,n+4)\times  \frac{{\cal I}_{n+4}(\s,y)}{E_{n+1}}\nonumber.
\end{align}
The integral over $\ell_1^\mu$ and $\ell_2^\mu$ in \eqref{integralL} is invariant under shifting of these variables, but in this paper we will not concentrate on this integral or its convergence. 
The $\Gamma$ integration contour is defined by the $2n+5$ equations
\begin{align}
&\L=0,~~~C_A:=y_A^2-(\s_A^2-\L^2)=0,~~ A=1,\ldots, n+4, \nonumber\\
& E_i=\half\sum_{A=1\atop A\ne i}^{n+4} {k_i\cdot k_A \over \s_{iA}}\left({y_A\over y_i}+1 \right)=0, ~~~~~~ i=1,\ldots, n,  \label{LscatteringL}
\end{align}
the Faddeev--Popov determinants, $|n+2,n+3,n+4|$ and $\Delta_{\rm FP}(n+1,n+2,n+3,n+4)$ are the same as  in \eqref{FPone} and \eqref{FPtwo} and the ${\cal I}_{n+4}(\s,y)$ is the integrand as in \eqref{Tree_prescription}.

Note that, without loss of generality, we have fixed the punctures, $\{\s_{n+1},\s_{n+2},\s_{n+3},\s_{n+4} \}$, and the scattering equations, $\{ E_{n+2},E_{n+3},E_{n+4} \}$ corresponding to the off-shell particles.  It was done in order to avoid handling these massive particles and clearly the prescription in \eqref{prescriptionL}, together with its integration contour $\Gamma$ in \eqref{LscatteringL},  is totally identical to the one given in \eqref{Tree_prescription}, up to the factors $1/E_{n+1}$ and  $1/E^{\rm t}_{n}$ respectively.

\subsection{The $\L-$algorithm at Two Loops}\label{L-algorithm-2L}

As it was noted previously, the only difference among the prescriptions given in \eqref{Tree_prescription} and \eqref{prescriptionL} is the term 
\begin{align}
E_{n+1}=E_{\ell_1}=&\sum_{j=1}^n {k_{n+1}\cdot k_j \over 2\,\s_{(n+1)j}}\left({y_j\over y_{n+1}}+1 \right)+
 {k_{n+1}\cdot k_{n+3}+\frac{1}{2}(k_{n+1}^2 + k_{n+3}^2) \over 2\, \s_{(n+1)(n+3)}}\left({y_{n+3}\over y_{n+1}}+1 \right)\nonumber\\
 &+ 
{k_{n+1}\cdot k_{n+4}-\frac{1}{2}(k_{n+1}^2 + k_{n+4}^2) \over 2\, \s_{(n+1)(n+4)}}\left({y_{n+4}\over y_{n+1}}+1 \right), 
\end{align}
instead of the traditional one
\be
E_{n+1}^{\rm t}=\half\sum_{A=1\atop A\ne n+1}^{n+4} {k_{n+1}\cdot k_A \over \s_{(n+1)A}}\left({y_A\over y_{n+1}}+1 \right).
\en
In the original version of the $\L-$algorithm given in \cite{Gomez:2016bmv}, after performing the integration over $\L$ in \eqref{Tree_prescription}, the factor $1/E^{\rm t}_n$ becomes the propagator
\begin{equation}\label{propagator}
\frac{1}{E_n^{\rm t}}=\left[\half\sum_{j=1\atop j\ne n}^{n} {k_n\cdot k_j \over \s_{nj}}\left({y_j\over y_n}+1 \right)\right]^{-1}_{\L=0} \longrightarrow~~ \frac{1}{k_{34\cdots n_un}},
\end{equation}
where we have considered that the puntures $\{ \s_3,\s_4,\ldots, \s_{n_u},\s_n \}$ are on the same branch cut. 

As a result, in order to develop the $\L-$algorithm at two loops, the key point that we must figure out is to know the behaviour of the factor, $1/E_{n+1}$, when $\L=0$.  From the gauge fixing $\{\s_{n+1},\s_{n+2},\s_{n+3},\s_{n+4} \}=\{\s_{\ell_1},\s_{-\ell_1},\s_{\ell_2},\s_{-\ell_2} \}$, we have three permissable configurations:
\begin{itemize}

\item {\bf $\s_{\ell_1}$ and $\s_{-\ell_1}$ on the same branch cut (upper).}\\
Without loss of generality, let us consider the punctures, $\{ \s_1,\ldots, \s_{n_u},\s_{\ell_1},\s_{-\ell_1}\}$ on the upper sheet, i.e. $y_i=\sqrt{\s_i^2-\L^2}$, $i\in\{1,\ldots, n_u, \ell_1, -\ell_1\}$, and the rest on the lower sheet, specifically $y_i=-\sqrt{\s_i^2-\L^2}$, $i\in\{n_u+1,\ldots, n, \ell_2, -\ell_2\}$. In this case, $E_{\ell_1}$ becomes
\begin{align}
&E_{n+1}\Big|_{\L=0}=E_{\ell_1}\Big|_{\L=0}=\sum_{j=1}^{n_u} {\ell_1\cdot k_j \over 2\,\s_{\ell_1j}}\left({\s_j\over \s_{\ell_1}}+1 \right)+
\sum_{j=n_u+1}^{n} {\ell_1\cdot k_j \over 2\,\s_{\ell_1j}}\left({-\s_j\over \s_{\ell_1}}+1 \right)
\nonumber\\
&~~+
 {\ell_1\cdot \ell_2+\frac{1}{2}(\ell_{1}^2 + \ell_{2}^2) \over 2\, \s_{\ell_1\ell_2}}\left({-\s_{\ell_2}\over \s_{\ell_1}}+1 \right)
 + 
{-\ell_1\cdot \ell_2-\frac{1}{2}(\ell_{1}^2 + \ell_{2}^2) \over 2\, \s_{\ell_1,-\ell_2}}\left({-\s_{-\ell_2}\over \s_{\ell_1}}+1 \right)\\
&=\sum_{j=1}^{n_u} {\ell_1\cdot k_j \over \s_{\ell_1j}}+
 {\ell_1\cdot k_0^{\rm upper} \over \s_{\ell_1}-\s_0},~~{\rm where} ~ k_0^{\rm upper}=k_{n_u+1}+\cdots + k_n, ~ {\rm and} ~\s_0=0.   \nonumber
\end{align} 
Using the scattering equations on the upper sheet, namely 
\begin{equation}
E_i\Big|_{\L=0}=\sum_{j\neq i}^{n_u}\frac{k_i\cdot k_j}{\s_{ij}} +\frac{k_i\cdot \ell_1}{\s_{i\ell_1}}+\frac{k_i\cdot (-\ell_1)}{\s_{i,-\ell_1}}+\frac{k_i\cdot k_0^{\rm upper}}{\s_i-\s_0}, ~~~i=1,\ldots, n_u,
\end{equation}
it is straightforward to verify the following identity
\begin{equation}
\sum_{i=1}^{n_u}\frac{(\s_i-\s_0)(\s_i-\s_{-\ell_{1}})}{(\s_{-\ell_{1}}-\s_0)}E_i+\frac{(\s_{\ell_1}-\s_0)(\s_{\ell_1}-\s_{-\ell_{1}})}{(\s_{-\ell_{1}}-\s_0)}E_{\ell_1}=k_{12\cdots n_u}.
\end{equation}
Therefore, on the support of the upper scattering equations $E_i=0,\, i=1,\ldots, n_u$, the factor $1/E_{\ell_1}$ is read as
\begin{equation}
\frac{1}{E_{\ell_1}}\Big|_{\L=0}=\frac{1}{k_{12\cdots n_u}}\times\left( \frac{\s_{\ell_{1}}\, \s_{\ell_1,-\ell_{1}}}{\s_{-\ell_1}} \right).
\end{equation}
In order to obtain the correct Faddeev--Popov determinant, as well on the upper as on the lower sheet  \cite{Gomez:2016bmv}, the term $\left( \frac{\s_{\ell_{1}}\, \s_{\ell_1,-\ell_{1}}}{\s_{-\ell_1}} \right)$  should be combined with the $\L-$expanssion of  $|-\ell_1,\ell_2,-\ell_2|\times \Delta_{\rm FP}(\ell_1,-\ell_1,\ell_2,-\ell_2)$.   Finally, we have achieved to the following rule 
\begin{equation}\label{ruleI}
\frac{1}{E_{\ell_1}}~~\rightarrow~~\frac{1}{k_{12\cdots n_u}},\qquad\textbf{ Rule\,I},
\end{equation}
where $\{ \s_1,\ldots, \s_{n_u},\s_{\ell_1},\s_{-\ell_1}\}$  are on the same branch cut. 
After this rule, the $\L-$algorithm can be performed in its usual way. 

It is important to remark that the punctures $\{\s_{\ell_1}, \s_{-\ell_1}\}$, or $\{\s_{\ell_2}, \s_{-\ell_2}\}$,  cannot be alone on the same branch cut, as it was discussed in  \cite{Cardona:2016wcr}. One can note that besides to this issue there is another one when the puntures,  $\{\s_{\ell_1}, \s_{-\ell_1}, \s_i\}$ with $k_i^2=0$,  are solely on the same branch cut. In this case, one should regularize the momentum conservation constraint and after checking that in fact this configuration vanishes. We will give an example of this subject.

\item {\bf $\s_{\ell_1}$ and $\s_{\ell_2}$ on the same branch cut (upper).}\\
Let us consider the punctures $\{ \s_1,\ldots, \s_{n_u},\s_{\ell_1},\s_{\ell_2}\}$ on the same sheet, for instance the upper sheet.
In this case, $E_{\ell_1}$ turns into
\begin{align}
&E_{n+1}\Big|_{\L=0}=E_{\ell_1}\Big|_{\L=0}=\sum_{j=1}^{n_u} {\ell_1\cdot k_j \over 2\,\s_{\ell_1j}}\left({\s_j\over \s_{\ell_1}}+1 \right)+
\sum_{j=n_u+1}^{n} {\ell_1\cdot k_j \over 2\,\s_{\ell_1j}}\left({-\s_j\over \s_{\ell_1}}+1 \right)
\nonumber\\
&~~+
 {\ell_1\cdot \ell_2+\frac{1}{2}(\ell_{1}^2 + \ell_{2}^2) \over 2\, \s_{\ell_1\ell_2}}\left({\s_{\ell_2}\over \s_{\ell_1}}+1 \right)
 + 
{-\ell_1\cdot \ell_2-\frac{1}{2}(\ell_{1}^2 + \ell_{2}^2) \over 2\, \s_{\ell_1,-\ell_2}}\left({-\s_{-\ell_2}\over \s_{\ell_1}}+1 \right)\\
&=\sum_{j=1}^{n_u} {\ell_1\cdot k_j \over \s_{\ell_1j}}+
 {\ell_1\cdot\ell_2 + \frac{1}{2}(\ell_{1}^2 + \ell_{2}^2) \over \s_{\ell_1 \ell_2}}
 +
 {\ell_1\cdot (k_{n_u+1}+\cdots + k_n) - \ell_1\cdot\ell_2- \frac{1}{2}(\ell_{1}^2 + \ell_{2}^2) \over \s_{\ell_1}-\s_0},   \nonumber
\end{align}
where $\s_0=0$.

Using the scattering equations on the upper sheet, namely 
\begin{equation}
E_i\Big|_{\L=0}=\sum_{j\neq i}^{n_u}\frac{k_i\cdot k_j}{\s_{ij}} +\frac{k_i\cdot \ell_1}{\s_{i\ell_1}}+\frac{k_i\cdot \ell_2}{\s_{i\ell_2}}+\frac{k_i\cdot k_0^{\rm upper}}{\s_i-\s_0}, ~~~i=1,\ldots, n_u,
\end{equation}
where $ k_0^{\rm upper}=k_{n_u+1}+\cdots + k_n+(-\ell_1)+(-\ell_2)$, one can verify the following identity
\begin{align}
&\sum_{i=1}^{n_u}\frac{(\s_i-\s_0)(\s_i-\s_{\ell_{2}})}{(\s_{\ell_{2}}-\s_0)}E_i+\frac{(\s_{\ell_1}-\s_0)(\s_{\ell_1}-\s_{\ell_{2}})}{(\s_{\ell_{2}}-\s_0)}E_{\ell_1}\nonumber\\
&=
\frac{1}{2}(\ell_{1}^2 + \ell_{2}^2) +\ell_1\cdot\ell_2+(\ell_1+\ell_2)\cdot(k_1+\cdots +k_{n_u})+k_{12\cdots n_u}\nonumber\\
&=
\frac{1}{2}(\ell_1 + \ell_2+k_1+\cdots k_n )^2.
\end{align}
Thus, on the support of the upper scattering equations, $E_i=0,\, i=1,\ldots, n_u$, we obtain the second rule
\begin{equation}\label{ruleII}
\frac{1}{E_{\ell_1}}~~\rightarrow~~\frac{1}{\frac{1}{2}(\ell_1 + \ell_2+k_1+\cdots k_n )^2},\qquad\textbf{ Rule\,II},
\end{equation}
and after this rule, the $\L-$algorithm can be performed in its usual way. 

\item {\bf $\s_{\ell_1}$ and $\s_{-\ell_2}$ on the same branch cut (upper).}\\
Following the same procedures described previously to get the {\bf Rules I} and {\bf II}, in \eqref{ruleI} and \eqref{ruleII}, it is straightforward to achieve the third rule
\begin{equation}\label{ruleIII}
\frac{1}{E_{\ell_1}}~~\rightarrow~~\frac{1}{-\frac{1}{2}(\ell_{1}^2 + \ell_{2}^2) -\ell_1\cdot\ell_2+(\ell_1-\ell_2)\cdot(k_1+\cdots +k_{n_u})+k_{12\cdots n_u}},\qquad\textbf{ Rule\,III},
\end{equation}
where we have used the support of the scattering equations
\begin{equation}
E_i\Big|_{\L=0}=\sum_{j\neq i}^{n_u}\frac{k_i\cdot k_j}{\s_{ij}} +\frac{k_i\cdot \ell_1}{\s_{i\ell_1}}+\frac{k_i\cdot (-\ell_2)}{\s_{i,-\ell_2}}+\frac{k_i\cdot k_0^{\rm upper}}{\s_i-\s_0}, ~~~i=1,\ldots, n_u,
\end{equation}
with $ k_0^{\rm upper}=k_{n_u+1}+\cdots + k_n+(-\ell_1)+\ell_2$ and $\s_0=0$.
\end{itemize}


\bibliographystyle{JHEP}
\bibliography{mybib}

\providecommand{\href}[2]{#2}\begingroup\raggedright\begin{thebibliography}{10}

\bibitem{Witten:2003nn}
E.~Witten, \emph{\emph{Perturbative gauge theory as a string theory in twistor
  space}}, \href{http://dx.doi.org/10.1007/s00220-004-1187-3}{\emph{Commun.
  Math. Phys.} {\bf 252} (2004) 189--258},
  [\href{http://arxiv.org/abs/hep-th/0312171}{{\tt hep-th/0312171}}].

\bibitem{Cachazo:2013gna}
F.~Cachazo, S.~He and E.~Y. Yuan, \emph{\emph{Scattering equations and
  Kawai-Lewellen-Tye orthogonality}},
  \href{http://dx.doi.org/10.1103/PhysRevD.90.065001}{\emph{Phys.Rev.} {\bf
  D90} (2014) 065001}, [\href{http://arxiv.org/abs/1306.6575}{{\tt
  1306.6575}}].

\bibitem{Cachazo:2013hca}
F.~Cachazo, S.~He and E.~Y. Yuan, \emph{\emph{Scattering of Massless Particles
  in Arbitrary Dimensions}},
  \href{http://dx.doi.org/10.1103/PhysRevLett.113.171601}{\emph{Phys.Rev.Lett.}
  {\bf 113} (2014) 171601}, [\href{http://arxiv.org/abs/1307.2199}{{\tt
  1307.2199}}].

\bibitem{Cachazo:2013iaa}
F.~Cachazo, S.~He and E.~Y. Yuan, \emph{\emph{Scattering in Three Dimensions
  from Rational Maps}},
  \href{http://dx.doi.org/10.1007/JHEP10(2013)141}{\emph{JHEP} {\bf 10} (2013)
  141}, [\href{http://arxiv.org/abs/1306.2962}{{\tt 1306.2962}}].

\bibitem{Cachazo:2013iea}
F.~Cachazo, S.~He and E.~Y. Yuan, \emph{\emph{Scattering of Massless Particles:
  Scalars, Gluons and Gravitons}},
  \href{http://dx.doi.org/10.1007/JHEP07(2014)033}{\emph{JHEP} {\bf 1407}
  (2014) 033}, [\href{http://arxiv.org/abs/1309.0885}{{\tt 1309.0885}}].

\bibitem{Cachazo:2014xea}
F.~Cachazo, S.~He and E.~Y. Yuan, \emph{\emph{Scattering Equations and
  Matrices: From Einstein To Yang-Mills, DBI and NLSM}},
  \href{http://arxiv.org/abs/1412.3479}{{\tt 1412.3479}}.

\bibitem{Cachazo:2014nsa}
F.~Cachazo, S.~He and E.~Y. Yuan, \emph{\emph{Einstein-Yang-Mills Scattering
  Amplitudes From Scattering Equations}},
  \href{http://dx.doi.org/10.1007/JHEP01(2015)121}{\emph{JHEP} {\bf 1501}
  (2015) 121}, [\href{http://arxiv.org/abs/1409.8256}{{\tt 1409.8256}}].

\bibitem{Cachazo:2016njl}
F.~Cachazo, P.~Cha and S.~Mizera, \emph{{Extensions of Theories from Soft
  Limits}}, \href{http://dx.doi.org/10.1007/JHEP06(2016)170}{\emph{JHEP} {\bf
  06} (2016) 170}, [\href{http://arxiv.org/abs/1604.03893}{{\tt 1604.03893}}].

\bibitem{Roiban:2004yf}
R.~Roiban, M.~Spradlin and A.~Volovich, \emph{{On the tree level S matrix of
  Yang-Mills theory}},
  \href{http://dx.doi.org/10.1103/PhysRevD.70.026009}{\emph{Phys. Rev.} {\bf
  D70} (2004) 026009}, [\href{http://arxiv.org/abs/hep-th/0403190}{{\tt
  hep-th/0403190}}].

\bibitem{Cachazo:2012da}
F.~Cachazo and Y.~Geyer, \emph{{A `Twistor String' Inspired Formula For
  Tree-Level Scattering Amplitudes in N=8 SUGRA}},
  \href{http://arxiv.org/abs/1206.6511}{{\tt 1206.6511}}.

\bibitem{Cachazo:2012kg}
F.~Cachazo and D.~Skinner, \emph{{Gravity from Rational Curves in Twistor
  Space}}, \href{http://dx.doi.org/10.1103/PhysRevLett.110.161301}{\emph{Phys.
  Rev. Lett.} {\bf 110} (2013) 161301},
  [\href{http://arxiv.org/abs/1207.0741}{{\tt 1207.0741}}].

\bibitem{Kawai:1985xq}
H.~Kawai, D.~C. Lewellen and S.~H.~H. Tye, \emph{{A Relation Between Tree
  Amplitudes of Closed and Open Strings}},
  \href{http://dx.doi.org/10.1016/0550-3213(86)90362-7}{\emph{Nucl. Phys.} {\bf
  B269} (1986) 1--23}.

\bibitem{Britto:2005fq}
R.~Britto, F.~Cachazo, B.~Feng and E.~Witten, \emph{\emph{Direct proof of
  tree-level recursion relation in Yang-Mills theory}},
  \href{http://dx.doi.org/10.1103/PhysRevLett.94.181602}{\emph{Phys.Rev.Lett.}
  {\bf 94} (2005) 181602}, [\href{http://arxiv.org/abs/hep-th/0501052}{{\tt
  hep-th/0501052}}].

\bibitem{Dolan:2013isa}
L.~Dolan and P.~Goddard, \emph{\emph{Proof of the Formula of Cachazo, He and
  Yuan for Yang-Mills Tree Amplitudes in Arbitrary Dimension}},
  \href{http://dx.doi.org/10.1007/JHEP05(2014)010}{\emph{JHEP} {\bf 1405}
  (2014) 010}, [\href{http://arxiv.org/abs/1311.5200}{{\tt 1311.5200}}].

\bibitem{Kalousios:2013eca}
C.~Kalousios, \emph{\emph{Massless scattering at special kinematics as Jacobi
  polynomials}},
  \href{http://dx.doi.org/10.1088/1751-8113/47/21/215402}{\emph{J.Phys.} {\bf
  A47} (2014) 215402}, [\href{http://arxiv.org/abs/1312.7743}{{\tt
  1312.7743}}].

\bibitem{Lam:2014tga}
C.~Lam, \emph{\emph{Permutation Symmetry of the Scattering Equations}},
  \href{http://dx.doi.org/10.1103/PhysRevD.91.045019}{\emph{Phys.Rev.} {\bf
  D91} (2015) 045019}, [\href{http://arxiv.org/abs/1410.8184}{{\tt
  1410.8184}}].

\bibitem{Cachazo:2016sdc}
F.~Cachazo and G.~Zhang, \emph{{Minimal Basis in Four Dimensions and Scalar
  Blocks}},  \href{http://arxiv.org/abs/1601.06305}{{\tt 1601.06305}}.

\bibitem{He:2016vfi}
S.~He, Z.~Liu and J.-B. Wu, \emph{{Scattering Equations, Twistor-string
  Formulas and Double-soft Limits in Four Dimensions}},
  \href{http://arxiv.org/abs/1604.02834}{{\tt 1604.02834}}.

\bibitem{Cachazo:2015nwa}
F.~Cachazo and H.~Gomez, \emph{{Computation of Contour Integrals on ${\cal
  M}_{0,n}$}}, \href{http://dx.doi.org/10.1007/JHEP04(2016)108}{\emph{JHEP}
  {\bf 04} (2016) 108}, [\href{http://arxiv.org/abs/1505.03571}{{\tt
  1505.03571}}].

\bibitem{Cachazo:2016ror}
F.~Cachazo, S.~Mizera and G.~Zhang, \emph{{Scattering Equations: Real Solutions
  and Particles on a Line}},  \href{http://arxiv.org/abs/1609.00008}{{\tt
  1609.00008}}.

\bibitem{Kalousios:2015fya}
C.~Kalousios, \emph{\emph{Scattering equations, generating functions and all
  massless five point tree amplitudes}},
  \href{http://dx.doi.org/10.1007/JHEP05(2015)054}{\emph{JHEP} {\bf 05} (2015)
  054}, [\href{http://arxiv.org/abs/1502.07711}{{\tt 1502.07711}}].

\bibitem{Dolan:2014ega}
L.~Dolan and P.~Goddard, \emph{\emph{The Polynomial Form of the Scattering
  Equations}}, \href{http://dx.doi.org/10.1007/JHEP07(2014)029}{\emph{JHEP}
  {\bf 1407} (2014) 029}, [\href{http://arxiv.org/abs/1402.7374}{{\tt
  1402.7374}}].

\bibitem{Huang:2015yka}
R.~Huang, J.~Rao, B.~Feng and Y.-H. He, \emph{\emph{An Algebraic Approach to
  the Scattering Equations}},  \href{http://arxiv.org/abs/1509.04483}{{\tt
  1509.04483}}.

\bibitem{Cardona:2015ouc}
C.~Cardona and C.~Kalousios, \emph{{Elimination and recursions in the
  scattering equations}},  \href{http://arxiv.org/abs/1511.05915}{{\tt
  1511.05915}}.

\bibitem{Cardona:2015eba}
C.~Cardona and C.~Kalousios, \emph{{Comments on the evaluation of massless
  scattering}},  \href{http://arxiv.org/abs/1509.08908}{{\tt 1509.08908}}.

\bibitem{Dolan:2015iln}
L.~Dolan and P.~Goddard, \emph{{General Solution of the Scattering Equations}},
   \href{http://arxiv.org/abs/1511.09441}{{\tt 1511.09441}}.

\bibitem{Sogaard:2015dba}
M.~Sogaard and Y.~Zhang, \emph{{Scattering Equations and Global Duality of
  Residues}},  \href{http://arxiv.org/abs/1509.08897}{{\tt 1509.08897}}.

\bibitem{Bosma:2016ttj}
J.~Bosma, M.~Sogaard and Y.~Zhang, \emph{{The Polynomial Form of the Scattering
  Equations is an H-Basis}},  \href{http://arxiv.org/abs/1605.08431}{{\tt
  1605.08431}}.

\bibitem{Zlotnikov:2016wtk}
M.~Zlotnikov, \emph{{Polynomial reduction and evaluation of tree- and
  loop-level CHY amplitudes}},  \href{http://arxiv.org/abs/1605.08758}{{\tt
  1605.08758}}.

\bibitem{Baadsgaard:2015ifa}
C.~Baadsgaard, N.~E.~J. Bjerrum-Bohr, J.~L. Bourjaily and P.~H. Damgaard,
  \emph{\emph{Scattering Equations and Feynman Diagrams}},
  \href{http://dx.doi.org/10.1007/JHEP09(2015)136}{\emph{JHEP} {\bf 09} (2015)
  136}, [\href{http://arxiv.org/abs/1507.00997}{{\tt 1507.00997}}].

\bibitem{Baadsgaard:2015voa}
C.~Baadsgaard, N.~E.~J. Bjerrum-Bohr, J.~L. Bourjaily and P.~H. Damgaard,
  \emph{\emph{Integration Rules for Scattering Equations}},
  \href{http://dx.doi.org/10.1007/JHEP09(2015)129}{\emph{JHEP} {\bf 09} (2015)
  129}, [\href{http://arxiv.org/abs/1506.06137}{{\tt 1506.06137}}].

\bibitem{Huang:2016zzb}
R.~Huang, B.~Feng, M.-x. Luo and C.-J. Zhu, \emph{{Feynman Rules of
  Higher-order Poles in CHY Construction}},
  \href{http://arxiv.org/abs/1604.07314}{{\tt 1604.07314}}.

\bibitem{Cardona:2016gon}
C.~Cardona, B.~Feng, H.~Gomez and R.~Huang, \emph{{Cross-ratio Identities and
  Higher-order Poles of CHY-integrand}},
  \href{http://dx.doi.org/10.1007/JHEP09(2016)133}{\emph{JHEP} {\bf 09} (2016)
  133}, [\href{http://arxiv.org/abs/1606.00670}{{\tt 1606.00670}}].

\bibitem{Mason:2013sva}
L.~Mason and D.~Skinner, \emph{\emph{Ambitwistor strings and the scattering
  equations}}, \href{http://dx.doi.org/10.1007/JHEP07(2014)048}{\emph{JHEP}
  {\bf 1407} (2014) 048}, [\href{http://arxiv.org/abs/1311.2564}{{\tt
  1311.2564}}].

\bibitem{Geyer:2015bja}
Y.~Geyer, L.~Mason, R.~Monteiro and P.~Tourkine, \emph{\emph{Loop Integrands
  for Scattering Amplitudes from the Riemann Sphere}},
  \href{http://dx.doi.org/10.1103/PhysRevLett.115.121603}{\emph{Phys. Rev.
  Lett.} {\bf 115} (2015) 121603}, [\href{http://arxiv.org/abs/1507.00321}{{\tt
  1507.00321}}].

\bibitem{Geyer:2015jch}
Y.~Geyer, L.~Mason, R.~Monteiro and P.~Tourkine, \emph{{One-loop amplitudes on
  the Riemann sphere}},
  \href{http://dx.doi.org/10.1007/JHEP03(2016)114}{\emph{JHEP} {\bf 03} (2016)
  114}, [\href{http://arxiv.org/abs/1511.06315}{{\tt 1511.06315}}].

\bibitem{Geyer:2016wjx}
Y.~Geyer, L.~Mason, R.~Monteiro and P.~Tourkine, \emph{{Two-Loop Scattering
  Amplitudes from the Riemann Sphere}},
  \href{http://arxiv.org/abs/1607.08887}{{\tt 1607.08887}}.

\bibitem{Cachazo:2015aol}
F.~Cachazo, S.~He and E.~Y. Yuan, \emph{{One-Loop Corrections from Higher
  Dimensional Tree Amplitudes}},  \href{http://arxiv.org/abs/1512.05001}{{\tt
  1512.05001}}.

\bibitem{He:2015yua}
S.~He and E.~Y. Yuan, \emph{{One-loop Scattering Equations and Amplitudes from
  Forward Limit}},
  \href{http://dx.doi.org/10.1103/PhysRevD.92.105004}{\emph{Phys. Rev.} {\bf
  D92} (2015) 105004}, [\href{http://arxiv.org/abs/1508.06027}{{\tt
  1508.06027}}].

\bibitem{Naculich:2014naa}
S.~G. Naculich, \emph{\emph{Scattering equations and BCJ relations for gauge
  and gravitational amplitudes with massive scalar particles}},
  \href{http://dx.doi.org/10.1007/JHEP09(2014)029}{\emph{JHEP} {\bf 1409}
  (2014) 029}, [\href{http://arxiv.org/abs/1407.7836}{{\tt 1407.7836}}].

\bibitem{Feng:2016nrf}
B.~Feng, \emph{{CHY-construction of Planar Loop Integrands of Cubic Scalar
  Theory}}, \href{http://dx.doi.org/10.1007/JHEP05(2016)061}{\emph{JHEP} {\bf
  05} (2016) 061}, [\href{http://arxiv.org/abs/1601.05864}{{\tt 1601.05864}}].

\bibitem{Chen:2016fgi}
T.~Wang, G.~Chen, Y.-K.~E. Cheung and F.~Xu, \emph{{A differential operator for
  integrating one-loop scattering equations}},
  \href{http://dx.doi.org/10.1007/JHEP01(2017)028}{\emph{JHEP} {\bf 01} (2017)
  028}, [\href{http://arxiv.org/abs/1609.07621}{{\tt 1609.07621}}].

\bibitem{Chen:2017edo}
G.~Chen, Y.-K.~E. Cheung, T.~Wang and F.~Xu, \emph{{A Combinatoric Shortcut to
  Evaluate CHY-forms}},  \href{http://arxiv.org/abs/1701.06488}{{\tt
  1701.06488}}.

\bibitem{Gomez:2016bmv}
H.~Gomez, \emph{{$\Lambda$ scattering equations}},
  \href{http://dx.doi.org/10.1007/JHEP06(2016)101}{\emph{JHEP} {\bf 06} (2016)
  101}, [\href{http://arxiv.org/abs/1604.05373}{{\tt 1604.05373}}].

\bibitem{Cardona:2016bpi}
C.~Cardona and H.~Gomez, \emph{{Elliptic scattering equations}},
  \href{http://dx.doi.org/10.1007/JHEP06(2016)094}{\emph{JHEP} {\bf 06} (2016)
  094}, [\href{http://arxiv.org/abs/1605.01446}{{\tt 1605.01446}}].

\bibitem{Cardona:2016wcr}
C.~Cardona and H.~Gomez, \emph{{CHY-Graphs on a Torus}},
  \href{http://dx.doi.org/10.1007/JHEP10(2016)116}{\emph{JHEP} {\bf 10} (2016)
  116}, [\href{http://arxiv.org/abs/1607.01871}{{\tt 1607.01871}}].

\bibitem{Baadsgaard:2015twa}
C.~Baadsgaard, N.~E.~J. Bjerrum-Bohr, J.~L. Bourjaily, S.~Caron-Huot, P.~H.
  Damgaard and B.~Feng, \emph{{New Representations of the Perturbative
  S-Matrix}},
  \href{http://dx.doi.org/10.1103/PhysRevLett.116.061601}{\emph{Phys. Rev.
  Lett.} {\bf 116} (2016) 061601}, [\href{http://arxiv.org/abs/1509.02169}{{\tt
  1509.02169}}].

\bibitem{Lam:2016tlk}
C.~S. Lam and Y.-P. Yao, \emph{{Evaluation of the Cachazo-He-Yuan gauge
  amplitude}}, \href{http://dx.doi.org/10.1103/PhysRevD.93.105008}{\emph{Phys.
  Rev.} {\bf D93} (2016) 105008}, [\href{http://arxiv.org/abs/1602.06419}{{\tt
  1602.06419}}].

\bibitem{Bjerrum-Bohr:2016axv}
N.~E.~J. Bjerrum-Bohr, J.~L. Bourjaily, P.~H. Damgaard and B.~Feng,
  \emph{{Manifesting Color-Kinematics Duality in the Scattering Equation
  Formalism}}, \href{http://dx.doi.org/10.1007/JHEP09(2016)094}{\emph{JHEP}
  {\bf 09} (2016) 094}, [\href{http://arxiv.org/abs/1608.00006}{{\tt
  1608.00006}}].

\bibitem{graph1}
J.~L. Gross and J.~Yellen, \emph{Graph Theory and Its Applications}.
\newblock Chapman and Hall, 2006.

\bibitem{graph2}
R.~Diestel, \emph{Graph Theory}.
\newblock Springer, third edition~ed., 2000.

\end{thebibliography}\endgroup
\end{document}